\documentclass[12pt]{article}
\usepackage{xcolor} 
\usepackage{tikz} 
\usepackage[utf8]{inputenc}
\usepackage{mathpazo}
\usepackage[ruled, algo2e]{algorithm2e}
\usepackage[section]{placeins}
\usepackage{multirow}
\usepackage{graphicx}
\usepackage{booktabs}
\usepackage{longtable}
\usepackage{algorithm}
\usepackage{algpseudocode}
\usepackage{framed}
\usepackage{lipsum}
\usepackage{mathtools, amsfonts}
\usepackage{siunitx}
\usepackage{amsmath, amssymb} 
\usepackage{bm} 
\usepackage[acronym]{glossaries} 

\usepackage{appendix}
\usepackage{afterpage}
\usepackage{array}
\usepackage{upgreek}
\usepackage{chngcntr}
\usepackage{bibentry}
\usepackage{longtable}
\usepackage{algpseudocode}
\usepackage{enumitem}

\newcolumntype{P}[1]{>{\centering\arraybackslash}p{#1}}
\SetKwProg{Fn}{Function}{}{}
\SetKwProg{Init}{initialise}{}{}
\SetKwComment{Comment}{/* }{ */}

\SetCommentSty{mycommfont}

\usepackage[colorlinks,allcolors=black,citecolor=blue,urlcolor=blue]{hyperref}
\usepackage[citestyle=apa,style=apa,backend=biber,url=false,uniquename=false, useprefix=true, doi=false]{biblatex}
\AtEveryBibitem{%
  \clearfield{note}%
  \clearfield{urlyear}
  \clearfield{urlmonth}
}
\usepackage{dsfont} 
\DefineBibliographyStrings{english}{%
  circa = {{}ca\adddot},
}

\SetKwProg{Fn}{Function}{}{}
\SetKwProg{Init}{initialise}{}{}
\SetKwComment{Comment}{/* }{ */}
\SetCommentSty{mycommfont}

\newcommand{\fundmat}{\vec{\Phi}(t, 0)}
\newcommand{\fundmats}{\vec{\Phi}(t, s)}
\newcommand{\fundmatso}{\vec{\Phi}(s, 0)}
\newcommand{\fundmatsu}{\vec{\Phi}(s, u)}
\newcommand{\fundmattv}{\vec{\Phi}(t, v)}

\newcommand{\boldbeta}{\boldsymbol{\beta}}

\newcommand{\Expec}{\mathbb{E}}

\newcommand{\Cov}{\operatorname{Cov}}
\newcommand{\diag}{\operatorname{diag}}
\newcommand{\Bias}{\operatorname{Bias}}

\newcommand{\boldeps}{\boldsymbol{\epsilon}}
\newcommand{\bc}{^{(bc)}}
\newcommand{\boldx}{\mathbf{x}}
\newcommand{\boldX}{\mathbf{X}}
\newcommand{\pkg}[1]{{\normalfont\fontseries{b}\selectfont #1}} \let\proglang=\textsf 
\title{An Understanding of Principal Differential Analysis}
\author{
Edward Gunning\thanks{Department of Biostatistics, Epidemiology and Informatics, University of Pennsylvania.}
\and 
Giles Hooker\thanks{Department of Statistics and Data Science, University of Pennsylvania.}}
\date{}

\begin{document}

\maketitle

\begin{abstract}
In functional data analysis, replicate observations of a smooth functional process and its derivatives offer a unique opportunity to flexibly estimate continuous-time ordinary differential equation models.
\textcite{ramsay_principal_1996} first proposed to estimate a linear ordinary differential equation from functional data in a technique called Principal Differential Analysis, by formulating a functional regression in which the highest-order derivative of a function is modelled as a time-varying linear combination of its lower-order derivatives. 
Principal Differential Analysis was introduced as a technique for data reduction and representation, using solutions of the estimated differential equation as a basis to represent the functional data.
In this work, we re-formulate PDA as a generative statistical model in which functional observations arise as solutions of a deterministic ODE that is forced by a smooth random error process.
This viewpoint defines a flexible class of functional models based on differential equations and leads to an improved understanding and characterisation of the sources of variability in Principal Differential Analysis.
It does, however, result in parameter estimates that can be heavily biased under the standard estimation approach of PDA. Therefore, we introduce an iterative bias-reduction algorithm that can be applied to improve parameter estimates.
We also examine the utility of our approach when the form of the deterministic part of the differential equation is unknown and possibly non-linear, where Principal Differential Analysis is treated as an approximate model based on time-varying linearisation.
We demonstrate our approach on simulated data from linear and non-linear differential equations and on real data from human movement biomechanics.
Supplementary \proglang{R} code for this manuscript is available at \url{https://github.com/edwardgunning/UnderstandingOfPDAManuscript}.
\end{abstract}

\section{Introduction}\label{sec:pda-intro}

\begin{figure}
	    \centering
	    \includegraphics[page=2,width=1\textwidth]{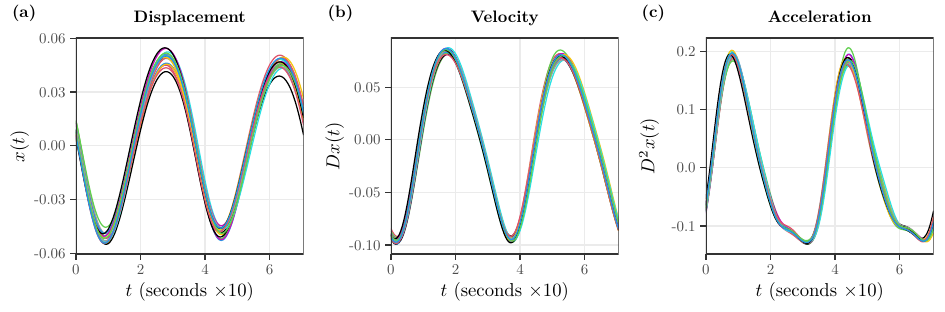}
		\caption{A sample of $20$ functional observations from a single runner collected during a treadmill run. Each observation corresponds to a single stride. \textbf{(a)} The vertical displacement of the runner's centre of mass in metres. \textbf{(b)} The vertical velocity of the runner's centre of mass. \textbf{(c)} The vertical acceleration of the runner's centre of mass. }
		\label{plot:running-data}
\end{figure}

One of the features unique to functional data analysis \parencite[FDA;][]{ramsay_functional_2005} is the ability to use derivatives in modelling. In FDA, each sampled observation is viewed as the realisation of a smooth functional process. Hence, each dataset provides replicate information about that process and its derivatives up to a suitable order $m$.
Figure \ref{plot:running-data} (a) displays the vertical displacement of a runner's centre of mass during a treadmill run, where each replicate function is a different stride. Figure 1 (b) and (c) contain the functions' first and second derivatives, which represent the velocity and acceleration, respectively, of the runner's centre of mass.
In many scientific fields, considerable interest lies in estimating ordinary differential equation (ODE) models that describe the relationship between a process and its derivatives to characterise its evolution over time \parencite{dattner_differential_2021}.

\textcite{ramsay_principal_1996} first proposed to utilise derivative information in FDA in an approach named \emph{Principal Differential Analysis} (PDA). In PDA, a time-varying linear ODE is estimated to describe the relationship between a sample of functions and their derivatives. An $m$th order ODE is constructed by modelling the $m$th-order derivative of a function as a time-varying linear combination of its lower-order derivatives. This formulation leads to a function-on-function regression model, i.e., a functional response variable (the $m$th order derivative) is regressed on a set of functional covariates (the lower-order derivatives), and falls naturally within the FDA framework. \textcite{ramsay_principal_1996} used PDA to estimate ODEs describing the force exerted by the thumb and forefinger during pinching and the motion of the lower lip during speaking. Since then, PDA has been used to study the dynamics of juggling \parencite{ramsay_functional_1999}, handwriting \parencite{ramsay_functional_2000, ramsay_functional_2009} and online-auction prices \parencite{wang_modeling_2008}. 

PDA was initially presented as a data-reduction technique. \textcite{ramsay_principal_1996} used the $m$ linearly independent solutions of the estimated ODE as basis functions to provide a low-dimensional representation of the observed functional data. From this perspective, PDA can be viewed as an alternative to functional principal component analysis (FPCA) because both techniques empirically determine a basis from the data \parencite[pp. 343-348]{ramsay_functional_2005}. The dimension reduction achieved by PDA may be preferable to FPCA in certain cases (e.g., the modelling of physical processes) because it takes the estimated relationship between derivatives into account.
In addition to data reduction, \textcite[p. 495]{ramsay_principal_1996} noted that the estimated ODE ``may also have a useful substantive interpretation". That is, the estimated parameters may be visualised individually or used collectively to qualitatively describe the system's behaviour. \textcite{ramsay_principal_1996} presented a brief stability analysis using the estimated parameters based on the theory of linear time-invariant ODEs which was further developed by \textcite{ramsay_functional_2009}. However, estimation of these parameters has received limited attention from a statistical perspective, so the properties of parameter estimates are not well understood.

In this work, we re-examine PDA as a generative statistical model in which the observations are solutions of a time-varying linear ODE that is forced by a smooth random error process.
This viewpoint facilitates a more complete characterisation of the sources of variability in PDA and provides a definite interpretation of the model parameters.
It allows us to define an additional set of basis functions to capture the variation in the data due to the smooth random error process.
However, it results in parameter estimates that can be severely biased.
Thus, we develop an iterative bias-reduction algorithm to reduce the bias in the parameter estimates.

Our second contribution is to examine the utility of PDA when the data arise from an unknown \emph{non-linear} ODE that is forced by a smooth random error process. Non-linear ODEs are required to describe many real-world phenomena, so assuming a known linear structure limits the applicability of PDA in practice. Therefore, we propose to view PDA as a time-varying linear approximation of an underlying non-linear ODE model, where the estimated parameters can be interpreted as elements of the system's Jacobian. This approach provides an understanding of PDA in settings where functional data are collected on a system but the form of the equations governing the system's dynamics is unknown. It also enables stability analysis in these settings, based on the theory of time-varying linearisation of non-linear systems. We demonstrate the approach on simulated data from a non-linear dynamical system and on kinematic data from human locomotion.

The remainder of this article is structured as follows. 
Section \ref{background} contains a brief description of the concepts from FDA and dynamical systems that underpin PDA and reviews PDA as initially proposed by \textcite{ramsay_principal_1996}. 
In Section \ref{sources-of-variation}, we build upon the original formulation of PDA as a data-reduction technique, extending it to account for a smooth stochastic disturbance to the underlying ODE and produce a generative statistical model.
Section \ref{sec:non-linear} presents a perspective of PDA as a linearised approximation to an unknown non-linear ODE model. Demonstrations of PDA on simulated data from ODE models and on real kinematic data from human locomotion are presented in Section \ref{examples}. We close with a discussion in Section \ref{sec:pda-discussion}.
\section{Background} \label{background}

\subsection{Functional Data and Derivatives}

We consider functional data $x_1 (t), \dots, x_N (t)$ which are realisations of a smooth random function $x(t)$ defined on an interval $[0, T] \subset \mathds{R}$. Often, $t$ represents time but it may also be, e.g., spatial distance \parencite{dallarosa_principal_2014}. We initially introduce $x(t)$ as a univariate function ($x: [0, T] \Rightarrow \mathds{R}$) and will use $\mathbf{x} (t)$ to represent multivariate (or vector-valued) functional data ($\mathbf{x}: [0, T] \Rightarrow \mathds{R}^d, \ d>1$) in later sections. A defining characteristic of FDA is that each sampled observation $x_i(t)$ is an entire function. FDA is based upon combining information between and within the $N$ replicate functions \parencites[][pp. 379-380]{ramsay_functional_2005}[][]{morris_functional_2015}.

It is also typically assumed that $x(t)$ possesses a number of derivatives. We use \emph{operator notation} $D^m x(t)$ to represent $\mathrm{d}^m x(t) / \mathrm{d}t^m$, which emphasises differentiation as an operator $D$ transforming $x(t)$ to produce a new function $D x (t)$ \parencites[][p. 20]{ramsay_functional_2005}[][p. 17]{ramsay_dynamic_2017}. Functional data then also provide replicate observations of $D^m x(t)$ up to some suitable order $m$. Information across replicates can be combined to estimate ordinary differential equation (ODE) models that describe relationships between $x(t)$ and one or more of its derivatives. This area is known as the \emph{dynamic modelling} (or \emph{dynamics}) of functional data \parencite[Chapters 17-19]{ramsay_functional_2005}.

\subsection{Ordinary Differential Equations}

We define an $m$th-order ODE for $x(t)$ as a relation between $t$, $x (t)$ and $D x(t), \dots,\allowbreak D^m x(t)$ and possibly some external ``forcing" function $f(t)$:
$$
D^m x(t) = g(t, x(t), D x(t), \dots, D^{m-1} x(t), f(t)).
$$
In applied mathematics, $x(t)$ is often called the ``state variable" of the ODE. When the form of the ODE does not depend explicitly on $t$ (i.e., $g(t, x(t),\allowbreak D x(t),\allowbreak \dots, \allowbreak D^{m-1} x(t), f(t))\allowbreak = \allowbreak g(x(t), D x(t), \dots, D^{m-1} x(t), f(t))$, we say that the ODE is time invariant (or autonomous) and when it does depend explicitly on $t$ we refer to it as time varying (or non-autonomous). For multivariate functions, we obtain systems of ODEs.

\begin{figure}
	    \centering
	    \includegraphics[page=1,width=1\textwidth]{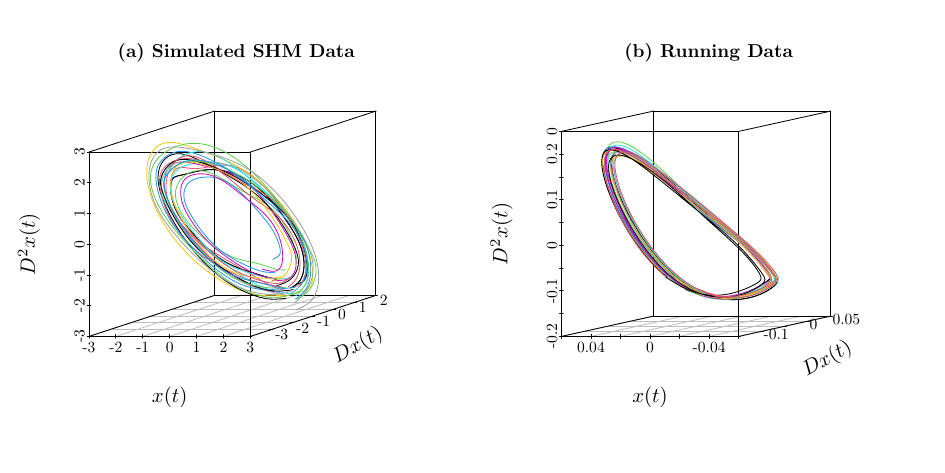}  
		\caption{Three-dimensional phase-plane plots. \textbf{(a)} Simulated data from the simple harmonic motion (SHM) model presented in Section \ref{sources-of-variation}. \textbf{(b)} The running data presented in Figure \ref{plot:running-data}.}
		\label{plot:phase-plane-3d}
\end{figure}

As an initial exploratory step, dynamics can be examined graphically by plotting $x(t)$ against one or more of its derivatives to produce a \emph{phase-plane plot} \parencites[pp. 13-14, 29-34]{ramsay_functional_2005}[][]{ramsay_functional_2002}.
Figure \ref{plot:phase-plane-3d} (a) displays a three-dimensional phase-plane plot of data simulated from the simple harmonic motion (SHM) model, defined by the second-order ODE
\begin{equation} \label{SHM-deterministic}
    D^2 x(t) = - x(t).
\end{equation}
We have added some smooth random noise to the ODE for each observation (the consequences of doing so are examined in Section \ref{sources-of-variation}).
As expected, the linear relationship between $D^2 x(t)$ and $x(t)$ is evident. 
Moreover, because solutions to \eqref{SHM-deterministic} are given by linear combinations of $\sin(t)$ and $\cos(t)$ functions, the observations exhibit almost-circular orbits in the $(x(t) , D x(t))$ and $(D x(t) , D^2 x(t))$ planes.
In practice, dynamic relations are likely to be more complex than the SHM model -- they can be non-linear, multivariate, involve more than two derivatives and vary with $t$. Figure \ref{plot:phase-plane-3d} (b) displays a three-dimensional phase-plane plot of the running data introduced in Figure \ref{plot:running-data}. The dynamics bear a resemblance to those seen in the simulated SHM data, i.e., a negative relationship between $D^2 x(t)$ and $x(t)$ and periodic orbits in the $(x(t) , D x(t))$ and $(D x(t) , D^2 x(t))$ planes. 
However, curved features are also visible and there is a departure from a purely circular orbit, which might indicate the presence of more complex behaviour (e.g., non-linearity) in certain places.

\subsection{Principal Differential Analysis}

\subsubsection{Estimating a Differential Equation From Functional Data}

\textcite{ramsay_principal_1996} first introduced PDA as estimating the linear time-varying ODE of order $m$
\begin{equation} \label{linear-ode-pda}
D^m x(t) = - \beta_0 (t) x(t) - \dotso - \beta_{m-1} (t) D^{m-1} x(t) + f(t).
\end{equation}
The ODE is linear because $D^m x(t)$ is a linear function of $x(t), \dots, D^{m-1} x(t)$ and time varying because the coefficients $\beta_0 (t), \dots, \beta_{m-1} (t)$ are functions of $t$. The ODE is \emph{homogeneous} if the forcing function $f(t) = 0$ and \emph{nonhomogeneous} if $f(t) \neq 0$. The forcing function represents an external input to the system defined by the homogeneous part of the equation \parencite{ramsay_principal_2006}.

Given $x_1 (t), \dots, x_N (t)$, their derivatives up to order $m$ and associated measurements of the forcing functions\footnote{It is possible that a common forcing function is assumed or known, in which case $f_i (t) = f(t), \forall i$.}\footnote{If a function $\eta(t)$ is measured and the external forcing effect is of the form$f(t) = \alpha (t) \eta(t)$, where $\alpha (t)$ is unknown, it can be estimated along with the other parameters in the model by removing $f_i(t)$ from the LHS of the equation and adding $\alpha (t) \eta_i(t)$ to the RHS.} $f_1(t), \dots, f_N (t)$, the ODE model can be formulated as a \emph{functional concurrent linear regression model} \parencite[FCLM;][]{ramsay_functional_2005}
\begin{equation}\label{eq:fclm-pda}
    D^m x_i (t) - f_i(t)  = - \beta_0 (t) x_i(t) - \dotso - \beta_{m-1} (t) D^{m-1}x_i(t) + \epsilon _i (t).
\end{equation}
That is, the functional response variable $D^m x(t) - f(t)$ only depends on the functional covariates $x(t),\dots, D^m x(t)$ at their ``current" time $t$ through the regression coefficient functions $\beta_0(t), \dots, \beta_m(t)$. Note that unlike conventional FCLM models, model \eqref{eq:fclm-pda} does \emph{not} include an intercept function.
The smooth random error function $\epsilon_i (t)$ represents the lack of fit of the $i$th replicate observation to the differential equation (we expand on the importance of this term in the next section).

The PDA model is typically estimated using (penalised) ordinary least squares (OLS) approaches, which aim to minimise the integrated sum of squared errors
\begin{equation}\label{eq:ISSE}
    \text{ISSE} = \sum_{i=1}^N \int_0^T \epsilon_i(t)^2\mathrm{d}t,
\end{equation}
often with a regularisation penalty added to enforce smoothness of the estimated functional response or regression coefficient functions \parencite[pp. 339-240]{ramsay_functional_2005}.
In this way, PDA can be seen as estimating the ``closest fitting" ODE to the data (adhering to certain smoothness constraints).
When the functional data are smoothed and observed on a fine grid, the criterion \eqref{eq:ISSE} can be minimised \emph{pointwise} by computing the standard OLS solution to \eqref{eq:fclm-pda} at each $t$ and smoothing or interpolating the pointwise parameter estimates to produce smooth regression coefficient functions \parencites[p. 236; pp. 338-339]{ramsay_functional_2005}{fan_two-step_2000}.
Alternatively, the functional data and the regression coefficient functions can both be represented by basis function expansions (e.g., B-splines, Fourier), so that the problem reduces to estimating the basis coefficients of the regression coefficient functions. In this case, the integral \eqref{eq:ISSE} can be computed numerically and penalties can be added to regularise the regression coefficient functions \parencite[pp. 255-256]{ramsay_functional_2005}.
It is also possible to fix certain $\beta_j (t) = 0$ by dropping $D^j x(t)$ from the model or formulate it such that the coefficients are constant (i.e., $\beta_j (t) = \beta_j, \ \forall t \in [0, T]$). Such choices may be implied by physical interpretations of the ODE \parencite{ramsay_functional_2000}.

\subsubsection{Data Reduction Based on ODE Solutions}

An $m$th-order linear ODE has $m$ linearly independent functions as solutions, i.e., any function $x(t)$ that satisfies the ODE in Equation \eqref{linear-ode-pda} can be written exactly as a weighted sum of these functions.
\textcite{ramsay_principal_1996} proposed to solve the ODE estimated in PDA to obtain estimates of these solutions, and then use them as basis functions to represent the observed functional data. 
More specifically, solutions of Equation \eqref{linear-ode-pda} can be computed by re-writing the $m$th-order ODE as a system of $m$ first-order ODEs. This is achieved by creating the $m$-vector function 
$$
\Tilde{\mathbf{x}} (t)
= 
\begin{pmatrix}
x(t) \\
D x(t) \\
\vdots \\
D^{m-1} x(t)
\end{pmatrix},
$$
leading to the system of first-order ODEs
$$
D \Tilde{\mathbf{x}} (t) = \mathbf{B} (t) \Tilde{\mathbf{x}} (t) + \Tilde{\mathbf{f}}(t),
$$
where
$$
\mathbf{B} (t) = \begin{pmatrix}
0 & 1 & \hdots & 0 \\
0 & 0 & \hdots & 0 \\
\vdots & \vdots & \vdots \ \vdots \ \vdots & \vdots \\
0 & 0 & \hdots & 1 \\
- \beta_0 (t) & - \beta_1 (t) & \hdots & - \beta_{m-1} (t)
\end{pmatrix},
$$
and $\Tilde{\mathbf{f}}(t) = (0, 0, \dots, f(t))^\top$. Then, solutions of the system are given by
\begin{equation} \label{linear-ode-solution}
    \Tilde{\mathbf{x}} (t) = \fundmat \ \Tilde{\mathbf{x}}_0 + \int _{0} ^ {t} \fundmats \Tilde{\mathbf{f}}(s)\mathrm{d}s,
\end{equation}
where $\Tilde{\mathbf{x}}_0 = (x(0), Dx(0), \dots, D^{m-1} x(0))^\top$ is a vector of initial conditions for $\Tilde{\mathbf{x}}(t)$ and $\fundmats$ is the \emph{state transition matrix} (or \emph{fundamental matrix}) \parencites[p. 19]{kirk_optimal_2012}[p. 56]{bressan_introduction_2007}. 

The state transition matrix represents how the state variable $\mathbf{\Tilde{x}}$ evolves from time $s$ to current time $t$ according to the homogeneous part of the ODE. For time-invariant systems (i.e., $\mathbf{B}(t)= \mathbf{B}$), the state transition matrix can be expressed as $\fundmats = e^{(t-s) \mathbf{B}}$, where
\begin{equation}\label{matexp}
    e^{t\mathbf{B}} = \sum_{k=0}^{\infty}\frac{t^k \mathbf{B}^k}{k!}.
\end{equation}
No analogous expression exists for time-varying systems, with the exception of first-order systems. Therefore, numerical methods are used to estimate the state transition matrix as the solution of the matrix ODE
$$
D \fundmats = \mathbf{B} (t) \fundmats
$$
with initial conditions $\vec{\Phi}(s, s) = \mathbf{I}_m$. In other words, the homogeneous part of the ODE is solved $m$ times with starting time $s$ and initial values $(1, \dots, 0)^\top, \dots, (0, \dots, 1)^\top$ to produce the columns of the state transition matrix.

Restricting attention to the homogeneous case (i.e., $f(t)=0$), \textcite{ramsay_principal_1996} proposed to use the elements of the state transition matrix as basis functions to represent the observed functional data (and their derivatives up to order $m-1$). 
We denote the $m$ solutions for $x(t)$, contained in the first row of $\fundmat$, by $u_1 (t), \dots, u_m (t)$, so that
$$
\mathbf{U}(t) = \begin{pmatrix}
u_1(t) & \dots & u_m (t) \\
D u_1(t) & \dots & D u_m (t)\\
\vdots  & \vdots & \vdots\\
D^{m-1} u_1(t) & \dots & D^{m-1} u_m(t)
\end{pmatrix} = \fundmat.
$$
Then, the basis representation for each individual functional observation $x_i(t)$ is given by the linear combination (or weighted sum)
\begin{equation} \label{pda-basis-function-representation}
x_i (t) = \sum_{k=1}^m x_{ik}^* u_k (t),
\end{equation}
where $x_{ik}^*$ are scalar coefficients.
In practice, estimates $\widehat{u}_1 (t), \dots, \widehat{u}_m (t)$ of $u_1 (t), \dots, u_m (t)$ are obtained by solving the ODE estimated by PDA, and the scalar basis coefficients $x_{ik}^*$ are estimated by OLS regression of the observed curves on the estimated basis functions.

The basis $\{\widehat{u}_k(t)\}_{k=1}^m$ is estimated directly from the data and provides a low-dimensional representation of each functional observation, hence it can be seen as similar, in some ways, to FPCA.
In contrast to FPCA, however, the basis has the interpretation of spanning the space of functions satisfying the closest fitting ODE.
More specifically, we can view them as capturing variation in the functional data due to \emph{initial conditions}. 
This can be seen by writing the solution of the (homogeneous) ODE as
$$
\mathbf{\Tilde{x}} (t) = \fundmat \ \mathbf{\Tilde{x}}_0 = \mathbf{U}(t) \ \mathbf{\Tilde{x}}_0,
$$
so that $x(t), \dots, D^{m-1}x(t)$ are linear combinations of the canonical basis functions and their derivatives, where the basis coefficients are the initial conditions\footnote{In practice, because each functional observation will not satisfy the ODE exactly, the initial values for each function will not match the estimated basis coefficients perfectly.}.
This perspective provides a starting point for viewing PDA as a generative statistical model that captures sources of variability in functional data according to an underlying ODE model.
However, as we demonstrate in the next section, another source of variability -- the lack of fit of each observation to the ODE -- needs to be accounted for to provide a complete characterisation.
\section{A Generative Model View of PDA} \label{sources-of-variation}

In addition to variation in initial conditions, we also observe variation in the fit to the ODE.
We restrict our focus to the homogeneous ODE
$$
D^m x(t) = - \beta_0 (t) x(t) - \dotso - \beta_{m-1} (t) D^{m-1} x(t),
$$
and the associated FCLM to be estimated using PDA
\begin{equation} \label{FCLM-PDA-homo}
    D^m x_i (t) = - \beta_0 (t) x_i(t) - \dotso - \beta_{m-1} (t) D^{m-1}x_i(t) + \epsilon _i (t).
\end{equation}
It has been suggested that the error term $\epsilon _i (t)$, representing the lack of fit of the $i$th observation to the ODE, can be viewed as a ``forcing" function \parencite{ramsay_principal_2006}. 
In the following sections, we demonstrate that if $\epsilon_i (t)$ is treated in this way -- as a \emph{smooth stochastic disturbance} to the ODE and hence a source of variation in $x_i (t)$ -- then:
\begin{enumerate}
    \item We obtain a complete characterisation of the sources of variability through a generative model for our data based on the underlying ODE and the smooth stochastic disturbance, 
    \item This enables us to derive a new set of basis functions, complementary to the canonical PDA basis functions, that capture variation in the data due to the stochastic disturbance, and
    \item From this perspective, the standard OLS estimation approach to PDA produces biased parameter estimates due to dependence among the covariates and error term in (\ref{FCLM-PDA-homo}), but the bias can be reduced with our proposed bias-reduction step.
\end{enumerate}

As a starting point for our generative model, we formally encode the smooth stochastic disturbance by specifying that the errors in \eqref{FCLM-PDA-homo} are independent copies of a zero-mean Gaussian process
$$
\epsilon (t) \sim \mathcal{GP}(0, C),
$$
where $C(s, t)$ is a smooth auto-covariance function. Smoothness in $C$ ensures smoothness in $D^m x(t)$ as well as satisfying conditions for numerical ODE solvers.
The smoothness assumption is important because, in practice, $x(t)$ is typically estimated from noisy measurements in a pre-smoothing step and then $Dx(t),\dots,D^{m}x(t)$ are calculated, where the pre-smoothing techniques assume smoothness of $x(t)$ and a number of its derivatives.
The Gaussian assumption for $\epsilon (t)$ is conventional in functional-response regression models.
However, it is not strictly necessary for the methodology presented hereafter, as only the first and second moments of the process are used.
Additionally, we assume that the stochastic disturbance is independent of initial conditions, which arise from an $m$-dimensional Gaussian distribution
$$
\Tilde{\mathbf{x}}_0 = \begin{pmatrix}
x (0) \\
\vdots \\
D^{m-1} x(0)
\end{pmatrix}
\sim
\mathcal{N}_m (\boldsymbol{\mu}_0, \boldsymbol{\Sigma}_0).
$$
Thus, we obtain a generative model for $x(t), \dots, D^{m-1} x(t)$ given by the ODE solution (\ref{linear-ode-solution}) (replacing $f(t)$ with $\epsilon (t)$) 
\begin{equation} \label{generative-ode-solution}
    \Tilde{\mathbf{x}} (t) = \fundmat \ \Tilde{\mathbf{x}}_0 + \int _{0} ^ {t} \fundmats \Tilde{\boldsymbol{\epsilon}}(s) \mathrm{d}s,
\end{equation}
where $\Tilde{\boldsymbol{\epsilon}}(t) = (0, \dots, 0, \ \epsilon (t))$.

\begin{figure}[h]
	    \centering
		\includegraphics[page=2,width=0.8\textwidth]{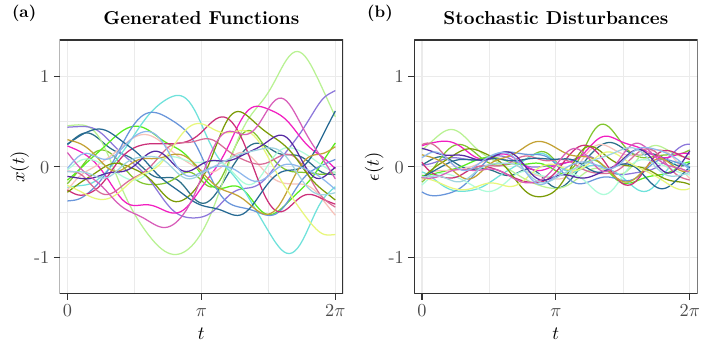}  
		\caption{20 simulated observations from model (\ref{SHM-data-generating-model}) with $\sigma = 0.25$ and $(x(0), \ Dx(0))^\top \sim \mathcal{N}_2(\mathbf{0}, \ 0.05 \ \mathbf{I}_2)$. \textbf{(a)}  The generated functions $x(t)$. \textbf{(b)} The stochastic disturbances $\epsilon(t)$.} 
		\label{plot:shm-dataset}
\end{figure}

The generative model \eqref{generative-ode-solution} defines a new class of flexible functional models based jointly solutions to an underlying deterministic ODE and a smooth stochastic disturbance.
As an example, we present a version of model \eqref{generative-ode-solution} where the deterministic part of the ODE is defined according to the well-known SHM equation and the smooth stochastic disturbance comes from a smooth Gaussian covariance kernel\footnote{This example involves a time-invariant ODE and a stationary stochastic process for the smooth stochastic disturbance. Although it is useful for demonstrating our proposed data-generating model, in general we do not assume time invariance of the underlying ODE or stationarity (or any parametric form) for the covariance of the stochastic disturbance. 
The ``replication" characteristic of functional data (i.e., that we have replicate observations of \emph{each} function and their derivatives at \emph{each} $t \in [0, T]$) enables the estimation of time-varying systems and unstructured auto-covariance functions.}
\begin{equation} \label{SHM-data-generating-model} 
    D^2 x(t) = - x(t) + \epsilon (t),
\end{equation}
for $t \in [0, 2 \pi]$ and where $\epsilon (t) \sim \mathcal{GP}(0,C)$ with $C(s, t) = C(|t-s|) = \sigma^2\phi(l \ |t-s|)$ where $\phi$ is a standard Gaussian density. For demonstration, we fix the parameters $l = 2$ and $\sigma = 0.25$.
To generate the functions, we first simulate $\epsilon (t)$ and then compute $x(t)$ numerically as the solution of (\ref{SHM-data-generating-model}) with initial conditions drawn, independently of $\epsilon (t)$, from a bivariate normal distribution $\mathcal{N}_2(\boldsymbol{\mu}_0, \boldsymbol{\Sigma}_0)$, with $\boldsymbol{\mu}_0 = \mathbf{0}$ and $\boldsymbol{\Sigma}_0 = 0.05 \  \mathbf{I}_2$. Figure \ref{plot:shm-dataset} shows 20 simulated observations from this setting. 


\subsection{Basis Functions}

We examine the moments of our generative model \eqref{generative-ode-solution}, allowing us to augment the canonical basis produced by PDA to fully capture all sources of variation in the data.
The expectation of model (\ref{generative-ode-solution}) is
$$
\Expec[\Tilde{\mathbf{x}} (t)] = \fundmat  \boldsymbol{\mu}_0 = \mathbf{U} (t)  \boldsymbol{\mu}_0,
$$
which shows that the canonical PDA basis functions are sufficient to represent the mean function. However, the covariance is
$$
\Cov(\Tilde{\mathbf{x}} (s), \Tilde{\mathbf{x}} (t)) 
$$
$$
=  \fundmatso \boldsymbol{\Sigma}_0 \fundmat^\top + \   \int_0^s \int_0^t \fundmatsu \mathbf{C}(u, v) \fundmattv^\top \mathrm{d}v \mathrm{d}u \ 
$$
$$
= \underbrace{\mathbf{U} (s) \boldsymbol{\Sigma}_0 \mathbf{U} (t) ^\top}_{\text{Variation due to initial conditions}} \ + \ \ \underbrace{\int_0^s \int_0^t \fundmatsu \mathbf{C}(u, v) \fundmattv^\top \mathrm{d}v \mathrm{d}u}_{\text{Variation due to stochastic disturbance}}.
$$
The first term represents variation due to initial conditions and can be written in terms of the canonical PDA basis functions. The second term captures variation due to the stochastic disturbance. 
Drawing on terminology from control theory, we may call the first term the ``zero-input covariance" and the second term the ``zero-state covariance". 
The decomposition illustrates that the stochastic disturbance must be accounted for to fully describe variation around the mean function. We propose to use the eigenfunctions of the zero-state covariance as a basis to capture variation due to the stochastic disturbance. 
Only estimates of $\beta_0 (t), \dots, \beta_{m-1} (t)$ and $C(s, t)$, which are calculated during PDA, are required to compute the eigenfunctions.

\begin{figure}
	    \centering
		\includegraphics[page=7,width=1\textwidth]{Figures/Understanding-of-PDA-Figures.pdf}
        \centering
        \includegraphics[width=0.9\textwidth]{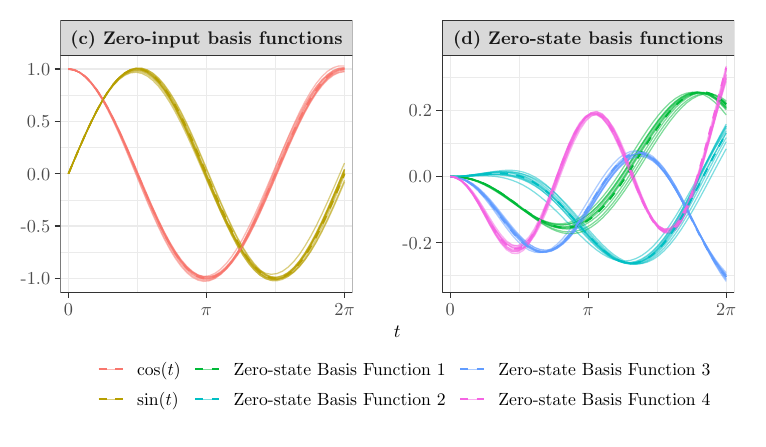}      
		\caption{\textbf{Top row:} The two covariance surfaces comprising $\Cov(x(s), x(t))$ for the example SHM model (\ref{SHM-data-generating-model}). \textbf{(a)} The zero-input covariance. \textbf{(b)} The zero-state covariance. \textbf{Bottom row}: Basis functions of the two covariance surfaces. Estimated basis functions, calculated from 10 simulated datasets of size $N=500$ from model (\ref{SHM-data-generating-model}) with three iterations of the bias correction applied, are represented by solid lines. The true functions are overlaid as dashed lines. \textbf{(c)} The zero-input basis functions. \textbf{(d)} The zero-state basis functions.
  }
		\label{plot:SHM-covariance-surfaces}
\end{figure}

The top panel of Figure \ref{plot:SHM-covariance-surfaces} shows the true zero-state and zero-input covariance functions for the simple second-order SHM model (\ref{SHM-data-generating-model}). The zero-input covariance is a periodic covariance function with stationary variance as it is constructed from the canonical basis functions $\sin(t)$ and $\cos(t)$ and a diagonal covariance matrix of initial conditions. In contrast, the zero-state covariance starts at $0$ for $t = s = 0$ and varies as it moves away from the origin, reflecting that it is constructed from a double integral, starting at zero, to $s$ and $t$. The bottom panel of Figure \ref{plot:SHM-covariance-surfaces} shows the basis functions for each covariance term: (c) contains the canonical PDA basis functions obtained from the state transition matrix, and (d) contains the first four eigenfunctions of the zero-state covariance.

\begin{figure}
	    \centering
		\includegraphics[width=1\textwidth]{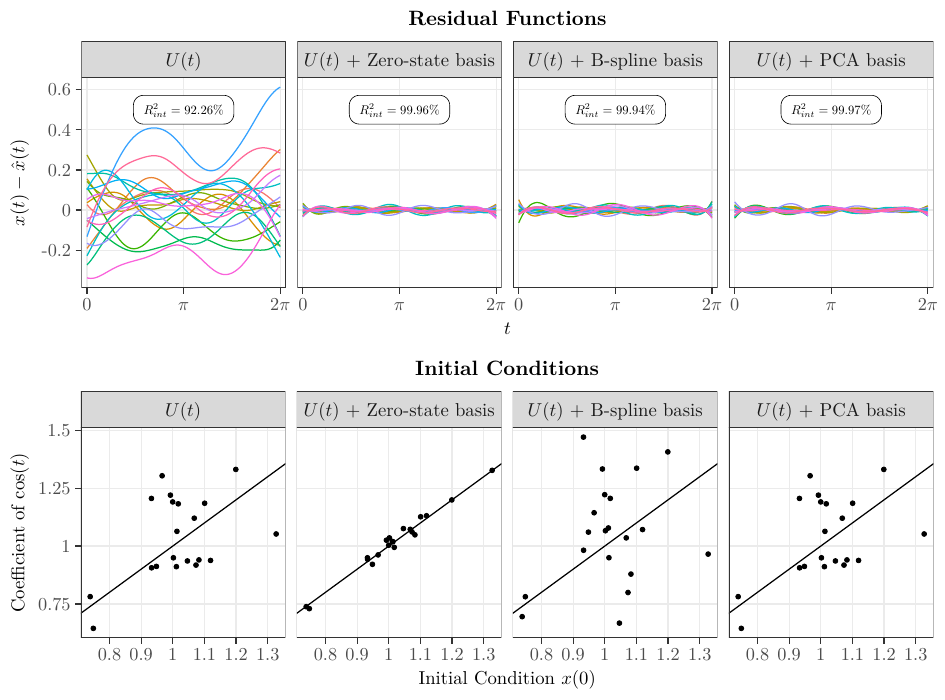}
		\caption{The results of representing the 20 example functions in Figure \ref{plot:shm-dataset} (a) as linear combinations of four different bases: 1) The canonical PDA basis functions $\cos(t)$ and $\sin(t)$, 2) the proposed basis of the canonical basis functions and the first four eigenfunctions of the zero-state covariance function, 3) the canonical basis functions and four cubic B-spline basis functions and 4) the canonical basis functions and the first four functional principal components of the residuals from the initial fit to the canonical basis functions. OLS regression was used to estimate the coefficients of the linear combination. \textbf{Top panel:} The residuals from each of the fits. The $R^2_{int}$ value is defined as $\frac{1}{2 \pi} \int_0^{2 \pi} R^2 (t) \mathrm{d}t$ where $R^2 (t) = \Sigma_{i=1}^N \left(x_i(t)-\widehat{x}_i (t)\right)^2 \ / \  \Sigma_{i=1}^N x_i(t)^2$. \textbf{Bottom panel:} The estimated coefficient of the first canonical basis function $\cos(t)$ plotted against the initial value $x(0)$.}
		\label{plot:SHM-bases-fits}
\end{figure}

To highlight the utility of our proposed basis, we represent the sample of 20 functions presented earlier in Figure \ref{plot:shm-dataset} (a) as linear combinations of 1) the canonical PDA basis functions, and 2) our proposed basis of both the canonical basis functions and the eigenfunctions of the zero-state covariance matrix. We use the first four eigenfunctions of the zero-state covariance for the demonstration\footnote{We use the true basis functions, rather than estimates, for the purpose of this demonstration.}. Consequently the combined basis has four extra basis functions, so we also use two other bases for a fairer comparison, combining the canonical basis functions with: 3) four cubic B-spline basis functions, and 4) the first four functional principal components of the residuals from the initial fit to the canonical basis functions. The results are shown in Figure \ref{plot:SHM-bases-fits}.

The residuals from the fits are shown in the top panel of Figure \ref{plot:SHM-bases-fits} -- large residuals in the first plot indicate that the canonical basis functions alone are not sufficient to represent the variation in the data. The additional variation, induced by the stochastic disturbance, is captured by the more flexible bases. On the bottom panel, we show the coefficients of the first canonical PDA basis function plotted against the initial values. Under the model (\ref{SHM-data-generating-model}), the coefficient of the first canonical basis function $\cos(t)$ should represent the initial value $x(0)$. These two quantities align most closely for our proposed basis (the second plot), thus it partitions the sources of variation in the data most accurately.

\subsection{Bias} \label{sec:bias}

We now consider the problem of reliably estimating the parameters in our generative model \eqref{generative-ode-solution}.  
In this section, we show that under our proposed data-generating model, the conventional exogeneity assumption in linear regression models (i.e., that covariates are orthogonal to the random error) that guarantees unbiased and consistent OLS parameter estimates is violated.
This arises because of the dependence among $x(t), \dots, D^{m-1} x(t)$ and $\epsilon (t)$, since $\epsilon (t)$ appears in the generative model for $x(t)$ and its derivatives.
We demonstrate, using simulated data from the model \eqref{SHM-data-generating-model}, how this bias arises and we propose an iterative bias-reduction algorithm to improve parameter estimates.

\begin{figure}[h]
	    \centering
	    \includegraphics[page=3,width=1\textwidth]{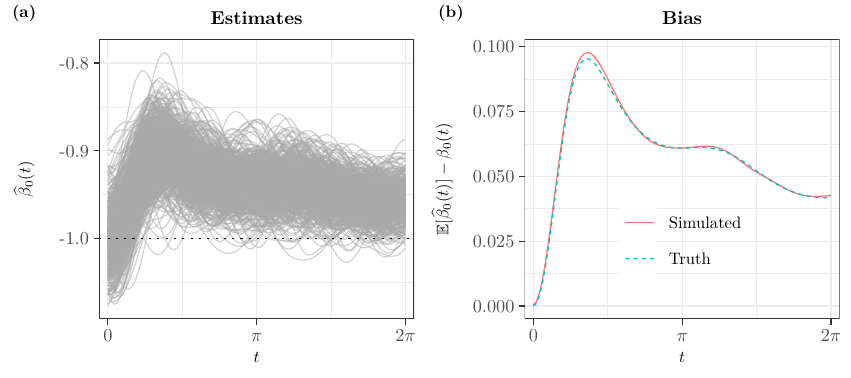}  
		\caption{\textbf{(a)} Estimates $\widehat{\beta}_0 (t)$ obtained from 500 generated datasets of size $N=500$. The dotted black line represents the estimand $\beta_0(t) = - 1$. \textbf{(b)} The estimate of the bias from the simulation (solid orange line) and the true bias computed numerically (dashed blue line).}
		\label{plot:bias-beta}
\end{figure}

For this demonstration, we generate $500$ datasets of size $N = 500$ from model \eqref{SHM-data-generating-model}, with all other parameters as defined thereafter.
Under the standard approach to PDA, the model we fit to the data is
$$
D^2 x(t) = \beta_0(t) x(t) + \epsilon (t),
$$
so the constant function $\beta_0 (t) = - 1$ is the target of the estimation. The OLS estimator for this single-parameter model is
$$
\widehat{\beta}_0 (t) = \frac{\sum_{i=1}^N x_i (t) D^2 x_i (t)}{\sum_{i=1}^N x_i (t) ^2} = \beta_0 (t) + \frac{\sum_{i=1}^N x_i (t) \epsilon _i(t)}{\sum_{i=1}^N x_i (t) ^2 },
$$
so the bias depends on the expectation of the second term.
Figure \ref{plot:bias-beta} (a) shows estimates of $\beta_0 (t)$ from the simulation. The estimates reflect that the OLS estimator is biased -- they are not centered around the dotted line $\beta_0 (t) = - 1$. Figure \ref{plot:bias-beta} (b) shows the bias obtained via simulation and the true bias (i.e., the expectation of Equation \eqref{eq:conditional-bias}) computed numerically.

Our solution is to use the ODE solution to estimate the unknown part of the bias. Given $x_1 (t), \dots, x_N (t)$, the conditional bias is 
\begin{equation}\label{eq:conditional-bias}
    \text{Bias}(\widehat{\beta}_0 (t) | x_1(t), \dots, x_N (t)) = \frac{\frac{1}{N} \sum_{i=1}^N x_i(t) \Expec[\epsilon _i(t) | x_i (t)]}{ \frac{1}{N} \sum_{i=1}^N x_i (t) ^2 }.
\end{equation}
Because the numerator is unknown, we replace it with an estimate of its expectation $\Expec[x(t)\epsilon(t)]$. By re-writing $x(t)$ in terms of $\epsilon(t)$ using (\ref{generative-ode-solution}), we obtain
$$
\Expec[x(t) \epsilon (t)] = \left[ \int _{0} ^ {t} \fundmats \mathbf{C} (s, t)  \mathrm{d}s\right]_{1,2},
$$
which only depends on the matrices
$$
\mathbf{C} (s, t) = 
\begin{pmatrix}
0 & 0 \\
0 & C(s, t)
\end{pmatrix}
\text{ and }
\mathbf{B} (t) = 
\begin{pmatrix}
0 & 0 \\
\beta_0 (t) & 0
\end{pmatrix},
$$
because $\mathbf{B} (t)$ is used to compute $\fundmats$. Therefore, to estimate $\Expec[x(t)\epsilon(t)]$ we only need estimates of $\beta_0 (t)$ and $C(s, t)$. Our proposed bias reduction algorithm, which can be re-applied iteratively using updated estimates of $\widehat{\beta_0} (t)$ and $\widehat{C} (s, t)$, is as follows:
\begin{framed}{\textbf{Bias Reduction Algorithm}}
    \begin{enumerate}
    \item Obtain initial OLS estimates $\widehat{\beta}_0 (t)$ and $\widehat{C} (s, t)$. Substitute these for $\beta_0 (t)$ and $C (s, t)$ to obtain an estimate $\widehat{\Expec}[x(t)\epsilon(t)]$ of $\Expec[x(t)\epsilon(t)]$.
    \item Calculate a bias-reduced estimate:
    $$
    \widehat{\beta} ^ {(bc)}_0(t) = \widehat{\beta}_0 (t) - \frac{\widehat{\Expec}[x(t)\epsilon(t)]}{ \frac{1}{N} \Sigma_{i=1}^N x_i (t) ^2}.
    $$
    \item If required, iterate through steps 1-2 using the bias-reduced estimate $\widehat{\beta}_0 ^ {(bc)}(t)$ in place of $\widehat{\beta}_0 (t)$ at step 1, calculating updated residuals and re-estimating $\widehat{C} (s, t)$.
\end{enumerate}
\end{framed}

\begin{figure}[h]
	    \centering
		\includegraphics[page=4,width=0.51\textwidth]{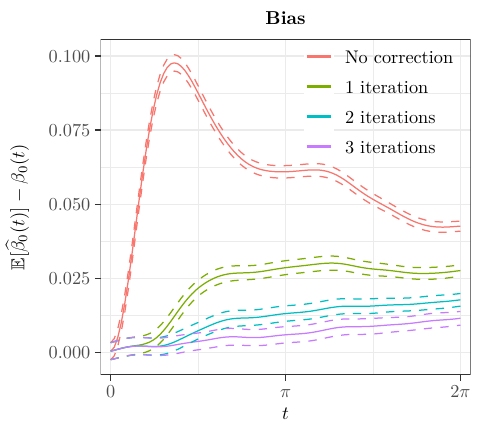}  
		\caption{Bias of the uncorrected estimator (the initial OLS estimate) and estimators using 1-3 iterations of the proposed bias correction. The solid line represents the estimate of the bias obtained from simulation and the dashed lines represent $95\%$ pointwise confidence intervals which quantify simulation uncertainty in the estimated bias.}
		\label{plot:bias-corrected-beta}
\end{figure}

Figure \ref{plot:bias-corrected-beta} shows the results of applying 1-3 iterations of the proposed correction to the 500 simulated datasets. The bias is significantly reduced after one iteration and is further reduced by the second and third iterations. Our intuition for the subsequent improvements of the second and third iterations is that updated estimates of $\beta (t)$ and $C(s, \ t)$ allow improved estimates of the bias to be obtained, which in turn leads to improved estimates of $\beta (t)$ and $C(s, \ t)$. 
Stopping criteria based on absolute or relative changes in parameter values could be used to choose the number of iterations.
However, in this work, we simply fix the maximum number of iterations as we have found that $10$ or fewer iterations is generally sufficient in the examples we consider.
The correction is more generally applicable to models of any order with any number of parameters. Detailed workings of the general correction and its numerical implementation are provided in Appendix \ref{sec:full-bias}.

\section{PDA as a Linear Approximation of Non-Linear Time-Invariant Dynamics} \label{sec:non-linear}

The properties of linear ODEs are well understood so they provide a useful starting point for building dynamic models from functional data \parencites[p. 313]{ramsay_functional_2005}[p. 194]{ramsay_functional_2009}. However, in many applications of interest \parencite[e.g., human movement;][]{van_emmerik_comparing_2016}, we expect to encounter \emph{non-linear}, and typically time-invariant, dynamics. Still, linear approximations of non-linear models around certain points of interest (e.g., equilibrium points or points on a limit cycle) are used to study local properties and behaviour. It is also common to view linear regression models as local approximations of more general functions within a given range of covariate values \parencites{berk_assumption_2021, betancourt_taylor_2022}. This motivates the question
\begin{quote}
    \emph{Can PDA provide a useful linear approximation to non-linear ODE models?}
\end{quote}

\subsection{Background}

To simplify our exposition, we consider second-order time-invariant ODEs of the form
$$
D^2 x(t) = g(x(t), D x(t)).
$$
Following the convention introduced in Section \ref{background}, the second-order ODE can be re-written as the coupled system of first-order equations
$$
Dx(t) = y(t) 
$$
$$
Dy (t) = g(x(t), y(t)).
$$
Importantly, we do not restrict $g$ to be a linear function of $x(t)$ and $y(t)$, nor do we assume it has any particular parametric form. This reflects many modern settings where data are collected on a system but the underlying dynamic equations governing the system are unknown \parencite{brunton_discovering_2016}.
The only assumption we make regarding the form of $g$ is that it is sufficiently smooth (i.e., continuously differentiable) around a given operating trajectory of $(x(t), y(t))^\top$ such that a Taylor approximation around that trajectory is valid.



\begin{figure}[h]
	    \centering
		\includegraphics[page=8,width=0.5\textwidth]{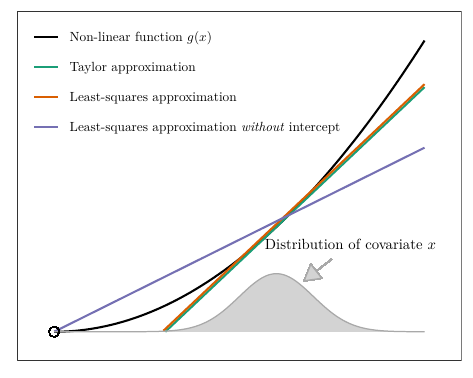}  
		\caption{Approximation of the non-linear function $g(x)$ (black curve) using linear approximations. The operating point for the Taylor approximation is at the mean of the covariate distribution.}
		\label{plot:intercept-demo}
\end{figure}

Our starting point for a linear approximation of $g$ is given by its first-order Taylor approximation about an operating point $(x^*, y^*)$ 
$$
g(x(t), y(t)) \approx g(x^*, y^*) + \frac{\partial g( x^*, y^*)}{\partial x} (x(t) - x^*) + \frac{\partial g( x^*, y^*)}{\partial y} (y(t) - y^*).
$$
Re-arranging gives
$$
g(x(t), y(t)) \approx \underbrace{g(x^*, y^*) - \frac{\partial g( x^*, y^*)}{\partial x} x^* - \frac{\partial g( x^*, y^*)}{\partial y} y^*}_{\alpha(t)} +  \underbrace{\frac{\partial g( x^*, y^*)}{\partial x}}_{\beta_0 (t)} x(t) + \underbrace{\frac{\partial g( x^*, y^*)}{\partial y}}_{\beta_1 (t)} y(t),
$$
which immediately suggests that a PDA model aiming to provide a linear approximation to an unknown non-linear function should include an intercept $\alpha (t)$. Figure \ref{plot:intercept-demo} illustrates this for a simple linear regression problem -- excluding an intercept forces the fitted line through the origin and moves the fitted slope away from that of the Taylor expansion.

From this perspective, we view PDA as estimating of the Jacobian of the system
$$
\mathbf{J}(x(t), y(t))
=
\begin{pmatrix}
0 & 1 \\
\frac{\partial g( x(t), y(t))}{\partial x} & \frac{\partial g( x(t), y(t))}{\partial y}
\end{pmatrix}
=
\nabla_{(x, y)} \mathbf{G} (x(t), y(t)),
$$
along a trajectory of $(x(t), y(t))^\top$, where $\mathbf{G} (x(t), y(t))^\top = \left(y(t), g(x(t), y(t) \right)^\top$. As we discuss in the following sections, this viewpoint allows the techniques introduced in Section \ref{sec:bias} to be used to reduce bias in the parameter estimates and provide insights into the system.

\subsection{Model}

We introduce a lack of fit to the non-linear model in the same way as the linear model in Section \ref{sources-of-variation}, viewing the observations $(x_i (t), y_i (t))^\top$ as the solution of the non-linear second-order ODE with a stochastic disturbance
$$
D y_i(t) = g(x_i(t), y_i(t)) + \epsilon_i(t),
$$
where $\epsilon_i(t)$ is a realisation of a mean-zero Gaussian process with a smooth covariance function $C(s, t)$. For non-linear models, there is, in general, no analytic expression for the solution so it is not possible to write the generative model as we have done for the linear case in \eqref{generative-ode-solution}.

However, when variation in $Dy(t)$ is sufficiently small such that, at each $t$, the error in the first-order Taylor approximation around the mean function $(\Bar{x} (t), \Bar{y} (t))^\top$ is negligible, we can view the model as
$$
D y_i(t) \approx \alpha(t) + \beta_0 (t) x_i(t) + \beta_1 (t) y_i(t) + \epsilon_i(t),
$$
where
$$
\alpha (t) = g(\Bar{x} (t), \Bar{y} (t)) - \frac{\partial g(\Bar{x} (t), \Bar{y} (t))}{\partial x} \Bar{x} (t) - \frac{\partial g( \Bar{x} (t), \Bar{y} (t))}{\partial y} \Bar{y} (t),
$$
$$
\beta_0 (t) = \frac{\partial g(\Bar{x} (t), \Bar{y} (t))}{\partial x} \quad \text{and} \quad \beta_1 (t) = \frac{\partial g(\Bar{x} (t), \Bar{y} (t))}{\partial y}.
$$
In practice, the observations $(x_i (t), y_i(t))^\top$ could be centered around their time-varying mean function $(\Bar{x} (t), \Bar{y} (t))^\top$ so that $\alpha(t) = g(\Bar{x} (t), \Bar{y} (t))$, which simplifies the interpretation of the intercept term. 

The linearised ODE model leads to the generative model 
\begin{equation} \label{eq:tv-linear-approx}
\begin{pmatrix} 
x(t) \\
y(t) 
\end{pmatrix}
\approx
\fundmat 
\begin{pmatrix} 
x_0 \\
y_0 
\end{pmatrix}
+
\int_0^t
\fundmats
\begin{pmatrix} 
0 \\
\alpha(s) 
\end{pmatrix}
\mathrm{d}s
+
\int_0^t
\fundmats
\begin{pmatrix} 
0 \\
\epsilon(s) 
\end{pmatrix}
\mathrm{d}s,
\end{equation}
where $(x_0, y_0)^\top$ is the initial state, and the state transition matrix $\fundmats$ is the solution of the matrix ODE
$$
D \fundmats = \mathbf{J}(\Bar{x}(t), \Bar{y}(t)) \fundmats
$$
with initial conditions $\vec{\Phi}(s, s) = \mathbf{I}_m$.

\begin{figure}[h]
	    \centering
		\includegraphics[page=9,width=0.8\textwidth]{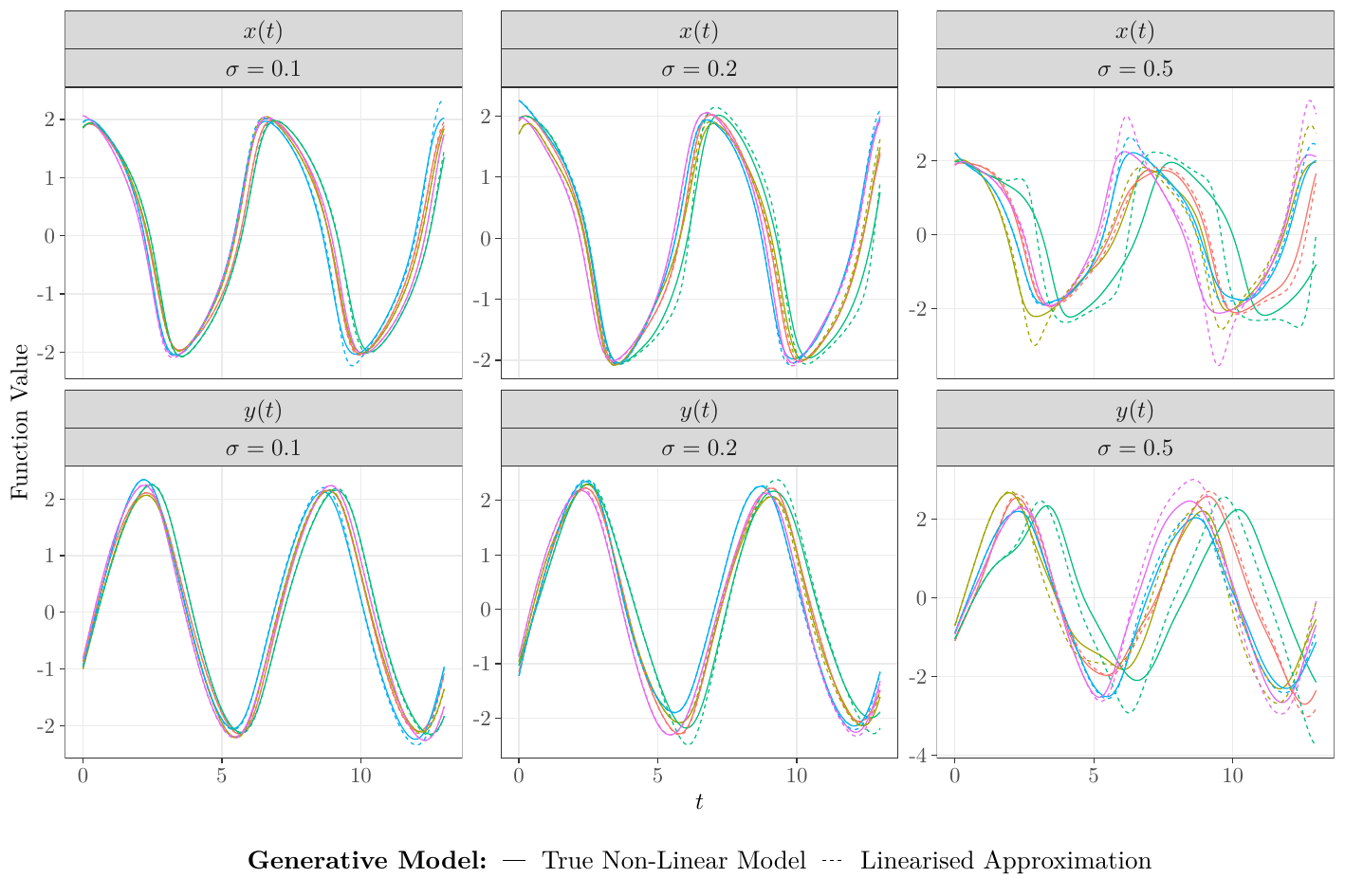}  
		\caption{Comparison of the true non-linear and time-varying linearised generative Van der Pol models. The same stochastic disturbance and random initial conditions are used to generate pairs of observations from the non-linear and linearised models, represented by the solid and dashed lines, respectively. The stochastic disturbances are drawn from $\mathcal{GP}(0, C)$, where $C(s,t) = \sigma^2 \phi (2 |s-t|)$, so that each column represents comparisons of the non-linear and linearised models with stochastic disturbances of increased amplitude. In all cases, the initial conditions are drawn independently of the stochastic disturbance from a bivariate Gaussian distribution with mean $\boldsymbol{\mu}_0 \approx (1.99, -0.91)^\top$ (a point approximately on the Van der Pol limit cycle) and covariance $\boldsymbol{\Sigma}_0 = \diag\{0.05, 0.05\}$. The non-linearity parameter $\mu$ is fixed at 1.}
		\label{plot:LNA-demo-01}
\end{figure}

\begin{figure}[h]
	    \centering
		\includegraphics[page=10,width=0.8\textwidth]{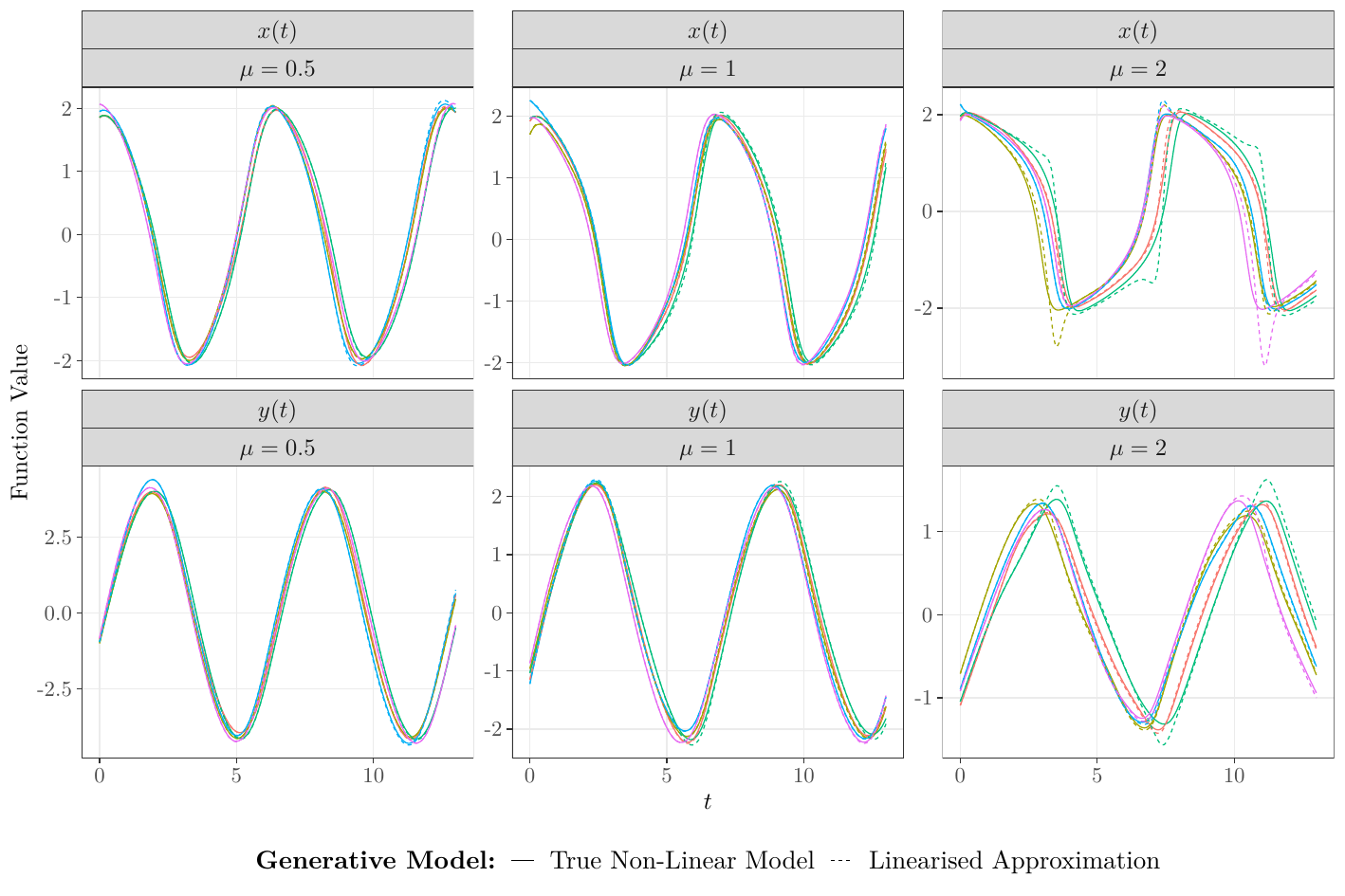}  
		\caption{Comparison of the true non-linear and time-varying linearised generative Van der Pol models. The same stochastic disturbance and random initial conditions are used to generate pairs of observations from the non-linear and linearised models, represented by the solid and dashed lines, respectively. The stochastic disturbances are drawn from $\mathcal{GP}(0, C)$, where $C(s,t) = (0.1)^2 \phi (2 |s-t|)$. In all cases, the initial conditions are drawn independently of the stochastic disturbance from a bivariate Gaussian distribution with mean $\boldsymbol{\mu}_0 \approx (1.99, -0.91)^\top$ (a point approximately on the Van der Pol limit cycle) and covariance $\boldsymbol{\Sigma}_0 = \diag\{0.05, 0.05\}$. Each column contains functions generated using a different value of the non-linearity parameter $\mu$, where a larger value represents a higher degree of non-linearity.}
		\label{plot:LNA-demo-02}
\end{figure}

The adequacy of the time-varying linearised model \eqref{eq:tv-linear-approx} as an approximation of the true generative model will, in general, depend on the scale of perturbations to $x(t)$ and $y(t)$ (through the stochastic disturbance and initial conditions), and on the degree of non-linearity in the underlying function $g$. We illustrate the dependence of the approximation on the scale of the perturbations in Figure \ref{plot:LNA-demo-01}. Pairs of observations generated by the true and linearised versions of the Van der Pol model, sharing identical stochastic disturbances and initial conditions, are indicated by the solid and dashed lines, respectively. When the amplitude of the stochastic disturbances is small ($\sigma = 0.1$, first column), the linearised model is an extremely good approximation for the true model, i.e., the solid and dashed lines are practically indistinguishable from one another. As the amplitude of the stochastic disturbance is increased, the quality of the approximation decreases, as can be seen by the disagreement between the solid and dashed lines for $\sigma = 0.5$ (third column). An analogous illustration for the effect of non-linearity, controlled by the parameter $\mu$ in the Van der Pol model, is contained in Figure \ref{plot:LNA-demo-02}.

An equally-pertinent, practical challenge is in determining when this interpretation of PDA is appropriate and meaningful in real data analysis settings.
We have in mind the speaking, handwriting and juggling examples considered in \textcite{ramsay_principal_1996, ramsay_functional_2000, ramsay_functional_1999} and the treadmill-running example that we present in Section \ref{examples}. 
In these datasets, observations represent repetitions of a stable physical process from the same subject that are comparable on a point-by-point scale, such that the average function represents what could be considered a stable limit cycle. 
Therefore, it is conceivable that each observation might deviate smoothly around the average trajectory due to variation initial conditions or a random disturbance to the process dynamics.
For other settings, the utility of PDA as an approximate model might be more difficult to justify without additional assumptions.

\subsection{Estimation}
For estimation, the linear approximation is treated as the data-generating model so that estimates are obtained in the same way as in the linear case -- initial estimates are obtained by OLS and then the bias-reduction algorithm proposed in Section \ref{sec:bias} can be applied. Hence, both the initial parameter estimates and the estimates of the bias both rely on the linear approximation.

\subsection{Consequences}

\subsubsection{Including an Intercept in PDA}

A consequence of this viewpoint of PDA, that differs from the methodology initially proposed by \textcite{ramsay_principal_1996}, is that an intercept should be included in the model if there is uncertainty regarding the form of the underlying ODE. When the underlying ODE is truly linear, such as the SHM model \eqref{SHM-data-generating-model}, the bias-reduced estimates should, on average, estimate the intercept at zero. Appendix \ref{sec:additional-hm-simulations} contains three simulated examples from linear ODE models to demonstrate this. On the other hand, when the underlying ODE is non-linear, including an intercept is necessary to obtain parameter estimates that are useful for understanding the underlying system from a local-approximation perspective.

\subsubsection{Qualitative Analysis}
The (approximate) generative model \eqref{eq:tv-linear-approx} allows new observations to be simulated and their qualitative behaviour to be examined.
Given a fitted model, it is helpful to simulate datasets of replicate trajectories and determine whether the simulated trajectories resemble the observed data in terms of their shape and variability \parencite[][p. 783]{ionides_comment_2007}.

To build on the idea of assessing local stability through PDA \parencite{ramsay_principal_1996}, estimates of the Jacobian $\mathbf{J}$ can be used to assess the stability of the deterministic part of the system
$$
D \mathbf{x}(t)  = \mathbf{G}(x(t), y(t)), 
$$
where $\mathbf{x}(t) = (x(t), y(t))^\top$. In the field of non-linear time series, there has been a long-standing interest in estimating the sensitivity of a dynamical system to different conditions -- known as ``testing the chaos hypothesis" -- from noisy measurements of a single trajectory \parencite{sano_measurement_1985, eckmann_liapunov_1986, nychka_finding_1992, sandubete_dchaos_2021}.
They have generally addressed this challenge by estimating the system's \emph{maximal Lyapunov exponent}, which measures, for a small value of $\boldsymbol{\eta}$, the rate at which trajectories starting from $\mathbf{x}(0) = \mathbf{x}_0$ and $\mathbf{x}_0 + \boldsymbol{\eta}$ diverge. Informally, we could assess this property by simulating observations that vary randomly in initial conditions only (i.e., from one part of the data-generating model).
Beyond this, there is potential to use the estimate of $\mathbf{J}(\bar{x}(t), \bar{y}(t))$ obtained in PDA to perform a more conventional stability analysis by using it to compute the Lyapunov exponents of the deterministic part of the system. However, we leave a full examination of this approach to future work.

\section{Examples} \label{examples}

\subsection{Simulation: Simple Harmonic Motion} \label{shm-simulation}

\begin{figure}
    \centering
    \includegraphics[page=5,width=0.8\textwidth]{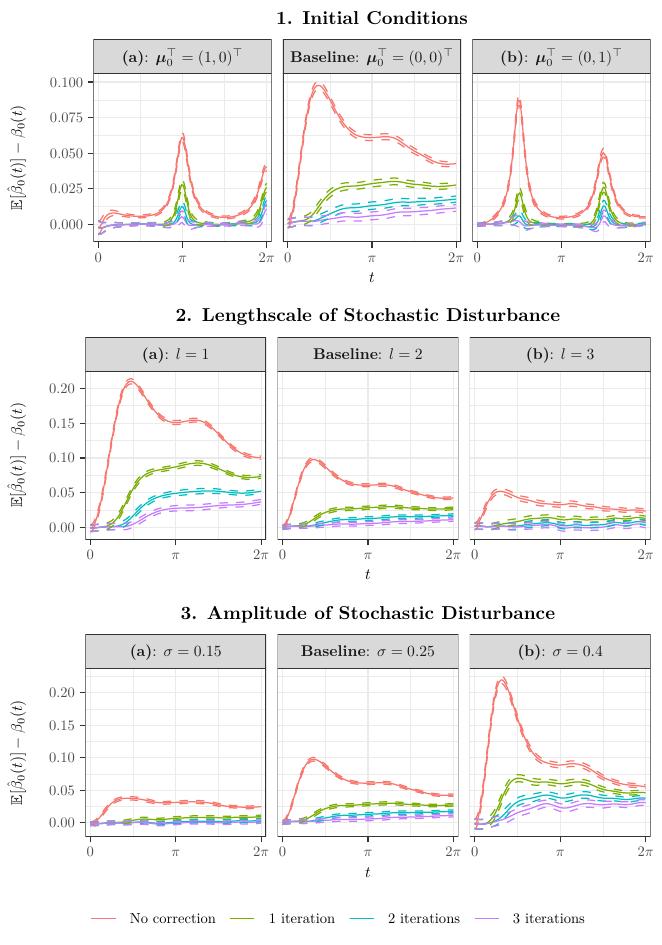}
    \caption{The results of the simulation to assess bias in the SHM model. The solid lines represent estimates of the bias in $\widehat{\beta}_0(t)$ and the dashed lines represent associated $95\%$ pointwise confidence intervals, which quantify simulation uncertainty in the estimated bias. Each row shows the results of varying a single parameter, while the other parameters are fixed at the baseline scenario $\boldsymbol{\mu}_0^\top = (0, 0)^\top$, $l=2$ and $\sigma=0.25$. Therefore, each row of the central column displays identical results, with the $y$-axis scale adjusted to facilitate within-row comparisons.}
    \label{fig:simulation-results}
\end{figure}

We simulate data from the SHM model \eqref{SHM-data-generating-model} and vary the parameters to assess estimation bias and the performance of the bias-reduction algorithm under different conditions. The parameter values used in the demonstration in Section \ref{sec:bias} are treated as the baseline scenario. We vary the following parameters one at a time while fixing the other parameters at baseline:
\begin{enumerate}
    \item Expectation of initial conditions: \textbf{(a)} $\boldsymbol{\mu}_0 = (1, 0)^\top$ and \textbf{(b)} $\boldsymbol{\mu}_0 = (0, 1)^\top$.
    \item Lengthscale of the stochastic disturbance: \textbf{(a)} $l=1$ and \textbf{(b)} $l = 3$.
    \item Amplitude of the stochastic disturbance: \textbf{(a)} $\sigma = 0.15$ and \textbf{(b)} $\sigma = 0.4$.
\end{enumerate}
For each scenario, we perform 200 simulations and apply three iterations of the bias-reduction algorithm. The results are shown in Figure \ref{fig:simulation-results}, where the baseline scenario is contained in the middle panel in each row for ease of comparison. Overall, the bias-reduction algorithm works well for each scenario and 2-3 iterations are sufficient to reduce most of the bias. Table \ref{tab:my-table} displays the Mean Integrated Squared Error (MISE) of $\widehat{\beta}_0 (t)$ for each scenario. Applying the bias-reduction algorithm leads to a reduced MISE in all scenarios. Improvements achieved by the second and third iterations appear to largely depend on the initial MISE, i.e., the second and third iterations appear to help when the MISE is initially large; a similar effect for the bias can be seen in Figure \ref{fig:simulation-results}. Some specific comments on each scenario are given below.

Varying the initial conditions has the effect of changing the ``shape" of the bias as a function of $t$ (Figure \ref{fig:simulation-results}, top row). This is because the initial conditions control the expectation of $x(t)$, which, in turn, affects the denominator in the bias, i.e.,  $\Expec[x(t)^2]$. Decreasing the lengthscale of the stochastic disturbance increases the magnitude of the bias (Figure \ref{fig:simulation-results}, middle row). An intuitive explanation for the observed effect is that decreasing the lengthscale makes the stochastic disturbances smoother, leading to more sustained forcing of the ODE and hence greater dependence between $x(t)$ and $\epsilon(t)$. Finally, increasing the forcing amplitude $\sigma$ increases the magnitude of the bias (Figure \ref{fig:simulation-results}, bottom row), we can understand this effect as also being due to increased dependence between $x(t)$ and $\epsilon(t)$.


\begin{table}[]
\resizebox{\textwidth}{!}{%
\begin{tabular}{@{}lllllll@{}}
\toprule
\multicolumn{3}{l}{}                  & \multicolumn{4}{l}{Mean Integrated Squared Error (Standard Error)}                    \\ \midrule
\multicolumn{3}{l}{No. of Iterations} & 0                & 1                & 2                & 3                \\ \midrule
\multicolumn{3}{l}{Scenario}          &                  &                  &                  &                  \\ \cmidrule(r){1-3}
\multirow{7}{*}{} &
  \multicolumn{2}{l}{Baseline} &
  0.0287 (0.0004) &
  0.0085 (0.0002) &
  0.0064 (0.0001) &
  0.0062 (0.0001) \\ \cmidrule(lr){2-3}
 &
  \multirow{2}{*}{\begin{tabular}[c]{@{}l@{}}1. Initial \\ \ \ \ Conditions\end{tabular}} &
  (a) &
  0.0036 (0.0001) &
  0.0016 (0.0001) &
  0.0014 ($< 0.0001$) &
  0.0014 ($< 0.0001$) \\ \cmidrule(lr){3-3}
  &                             & (b) & 0.0055 (0.0001) & 0.0015 (0.0001) & 0.0013 (0.0001) & 0.0013 (0.0001) \\ \cmidrule(lr){2-3}
  & \multirow{2}{*}{2. Lengthscale}  & (a) & 0.1364 (0.0014) & 0.0363 (0.0007) & 0.0153 (0.0005) & 0.0088 (0.0004) \\ \cmidrule(lr){3-3}
  &                             & (b) & 0.0115 (0.0003) & 0.0057 (0.0002) & 0.0058 (0.0002) & 0.0061 (0.0002) \\ \cmidrule(lr){2-3}
  & \multirow{2}{*}{3. Amplitude}   & (a) & 0.0071 (0.0002) & 0.0023 (0.0001) & 0.0022 (0.0001) & 0.0022 (0.0001) \\ \cmidrule(lr){3-3}
  &                             & (b) & 0.0844 (0.0014) & 0.0256 (0.0007) & 0.0169 (0.0005) & 0.0143 (0.0005) \\ \cmidrule(l){1-7} 
\end{tabular}
}
\caption{Results of the simulation from the SHM model \eqref{SHM-data-generating-model}. The Mean Integrated Squared Error (MISE) of $\widehat{\beta}_0(t)$ is shown under the different data-generating scenarios and after 0-3 iterations of the bias-reduction algorithm. Estimated Monte Carlo standard errors are shown in brackets to quantify simulation uncertainty in the estimated MISE due to using a finite number of repetitions. The parameter values $\boldsymbol{\mu}_0^\top = (0, 0)^\top$, $l=2$ and $\sigma=0.25$ are used in the baseline scenario. Initial conditions are varied at \textbf{(a)} $\boldsymbol{\mu}_0 = (1, 0)^\top$ and \textbf{(b)} $\boldsymbol{\mu}_0 = (0, 1)^\top$. The lengthscale of the stochastic disturbance is varied at \textbf{(a)} $l=1$ and \textbf{(b)} $l = 3$. The amplitude of the stochastic disturbance is varied at \textbf{(a)} $\sigma = 0.15$ and \textbf{(b)} $\sigma = 0.4$.}
\label{tab:my-table}
\end{table}

\subsection{Simulation: Van der Pol Equation} \label{vdp}

The Van der Pol (VdP) equation was initially formulated by \textcite{van_der_pol_frequency_1927} to describe electrical circuits employing vacuum tubes \parencite[see][p. 7]{RJ-2010-013}. It is a second-order, non-linear, time-invariant ODE 
$$
D^2 x(t) - \mu \left(1 - x(t)^2 \right) D x(t) + x(t) = 0, \quad \mu \geq 0,
$$
and can be understood as a non-linearly damped oscillator, where the parameter $\mu$ controls the amount of non-linear damping. SHM occurs as a special case when $\mu = 0$. As described by \textcite[][p. 8]{RJ-2010-013}, the VdP equation is commonly used as a test case for numerical ODE solvers because its solutions are \emph{stiff} for large values of $\mu$, i.e., they consist of parts that ``change very slowly, alternating with regions of very sharp changes" \parencite[see also][p. 70]{soetaert_solving_2012}. In addition, for $\mu > 0$ the ODE has a stable limit cycle so, as described in Section \ref{sec:non-linear}, it is natural to view PDA as an approximate model describing behaviour around the limit cycle trajectory.

To construct a data-generating process, we first convert the second-order ODE to an equivalent system of two coupled first-order ODEs
 \begin{align*}
     Dx(t) &= \mu \left(x(t) - \frac{x(t)^3}{3} - y(t)\right) \\
     D y(t) &= \frac{x(t)}{\mu},
 \end{align*}
which uses the transformation $y(t) = x(t) - \frac{x(t)^3}{3} + \frac{Dx(t)}{\mu}$ for $\mu > 0$. We consider values of $t \in [0, 13]$ which corresponds to approximately two periods of oscillations when $\mu$ is small ($\mu \leq 1$), and use an equally-spaced grid of length $200$ for discretisation. We add smooth Gaussian noise independently to $Dx(t)$ and $Dy(t)$ and generate $N$ independent realisations of function pairs $(x_i(t), y_i(t))^\top$ as solutions of the stochastically forced coupled ODEs. We use the same Gaussian process as in Section \ref{shm-simulation} to generate the smooth stochastic disturbance with the parameters $l = 2$ and $\sigma=0.1$ fixed; however, these parameters are varied in additional simulations contained in Appendix \ref{sec:additional-vdp}. Initial conditions are drawn as $(x_i(0), y_i(0))^\top \sim \mathcal{N}_2 (\boldsymbol{\mu}_0, \boldsymbol{\Sigma}_0)$, where $\boldsymbol{\mu}_0 \sim (1.99, -0.91)^\top$ is a point lying approximately on the stable limit cycle of the deterministic ODE. We primarily use $\boldsymbol{\Sigma}_0 = \diag \{0.05, 0.05 \}$ but slightly re-scale its elements as the behaviour of the system changes with $\mu$ to generate more ``realistic-looking" datasets.

\begin{figure}[h]
    \centering
    \includegraphics[page=1, width = 0.8\textwidth]{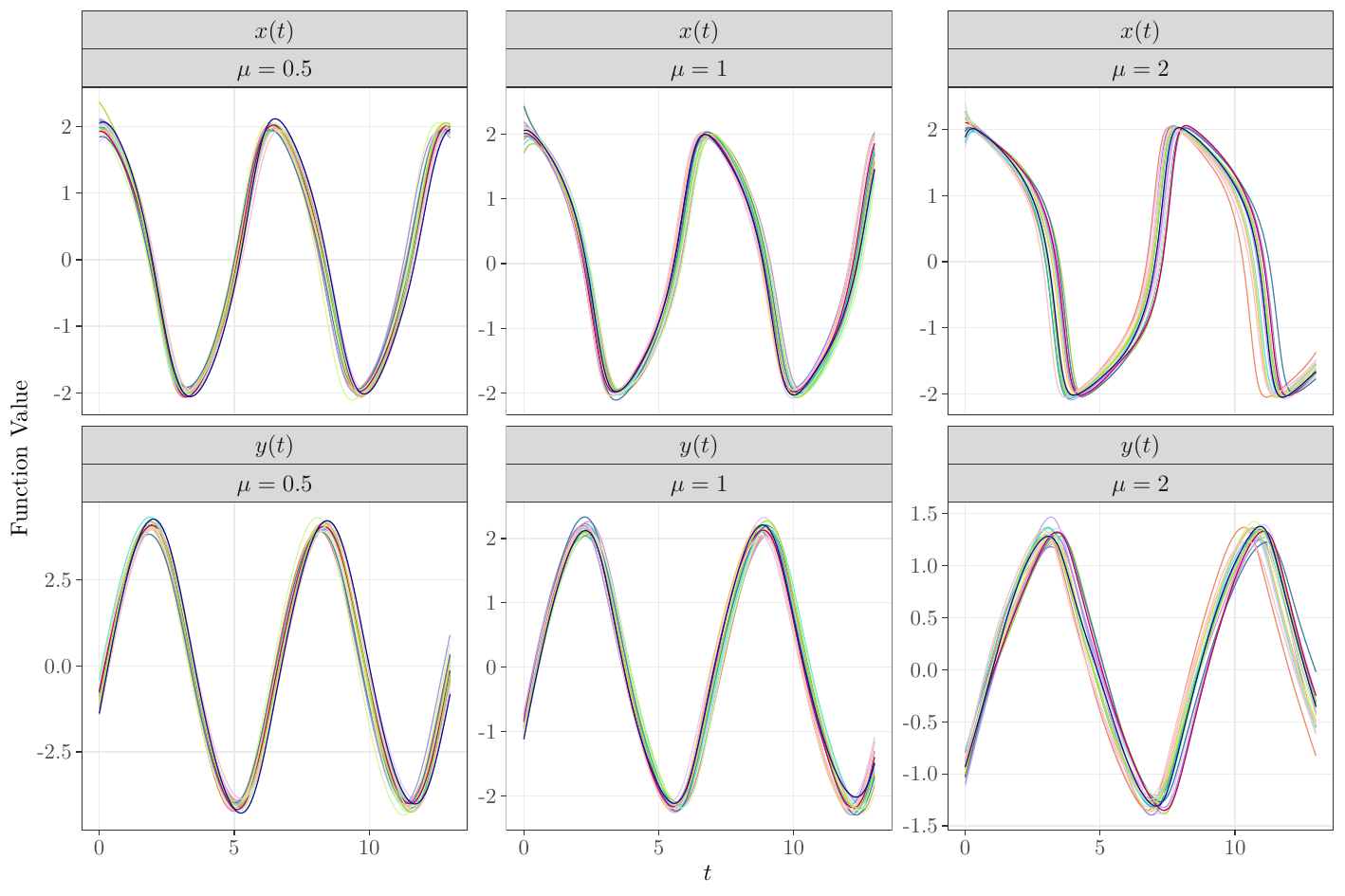}
    \caption{20 generated function pairs $(x_i(t), y_i(t))^\top$ from the VdP model at the values $\mu = 0.5$, $\mu = 1$ and $\mu = 2$.}
    \label{fig:vary-mu-example-dataset}
\end{figure}

In what follows, we are concerned with estimating the parameters of the linearised approximation to the coupled system of ODEs at three different values of $\mu$: $0.5, \ 1, \text{ and } 2$. Figure \ref{fig:vary-mu-example-dataset} displays 20 simulated observations from the data-generating model at the three chosen values of $\mu$. Qualitative changes in the system are evident and are most notable for $\mu = 2$ when the degree of non-linearity is largest. In this setting, the shape of the oscillations being produced is sharpest and the period of the oscillations is longest (i.e., approximately 1.5 periods are completed on $[0, 13]$ when $\mu = 2$, as opposed to two full periods for $\mu=0.5$ and $\mu=1$). The scale of $y(t)$ is also noticeably decreased for $\mu = 2$ because it is multiplied by $1/\mu$, and the functions qualitatively exhibit more phase (i.e., timing) variation.

Estimating the parameters involves fitting the coupled PDA models
\begin{align*}
    Dx_i (t) &= \alpha^{(x)} (t) + \beta_{xx} (t) x_i(t) + \beta_{xy} (t) y_i(t)  + \epsilon_i^{(x)} (t)  \\
    Dy_i(t) &= \alpha^{(y)}(t) + \beta_{yx} (t) x_i(t) + \beta_{yy} (t) y_i(t) + \epsilon_i^{(y)} (t).
\end{align*}
The model for $Dx_i(t)$ is a linearised approximation. Although $\beta_{xy}(t) = - \mu$ is the true parameter, the parameters $\beta_{xx} (t)$ and $\alpha^{(x)} (t)$ are based on partial derivatives from a linearised approximation
\begin{align*}
    \beta_{xx} (t) &=  \mu \left(1 - x_0(t)^2 \right) \\
    \alpha^{(x)} (t) &= \mu \left(x_0(t) - \frac{1}{3} x_0(t)^3 - y_0(t)\right) - \mu \left(1 - x_0(t)^2 \right) x_0(t) + \mu y_0 (t),
\end{align*}
about an operating trajectory $(x_0 (t), y_0 (t))^\top$, for which we choose the limit-cycle trajectory as a ``ground truth". The model for $Dy_i(t)$ is correctly specified, i.e., the true model is linear  with $\alpha^{(y)}(t) = 0$, $\beta_{yx} (t) = 1/\mu$ and $\beta_{yy} (t) = 0$. The coupled models are initially estimated separately using OLS and then the estimated parameters and estimates of the residual covariance functions are combined to estimate the bias using a version of the bias-reduction algorithm that is adapted for the coupled system; full details of the algorithm are provided in Appendix \ref{sec:additional-vdp}.

\begin{figure}
    \centering
    \includegraphics[width = 0.75\textwidth]{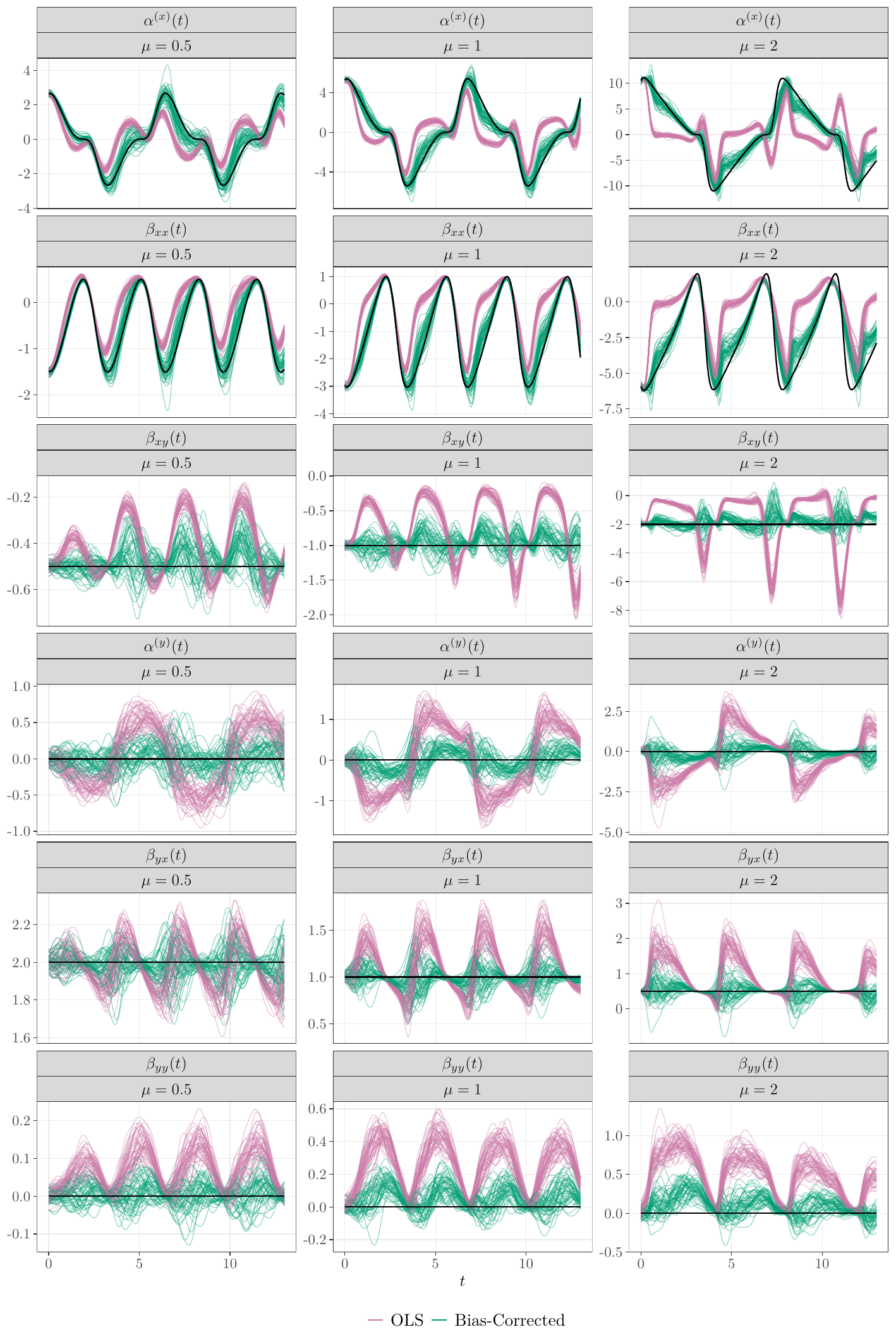}
    \caption{Results of the simulation for the VdP model. Each row corresponds to a different parameter and each column corresponds to a different value of $\mu$. The results from all 50 simulation replicates are shown, with the initial OLS estimates coloured pink and the bias-corrected estimates coloured dark green. The solid black lines indicate the true parameter values.}
    \label{fig:vary-mu-results}
\end{figure}

In all three scenarios ($\mu = 0.5, \ 1 \text{ and } 2$), $50$ datasets of size $N = 200$ are generated. The smaller number of simulation replicates is chosen because the model is more complex and applying the bias-reduction algorithm is more computationally demanding than in the simple one-parameter case. For each simulated dataset, we apply 10 iterations of the bias-reduction algorithm and compare the initial OLS estimates with final bias-reduced estimates. The larger number of iterations is chosen because there are now nine parameters being estimated and corrected. The results are displayed in Figure \ref{fig:vary-mu-results} -- the initial OLS estimates are plotted in pink, the bias-reduced estimates are in dark green and the true parameters are indicated by a solid black line.

The shapes of the true and estimated time-varying parameters $\alpha^{(x)}(t)$ and $\beta_{xx} (t)$ are noticeably different for the different values of $\mu$. Reflective of the qualitative differences in the generated functions, the parameters for $\mu = 2$ exhibit the sharpest changes. The bias in the OLS estimates is also the most severe for $\mu = 2$, e.g., the biased estimates of $\beta_{xy}(t)$ range between $0$ and $-8$ when the true parameter is the constant function $\beta_{xy}(t) = - 2$. The bias-reduction algorithm significantly reduces bias in all scenarios and produces the largest corrections to the initial parameter estimates when $\mu = 2$. Certain parts of $\alpha^{(x)}(t)$ and $\beta_{xx}(t)$ cannot be captured perfectly at $\mu = 2$, likely due to the linear approximation being less suitable as the degree of non-linearity in the underlying function increases. Overall, however, the interpretation of PDA as estimating a linearised approximation to the non-linear system and the performance of the bias-reduction algorithm are satisfactory in all three scenarios. Further insights into the conditions under which the approximation is reasonable are provided in Appendix \ref{sec:additional-vdp}, where simulations varying $\sigma$ and $l$ are presented.

\subsection{Running Data Analysis} \label{running}
We illustrate  our methodology on data collected in the Dublin City University (DCU) running injury surveillance (RISC) study, where biomechanical data from recreational runners were collected during a treadmill run to understand running technique and its links to injury (e.g., see work by \textcite{dillon_injury-resistant_2021, burke_comparison_2022}). A large dataset of kinematic variables was collected on 300 recreational runners using a three-dimensional motion analysis system. For this analysis, we focus on a single kinematic time series, the vertical position of a single runner's centre of mass (CoM), which is displayed in Figure \ref{plot:running-data}. 
This runner was chosen for the demonstration because their observed running kinematics were extremely stable over the full treadmill run.
This variable was chosen because previous models of running mechanics have been based on ``a simple spring-mass system, where a leg-spring supports the point mass representing the runner’s CoM" \parencite[][p. 1]{kulmala_running_2018}. Hence, it is natural to estimate a second-order ODE to describe the CoM trajectory.

\begin{figure}
    \centering
    \includegraphics[width = 1\textwidth]{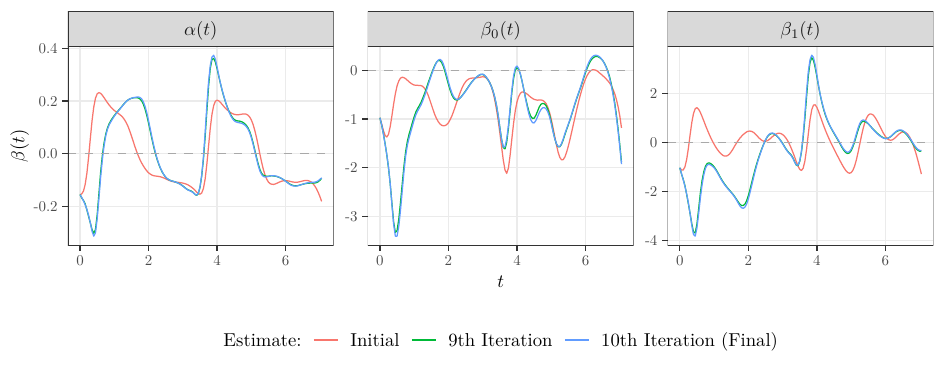}
    \caption{The parameter estimates obtained from PDA. The initial OLS estimates are shown in red and the estimates after applying $9$ and $10$ iterations of the bias-reduction algorithm are shown in green and blue, respectively.}
    \label{fig:subject-01-parameter-estimates}
\end{figure}

We treat the CoM trajectory from each individual stride as a replicate observation and denote it by $x_i(t)$, $i=1, \dots, 82$. We represent each smoothed $x_i(t)$ using a fifth-degree B-spline basis and calculate $Dx_i(t)$ and $D^2x_i(t)$ directly from the B-spline representation (full details of the data preparation are provided in Appendix \ref{sec:additional-running}).
We then fit the second-order PDA model
$$
D^2 x_i(t) = \alpha(t) + \beta_0 (t) x_i(t) + \beta_1 (t) Dx_i(t) + \epsilon_i(t), \quad t \in [0, 7.1],
$$
and apply $10$ iterations of the bias-reduction algorithm. On each iteration, we regularise the estimates by lightly post-smoothing $\widehat{\alpha}(t)$, $\widehat{\beta}_0(t)$, $\widehat{\beta}_1(t)$ and the residual covariance function estimate $\widehat{C}(s, t)$ using penalised splines \parencite{wood_fast_2011, xiao_fast_2016}.
Figure \ref{fig:subject-01-parameter-estimates} displays the parameter estimates from PDA. Applying the bias-reduction algorithm changes the shape, direction and magnitude of the parameters at different time points. Relative changes in the parameters between the $9$th and $10$th iterations are small (i.e., the blue and green lines are practically indistinguishable at most points), emphasising that stopping at $10$ iterations is reasonable. Comparisons between the initial estimates and the partial derivatives of a non-linear fit are contained in Appendix \ref{sec:additional-running}.

\begin{figure}
    \centering
    \includegraphics[width = 0.75\textwidth]{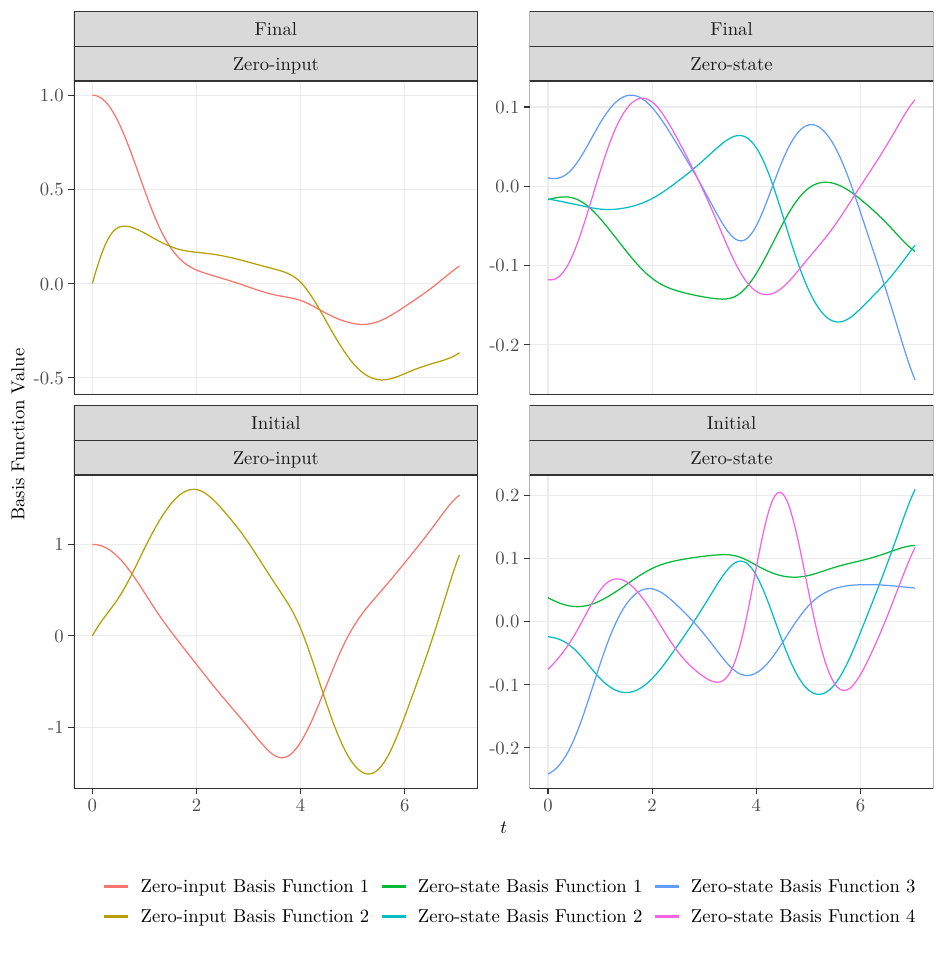}
    \caption{The basis functions of the zero-input covariance function (left column) and the basis functions of the zero-state covariance function (right column). 
    The basis functions in the top panel are calculated from the final parameters after applying the bias-reduction algorithm.
    The basis functions in the bottom panel are calculated from the initial parameters.}
    \label{fig:subject-01-pda-basis}
\end{figure}

\begin{figure}
    \centering
    \includegraphics[width = 0.75\textwidth]{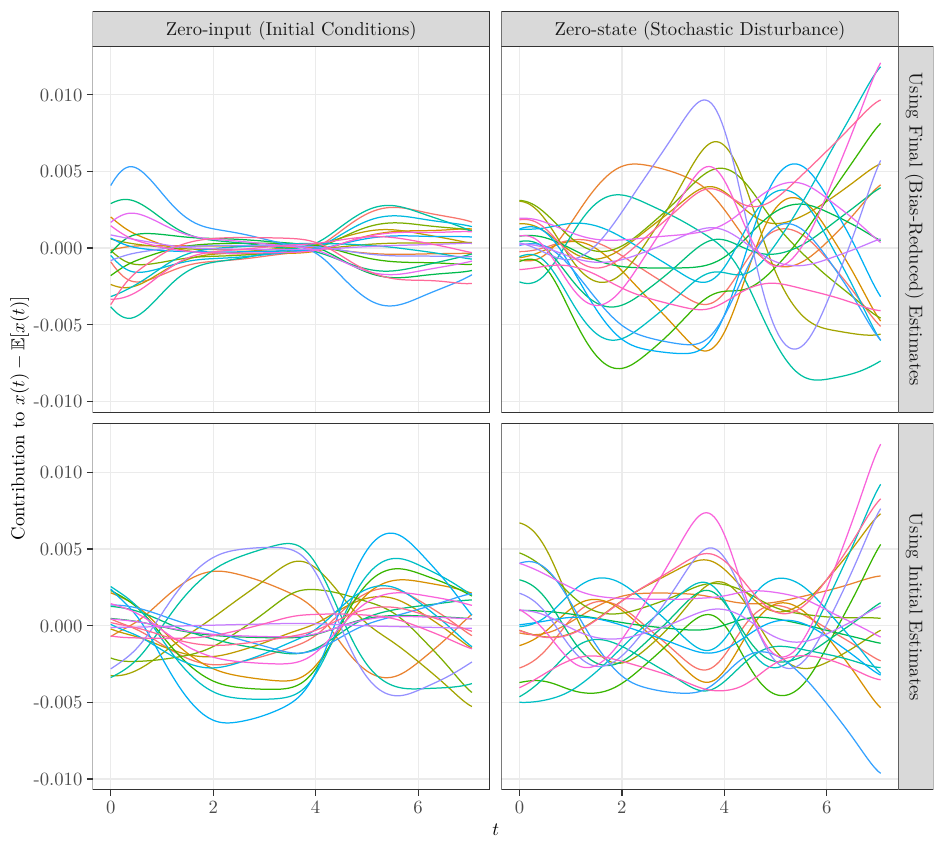}
    \caption{Decomposition of a random sample of $20$ mean-centred functional observations $x_i (t) - \Expec[x(t)]$ using the zero-input and zero-state basis functions. The top panel displays the decomposition based on the final parameters obtained after applying the bias-reduction algorithm.
    The bottom panel displays the decomposition based on the initial OLS parameter estimates.}
    \label{fig:subject-01-variance-decomposition}
\end{figure}

Figure \ref{fig:subject-01-pda-basis} displays the basis functions from PDA. 
The left column displays the zero-input basis functions that capture variation due to initial conditions (i.e., the canonical PDA basis functions) and the right column contains the zero-state basis functions that capture variation due to the stochastic disturbance. The top row displays these basis functions as calculated from the final bias-reduced parameter estimates.
For comparison, the bottom row displays these basis functions as calculated from the initial parameter estimates.
The basis functions computed from the initial parameter estimates are more oscillatory than those computed using the bias-reduced estimates.
Figure \ref{fig:subject-01-variance-decomposition} displays a decomposition of variation obtained by regressing a sample of $20$ mean-centered observations on the combined basis of zero-state and zero-input basis functions.
Using the basis functions computed from the final bias-reduced parameters (top row), the zero-input basis appears to capture variation in initial conditions that is pushed back towards the mean function.
In contrast, using the basis computed from the initial PDA parameter estimates, the deviations captured by the zero-input basis exhibit more periodic behaviour, reflective of the periodic nature of these zero-input basis functions (Figure \ref{fig:subject-01-pda-basis}, bottom left).
There is also more variability captured by the zero-state basis, i.e., due to the stochastic disturbance, when using the final bias-reduced parameters.

\section{Discussion}\label{sec:pda-discussion}

In this work, we have re-examined and extended Principal Differential Analysis, the technique for estimating ODEs from functional data first proposed by \textcite{ramsay_principal_1996}.
Rather than view PDA as a data-reduction technique as initially presented, we formulate it as a generative statistical model in which observations are solutions of a deterministic ODE that is forced by a smooth random error process.
The generative model arises from accounting for the lack of fit to the ODE and leads to a more complete characterisation of the sources of variability in PDA.
For data reduction, the generative model can be used to construct basis functions that capture variation due to the stochastic disturbance, complimenting the canonical basis functions produced by PDA.
We have demonstrated, however, that estimates of the model's parameters can be severely biased, so we have developed an iterative bias-reduction technique to improve them.
We have discussed the utility of PDA as an approximation when the deterministic ODE in the generative model is unknown and possibly non-linear.
We have demonstrated our methodology on simulated data from linear and non-linear ODE models and on real kinematic data from human movement biomechanics.
This work defines a new class of flexible functional models based on ODEs, opening up a number of possibilities for future work.
We briefly highlight some of these possibilities below.

We have primarily focused on formulating and estimating a generative model for PDA, so a definite avenue for future work is to quantify uncertainty in the estimated parameters.
Additionally, as we have focused on understanding and extending PDA methodologically, we have not touched on some issues that arise in practice.
As in \textcite{ramsay_principal_1996}, we have assumed that samples of smooth functions are observed and their derivatives up to a specified order are also available or can be calculated numerically with negligible error. 
However, functional data are often observed with measurement error and need to be smoothed before calculating derivatives and performing analysis. For this, there is a vast literature on non-parametric smoothing techniques \parencite[see, e.g.,][]{ramsay_functional_2005, ruppert_semiparametric_2003, fan_local_2017} which should suffice provided the functions are measured on a sufficiently dense grid.
In real data analysis settings, the order of derivative to model will also have to be chosen. 
In our simulations, we assumed that the order was known \emph{a priori} and we chose a second-order model for the running data heuristically based on analogies with a spring-mass model and inspection of the phase-plane plots in Figure \ref{plot:phase-plane-3d}.
Selecting between models of different orders is a potentially challenging task as models for successive derivatives are not nested and several models can be consistent with the same data-generating process \parencite[p. 52]{hooker_comments_2010}.
With this in mind, more work is needed on model selection and general diagnostics for PDA.

Another important piece of future work concerns the incorporation of registration into dynamic models for functional data. We have observed, for both SHM and VdP models, that the smooth stochastic disturbance produces functional observations that vary in both amplitude and phase (i.e., timing). This has important consequences for our view of PDA as a linear approximation to a non-linear ODE model. When functional data are misaligned, the mean function may no longer represent a suitable operating trajectory. Simply registering $x(t)$ using a warping function $h(t)$ would decouple the relationship between $x(t)$ and its derivatives because
$$
D x ( h(t)) = D h(t) \ \frac{\mathrm{d}x (h(t))}{\mathrm{d}h(t)} .
$$
\textcite[p. 190]{ramsay_functional_2009} suggested, rather than registering $x(t)$ and calculating its derivatives, to instead register $x(t)$ first and then register each of its derivatives using the same warping function. The resulting functions are no longer derivatives and anti-derivatives of one another, but the relationship between them remains intact. This approach, however, may not be directly compatible with our generative model and bias-reduction algorithm, which are formulated in terms of the unregistered functional data. Investigating whether our generative model could be used to inform registration would be a particularly interesting next step.

To close, we briefly mention connections between our methodology and some alternative approaches that could be taken.
While our generative model incorporates randomness through the stochastic disturbance, an alternative formulation has been proposed by \textcite{chow_comparison_2016} in which variability is introduced through random perturbations to the ODE parameters.
A key difference between this approach and our perspective of PDA as an approximate model described in Section \ref{sec:non-linear} is that it assumes the form of the deterministic ODE is known (potentially after some model exploratorion, see \textcite[p. 176]{chow_comparison_2016}) and focuses only on estimation of the parameters using a non-linear mixed effects model. 
However, it would be particularly interesting to formulate a model with both stochastic disturbances and random perturbations to parameters, for example, when extending the single-subject analysis in Section \ref{examples} to model data from multiple subjects.
In fact, both stochastic forcing functions and parameters were posited as extensions of PDA in earlier work by \textcite{ramsay_functional_2000}, but not pursued.
When working with non-linear ODEs as described in Section \ref{sec:non-linear}, it is also natural to ask whether we could model the full non-linear function $g(x, y)$ non-parametrically rather than use PDA to linearly approximate the system's Jacobian. 
A brief simulation for the Van der Pol model, contained in Appendix \ref{sec:additional-vdp}, revealed that the partial derivatives of such non-parametric estimates are subject to similar problems with bias to those we encounter in PDA.
Finally, PDA can be seen as the functional analogue of the classical gradient matching (or two-stage least squares) method, as non-parametric regression is used to estimate the function, then its derivatives are calculated and a second regression used to estimate the ODE \parencite{varah_spline_1982}. 
It is therefore an ``indirect method" as parameter estimation does not involve solving the ODE directly.
In contrast, a direct approach to estimating our generative model would involve parameterising the stochastic disturbance using basis functions and random effects. 
Examples for simpler and non-functional models are provided in \textcite[Chapter 5]{wrobel_functional_2019} and \textcite{hooker_forcing_2009}, respectively, but we leave the investigation of a direct approach for our model to future work.
\section*{Acknowledgment}
During the conception and development of this work, E.G. was supported in part by Science Foundation Ireland (SFI) (Grant No. 18/CRT/6049) and co-funded under the European Regional Development Fund.
This work emanated from an initial research visit by E.G. to G.H. at the Department of Statistics at the University of California, Berkeley, supported by the SFI Centre for Research Training in Foundations of Data Science and the host department; we gratefully acknowledge the support of both parties for this visit.
We are grateful to Prof. Kieran Moran and the RISC study group at Dublin City University for the dataset used in our example analysis.
\textbf{Note}: An early-stage, preliminary version of this work first appeared as a chapter in E.G.'s Ph.D. thesis.

\appendix
\section{Generative Model}\label{sec:generative-model-definitions}
\subsection{ODE Model}
We work with the following linear time-varying ODE model
\begin{equation}
D^m x(t) = - \beta_0 (t) x(t) - \dotso - \beta_{m-1} (t) D^{m-1} x(t) + \epsilon(t),
\end{equation}
for $t \in [0, T]$, where $\epsilon(t) \sim \mathcal{GP}(0, C)$.
The model is re-written in matrix form as
$$
D \widetilde{\mathbf{x}} (t) = \mathbf{B} (t) \widetilde{\mathbf{x}} (t) + \widetilde{\boldsymbol{\epsilon}}(t),
$$
where
$$
\widetilde{\mathbf{x}} (t)
= 
\begin{pmatrix}
x(t) \\
D x(t) \\
\vdots \\
D^{m-1} x(t)
\end{pmatrix},
\quad
\widetilde{\mathbf{x}}_0 = 
\begin{pmatrix}
x(0) \\
D x(0) \\
\vdots \\
D^{m-1} x(0)
\end{pmatrix},
$$
$$
\mathbf{B} (t) = \begin{pmatrix}
0 & 1 & 0 & \hdots & 0 \\
0 & 0 & 1 & \hdots & 0 \\
\vdots & \vdots & \vdots & \vdots \ \vdots \ \vdots & \vdots \\
0 & 0 & 0 & \hdots & 1 \\
- \beta_0 (t) & - \beta_1 (t) & - \beta_2 (t) & \hdots & - \beta_{m-1} (t)
\end{pmatrix}
\quad
\text{and}
\quad
\widetilde{\boldsymbol{\epsilon}}(t) = 
\begin{pmatrix}
0 \\
0 \\
\vdots \\
\epsilon(t)
\end{pmatrix}.
$$

\subsection{Solution and Generative Model}
The solution to the ODE, and hence the generative model for $\widetilde{\mathbf{x}}(t)$, is
\begin{equation}
    \widetilde{\mathbf{x}} (t) = \fundmat \ \widetilde{\mathbf{x}}_0 + \int _{0} ^ {t} \fundmats \widetilde{\boldsymbol{\epsilon}}(s) \mathrm{d}s,
\end{equation}
where $\fundmat$ is the state transition matrix, which contains the $m$ linearly independent solutions to the homogeneous equation $D \widetilde{\mathbf{x}} (t) = \mathbf{B} (t) \widetilde{\mathbf{x}} (t)$ in its columns and
$$
\mathbf{C} (s, t) =
\Expec \left[ \widetilde{\boldsymbol{\epsilon}}(s) \widetilde{\boldsymbol{\epsilon}}(t)^\top \right]
=
\begin{pmatrix}
0 & 0 & \hdots & 0 \\
0 & 0 & \hdots & 0 \\
\vdots & \vdots & \vdots \vdots \vdots & \vdots \\
0 & 0 & \hdots & C(s, t)
\end{pmatrix}.
$$
We assume that the vector of initial conditions is independent of $\epsilon(t)$ and follows an $m$-dimensional Gaussian distribution
$$
\widetilde{\mathbf{x}}_0 = \begin{pmatrix}
x (0) \\
\vdots \\
D^{m-1} x(0)
\end{pmatrix}
\sim
\mathcal{N}_m (\boldsymbol{\mu}_0, \boldsymbol{\Sigma}_0).
$$

\section{Moments of Generative Model}

\subsection{Expectation}
$$
\Expec[\widetilde{\mathbf{x}} (t)] = \fundmat \boldsymbol{\mu}_0
$$
\textbf{Workings:}
$$
\Expec\left[\fundmat \ \widetilde{\mathbf{x}}_0 + \int _{0} ^ {t} \fundmats \widetilde{\boldsymbol{\epsilon}}(s) \mathrm{d}s\right]
$$
$$
=\fundmat \ \underbrace{\Expec[\widetilde{\mathbf{x}}_0]}_{=\boldsymbol{\mu}_0} + \int _{0} ^ {t} \fundmats \underbrace{\Expec[\widetilde{\boldsymbol{\epsilon}}(s)]}_{=\mathbf{0}} \mathrm{d}s
$$
$$
= \fundmat \ \boldsymbol{\mu}_0.
$$

\subsection{Covariance}
$$
\Cov(\widetilde{\mathbf{x}} (s), \widetilde{\mathbf{x}} (t)) 
$$
$$
=  \fundmatso \boldsymbol{\Sigma}_0 \fundmat^\top + \   \int_0^s \int_0^t \fundmatsu \mathbf{C}(u, v) \fundmattv^\top \mathrm{d}v \mathrm{d}u \ 
$$
\textbf{Workings:} \\
\textbf{Step 1:}
$$
\widetilde{\mathbf{x}} (t) - \Expec[\widetilde{\mathbf{x}} (t)]
$$
$$
= \widetilde{\mathbf{x}} (t) - \fundmat \boldsymbol{\mu}_0
$$
$$
= \fundmat \ \widetilde{\mathbf{x}}_0 + \int _{0} ^ {t} \fundmats \widetilde{\boldsymbol{\epsilon}}(s) \mathrm{d}s - \fundmat \boldsymbol{\mu}_0
$$
$$
= \fundmat \ (\widetilde{\mathbf{x}}_0 - \boldsymbol{\mu}_0)  + \int _{0} ^ {t} \fundmats \widetilde{\boldsymbol{\epsilon}}(s) \mathrm{d}s.
$$
\textbf{Step 2:} 
\begin{align}
\left( \widetilde{\mathbf{x}} (s) - \Expec[\widetilde{\mathbf{x}} (s)]\right) \left(\widetilde{\mathbf{x}} (t) - \Expec[\widetilde{\mathbf{x}} (t)]\right)^\top
    =  &\left( \fundmatso \ (\widetilde{\mathbf{x}}_0 - \boldsymbol{\mu}_0) + \int _{0} ^ {s} \fundmatsu \widetilde{\boldsymbol{\epsilon}}(u) \mathrm{d}u \right) \\
    &\left( \fundmat \ (\widetilde{\mathbf{x}}_0 - \boldsymbol{\mu}_0) + \int _{0} ^ {t} \fundmattv \widetilde{\boldsymbol{\epsilon}}(v) \mathrm{d}v\right)^\top \\
    =  &\left(\fundmatso \ (\widetilde{\mathbf{x}}_0 - \boldsymbol{\mu}_0)  + \int _{0} ^ {s} \fundmatsu \widetilde{\boldsymbol{\epsilon}}(u) \mathrm{d}u\right) \\
    &\left( \left( \fundmat \ (\widetilde{\mathbf{x}}_0 - \boldsymbol{\mu}_0)\right)^\top  + \left(\int _{0} ^ {t} \fundmattv \widetilde{\boldsymbol{\epsilon}}(v) \mathrm{d}v\right)^\top \right) \\
    =  &\left(\fundmatso \ (\widetilde{\mathbf{x}}_0 - \boldsymbol{\mu}_0)  + \int _{0} ^ {s} \fundmatsu \widetilde{\boldsymbol{\epsilon}}(u) \mathrm{d}u\right) \\
    &\left((\widetilde{\mathbf{x}}_0 - \boldsymbol{\mu}_0)^\top \fundmat^\top  + \int _{0} ^ {t} \widetilde{\boldsymbol{\epsilon}}(v)^\top \fundmattv^\top  \mathrm{d}v\right),
\end{align} 
which is obtained from the standard identities for matrix transposes $(\mathbf{A} + \mathbf{B})^\top = \mathbf{A}^\top + \mathbf{B^\top}$ and $(\mathbf{A}\mathbf{B}\mathbf{C})^\top = \mathbf{C}^\top\mathbf{B}^\top\mathbf{A}^\top$. Using the distributivity of matrix multiplication (i.e., $(\mathbf{A} + \mathbf{B}) \mathbf{C} = \mathbf{A}\mathbf{C} + \mathbf{B}\mathbf{C}$) and re-arranging terms by bringing some inside the integrals, this is equal to  
\begin{align}
     & \ \fundmatso \ (\widetilde{\mathbf{x}}_0 - \boldsymbol{\mu}_0) (\widetilde{\mathbf{x}}_0 - \boldsymbol{\mu}_0)^\top \fundmat^\top
    +  \left(\int _{0} ^ {s} \fundmatsu \widetilde{\boldsymbol{\epsilon}}(u) \mathrm{d}u \right) (\widetilde{\mathbf{x}}_0 - \boldsymbol{\mu}_0)^\top \fundmat^\top \\
    + & \ \fundmatso \ (\widetilde{\mathbf{x}}_0 - \boldsymbol{\mu}_0) \int _{0} ^ {t} \widetilde{\boldsymbol{\epsilon}}(v)^\top \fundmattv^\top  \mathrm{d}v 
    + \int _{0} ^ {s} \int _{0} ^ {t}  \fundmatsu \widetilde{\boldsymbol{\epsilon}}(u) \widetilde{\boldsymbol{\epsilon}}(v)^\top \fundmattv^\top  \mathrm{d}v \mathrm{d}u.
\end{align}
\textbf{Step 3:} 
\begin{align}
    \ \Expec[(\widetilde{\mathbf{x}} (s) - \Expec[\widetilde{\mathbf{x}} (s)]) (\widetilde{\mathbf{x}} (t) - \Expec[\widetilde{\mathbf{x}} (t)])^\top]
    = & \ \fundmatso \ \underbrace{\Expec[(\widetilde{\mathbf{x}}_0 - \boldsymbol{\mu}_0) (\widetilde{\mathbf{x}}_0 - \boldsymbol{\mu}_0)^\top]}_{= \ \boldsymbol{\Sigma}_0} \fundmat^\top \\
    + & \ \int _{0} ^ {s} \fundmatsu \underbrace{\Expec[\widetilde{\boldsymbol{\epsilon}}(u)]}_{= \ \mathbf{0}} \mathrm{d}u \ \underbrace{\Expec[(\widetilde{\mathbf{x}}_0 - \boldsymbol{\mu}_0)^\top]}_{= \ \mathbf{0}} \fundmat^\top \\
    + & \ \fundmatso \ \underbrace{\Expec[(\widetilde{\mathbf{x}}_0 - \boldsymbol{\mu}_0)]}_{= \ \mathbf{0}} \int _{0} ^ {t} \underbrace{\Expec[\widetilde{\boldsymbol{\epsilon}}(v)^\top]}_{= \ \mathbf{0}} \fundmattv^\top  \mathrm{d}v \\
    + & \  \int _{0} ^ {s} \int _{0} ^ {t}  \fundmatsu \underbrace{\Expec[\widetilde{\boldsymbol{\epsilon}}(u) \widetilde{\boldsymbol{\epsilon}}(v)^\top]}_{= \ \mathbf{C} (u, v)} \fundmattv^\top  \mathrm{d}v \mathrm{d}u,
\end{align}
which is obtained from the linearity of expectation, that $\fundmats$ is fixed and that $\boldsymbol{\epsilon}(t)$ and $\widetilde{\mathbf{x}}_0$ are independent so, e.g.,  $\Expec[\boldsymbol{\epsilon}(t)\widetilde{\mathbf{x}}_0^\top] = \Expec[\boldsymbol{\epsilon}(t)] \Expec[\widetilde{\mathbf{x}}_0^\top]$, and the definitions in Section \ref{sec:generative-model-definitions}. Putting it all together, we obtain
$$
\Cov(\widetilde{\mathbf{x}} (s), \widetilde{\mathbf{x}} (t))  = \fundmatso  \boldsymbol{\Sigma}_0 \fundmat^\top +
\int _{0} ^ {s} \int _{0} ^ {t}  \fundmatsu \mathbf{C} (u, v) \fundmattv^\top  \mathrm{d}v \mathrm{d}u.
$$
\vfill
\section{Bias}\label{sec:full-bias}
\subsection{Bias in the Coefficients}
In this section, we provide the general form of the bias arising from dependence between the covariates in the PDA model and the stochastic disturbance. We define 
$$
D^{m} \mathbf{x} (t)
=
\begin{pmatrix}
D^{m} x_1 (t) \\
\vdots \\
D^{m} x_N (t)
\end{pmatrix}, 
\quad
\widetilde{\mathbf{X}} (t)
= 
\begin{pmatrix}
x_1 (t) & \dots & D^{m-1} x_1 (t) \\
\vdots & \vdots \vdots \vdots & \vdots \\
x_N (t) & \dots & D^{m-1} x_N (t) \\
\end{pmatrix} ,
$$
$$
\boldsymbol{\beta} (t) = 
\begin{pmatrix}
\beta_0 (t) \\
\vdots \\
\beta_{m-1} (t) \\
\end{pmatrix}
\quad \text{and} \quad
\boldsymbol{\epsilon} (t) 
=
\begin{pmatrix}
\epsilon_1 (t) \\
\vdots \\
\epsilon_N (t) \\
\end{pmatrix}.
$$
Therefore the PDA regression model is written in matrix form\footnote{When an intercept is included, a column of ones is added to  $\widetilde{\mathbf{X}}$.} as
$$
\underbrace{D^{m} \mathbf{x} (t)}_{N \times 1}
= 
\underbrace{\widetilde{\mathbf{X}} (t)}_{N \times m}
\underbrace{\boldsymbol{\beta} (t)}_{m \times 1}
+
\underbrace{\boldsymbol{\epsilon} (t)}_{N \times 1}.
$$
The least-squares estimator is
\begin{align*}
    \widehat{\boldsymbol{\beta}} (t) &= (\widetilde{\mathbf{X}} (t) ^\top \widetilde{\mathbf{X}} (t))^{-1} \widetilde{\mathbf{X}} (t)^\top D^{m} \mathbf{x} (t) \\
    &= (\widetilde{\mathbf{X}} (t) ^\top \widetilde{\mathbf{X}} (t))^{-1} \widetilde{\mathbf{X}} (t)^\top (\widetilde{\mathbf{X}} (t)
\boldsymbol{\beta} (t) + \boldsymbol{\epsilon} (t))\\
    &= \underbrace{(\widetilde{\mathbf{X}} (t) ^\top \widetilde{\mathbf{X}} (t))^{-1} (\widetilde{\mathbf{X}} (t)^\top \widetilde{\mathbf{X}} (t))}_{ = \ \mathbf{I}_{m}}
\boldsymbol{\beta} (t)
+
(\widetilde{\mathbf{X}} (t) ^\top \widetilde{\mathbf{X}} (t))^{-1} (\widetilde{\mathbf{X}} (t)^\top \boldsymbol{\epsilon} (t)) \\
&=  \boldsymbol{\beta} (t) + 
(\widetilde{\mathbf{X}} (t) ^\top \widetilde{\mathbf{X}} (t))^{-1} (\widetilde{\mathbf{X}} (t)^\top \boldsymbol{\epsilon} (t)).
\end{align*}
Therefore, the bias is
$$
\Bias\left(\widehat{\boldbeta} (t)\right) = \Expec \left[\left(\widetilde{\mathbf{X}} (t) ^\top \widetilde{\mathbf{X}} (t)\right)^{-1} \widetilde{\mathbf{X}} (t)^\top \boldsymbol{\epsilon} (t)\right].
$$

\subsection{General Form of the Bias Correction}\label{sec:general-bias-correction}
Given $\widetilde{\mathbf{X}} (t)$, the conditional bias is
\begin{align*}
    \Bias\left(\widehat{\boldsymbol{\beta}} (t) | \widetilde{\mathbf{X}} (t)\right) &= \Expec\left[\widehat{\boldsymbol{\beta}} (t) | \widetilde{\mathbf{X}} (t)\right] - \boldsymbol{\beta} (t)  \\
    &= \left(\widetilde{\mathbf{X}} (t) ^\top \widetilde{\mathbf{X}} (t)\right)^{-1} \left(\widetilde{\mathbf{X}} (t)^\top \Expec\left[\boldsymbol{\epsilon} (t)| \widetilde{\mathbf{X}} (t)\right]\right).
\end{align*}
Looking more closely, the numerator is
\begin{align*}
    \widetilde{\mathbf{X}} (t)^\top \Expec\left[\boldsymbol{\epsilon} (t)| \widetilde{\mathbf{X}} (t)\right]
&=
\underbrace{
\begin{pmatrix}
x_1 (t) & \dots & x_N (t) \\ 
\vdots & \vdots \vdots \vdots & \vdots \\
D^{m-1} x_1 (t)  & \dots & D^{m-1} x_N (t) \\
\end{pmatrix}
}_{m \times N}
\underbrace{\begin{pmatrix}
\Expec\left[\epsilon_1 (t) | \widetilde{\mathbf{X}} (t)\right] \\
\vdots \\
\Expec\left[\epsilon_N (t) | \widetilde{\mathbf{X}} (t)\right] \\
\end{pmatrix}}_{N \times 1} \\
&= 
\underbrace{
\begin{pmatrix}
\sum_{i=1}^N x_i (t) \Expec\left[\epsilon_i (t)|\widetilde{\mathbf{X}} (t)\right] \\
\vdots \\
\sum_{i=1}^N D^{m-1} x_i (t) \Expec\left[\epsilon_i (t)|\widetilde{\mathbf{X}} (t)\right]
\end{pmatrix}
}_{m \times 1} \\
&= N 
\cdot 
\begin{pmatrix}
\frac{1}{N} \sum_{i=1}^N x_i (t) \Expec\left[\epsilon_i (t)|\widetilde{\mathbf{X}} (t)\right] \\
\vdots \\
\frac{1}{N} \sum_{i=1}^N D^{m-1} x_i (t) \Expec\left[\epsilon_i (t)|\widetilde{\mathbf{X}} (t)\right]
\end{pmatrix}. 
\end{align*}
Therefore, we aim to replace this term with its expectation
$$
N \cdot 
\begin{pmatrix}
\Expec\left[x(t) \epsilon(t)\right] \\
\vdots \\
\Expec\left[D^{m-1} x(t) \epsilon(t)\right]
\end{pmatrix}.
$$
If $\boldsymbol{\beta}(t)$ and $\mathbf{C} (s, t)$ are known, then
\begin{align*}
    \widetilde{\mathbf{x}} (t) \widetilde{\boldsymbol{\epsilon}} (t) ^\top
    =&
    \begin{pmatrix}
x(t) \\
D x(t) \\
\vdots \\
D^{m-1} x(t)
\end{pmatrix}
\begin{pmatrix}
0, & 0, & \dots & \epsilon(t)  \\
\end{pmatrix} \\
=& \underbrace{
\begin{pmatrix}
0 & \dots & x(t) \epsilon (t) \\
\vdots & \vdots \vdots \vdots & \vdots \\
0 & \dots & D^{m-1} x(t) \epsilon (t)
\end{pmatrix}
}_{m \times m}.
\end{align*}
Using the ODE solution, this is also equal to
\begin{align*}
    &\left(\fundmat \ \widetilde{\mathbf{x}}_0 + \int _{0} ^ {t} \fundmats \widetilde{\boldsymbol{\epsilon}}(s)  \mathrm{d}s \right) \widetilde{\boldsymbol{\epsilon}} (t) ^\top \\
    =& \ \fundmat \ \widetilde{\mathbf{x}}_0 \ \widetilde{\boldsymbol{\epsilon}} (t) ^\top + \int _{0} ^ {t} \fundmats \widetilde{\boldsymbol{\epsilon}}(s) \widetilde{\boldsymbol{\epsilon}} (t) ^\top  \mathrm{d}s.
\end{align*}
Thus 
\begin{align}
    \Expec\left[\widetilde{\mathbf{x}} (t) \widetilde{\boldsymbol{\epsilon}} (t) ^\top\right] &= 
    \begin{pmatrix}
    0 & \dots & \Expec\left[x(t) \epsilon (t)\right] \\
    \vdots & \vdots \vdots \vdots & \vdots \\
    0 & \dots & \Expec\left[D^{m-1} x(t) \epsilon (t)\right]
    \end{pmatrix} \\
    &= \Expec\left[\fundmat \ \widetilde{\mathbf{x}}_0 \ \widetilde{\boldsymbol{\epsilon}} (t) ^\top + \int _{0} ^ {t} \fundmats \widetilde{\boldsymbol{\epsilon}}(s) \widetilde{\boldsymbol{\epsilon}} (t) ^\top  \mathrm{d}s\right] \\
    &= \fundmat \ \widetilde{\mathbf{x}}_0 \ \underbrace{\Expec\left[\widetilde{\boldsymbol{\epsilon}} (t) ^\top\right]}_{=\textbf{0}} + \int _{0} ^ {t} \fundmats \underbrace{\Expec \left[ \widetilde{\boldsymbol{\epsilon}}(s) \widetilde{\boldsymbol{\epsilon}} (t) ^\top \right]}_{= \textbf{C}(s, t)} \mathrm{d}s \\
    &= \int _{0} ^ {t} \fundmats \textbf{C}(s, t) \mathrm{d}s.
\end{align}
Therefore the bias is
$$
N \cdot \left(\widetilde{\mathbf{X}} (t) ^\top \widetilde{\mathbf{X}} (t)\right)^{-1} \left[  \int _{0} ^ {t} \fundmats \textbf{C}(s, t) \mathrm{d}s \right]_{\cdot \ m},
$$
where $_{\cdot \ m}$ denotes the $m$th column of the matrix. In practice, $\boldsymbol{\beta}(t)$ and $\mathbf{C} (s, t)$ are unknown so we plug in an estimate $\widehat{C}(s, t)$, which is calculated from the empirical residuals, for $C(s, t)$ and use the estimate $\widehat{\boldbeta} (t)$, rather than $\boldbeta (t)$ to calculate $\fundmats$. The following generalises the high-level algorithm sketch contained in the main text, while full details of the algorithm and its numerical implementation are contained in Section \ref{sec:numerical-implementation}.
When an intercept is included and a vector of ones is added as the first column of the design matrix $\widetilde{\mathbf{X}}$, then $0$ is added as the first entry of the vector $\left[  \int _{0} ^ {t} \fundmats \textbf{C}(s, t) \mathrm{d}s \right]_{\cdot \ m}$ because $\Expec\left[\sum_{i=1}^N \epsilon_i(t)\right] = N \cdot E[\epsilon(t)] = 0$.

\begin{framed}{\textbf{General Bias Reduction Algorithm}}
    \begin{enumerate}
    \item Obtain initial OLS estimates $\widehat{\boldbeta} (t)$ and $\widehat{C} (s, t)$. Substitute these for $\boldbeta (t)$ and $C (s, t)$ to obtain an estimate $\widehat{\Expec}[\widetilde{\mathbf{x}} (t) \widetilde{\boldsymbol{\epsilon}} (t) ^\top]$ of $\Expec\left[\widetilde{\mathbf{x}} (t) \widetilde{\boldsymbol{\epsilon}} (t) ^\top\right]$.
    \item Calculate a bias-reduced estimate:
    $$
    \widehat{\boldbeta}^ {(bc)}(t) = \widehat{\boldbeta}  (t) - N \cdot (\widetilde{\mathbf{X}} (t) ^\top \widetilde{\mathbf{X}} (t))^{-1} \widehat{\Expec}[\widetilde{\mathbf{x}} (t) \widetilde{\boldsymbol{\epsilon}} (t) ^\top]_{\cdot \ m}
    $$
    \item If required, iterate through steps 1-2 using the bias-reduced estimate $\widehat{\boldbeta}^ {(bc)}(t)$ in place of $\widehat{\boldbeta}(t)$ at step 1, calculating updated residuals and re-estimating $\widehat{C} (s, t)$.
\end{enumerate}
\end{framed}

\vfill

\subsection{Bias in the Covariance Function}
We have that
\begin{align*}
& \widehat{\boldsymbol{\epsilon}} (s) ^\top \widehat{\boldsymbol{\epsilon}} (t)\\
 =& \left(D^{m} \mathbf{x} (s) - \widetilde{\mathbf{X}} (s) \widehat{\boldsymbol{\beta}} (s)\right)^\top
\left(D^{m} \mathbf{x} (t) - \widetilde{\mathbf{X}} (t) \widehat{\boldsymbol{\beta}} (t)\right) \\
= & \left(\widetilde{\mathbf{X}} (s) \boldsymbol{\beta} (s) + \boldsymbol{\epsilon} (s)  - \widetilde{\mathbf{X}} (s) \widehat{\boldsymbol{\beta}} (s)\right)^\top
\left(\widetilde{\mathbf{X}} (t) \boldsymbol{\beta} (t) + \boldsymbol{\epsilon} (t)  - \widetilde{\mathbf{X}} (t) \widehat{\boldsymbol{\beta}} (t)\right) \\
= & \left(\widetilde{\mathbf{X}} (s) \left(\boldsymbol{\beta} (s) -  \widehat{\boldsymbol{\beta}} (s)\right) + \boldsymbol{\epsilon} (s)\right)^\top
\left(\widetilde{\mathbf{X}} (t) \left(\boldsymbol{\beta} (t) -  \widehat{\boldsymbol{\beta}} (t)\right) + \boldsymbol{\epsilon} (t)\right) \\
= & 
\left(\left(\boldsymbol{\beta} (s) -  \widehat{\boldsymbol{\beta}} (s)\right)^\top \widetilde{\mathbf{X}} (s)^\top  + \boldsymbol{\epsilon} (s)^\top\right)
\left(\widetilde{\mathbf{X}} (t) \left(\boldsymbol{\beta} (t) -  \widehat{\boldsymbol{\beta}} (t)\right) + \boldsymbol{\epsilon} (t)\right) \\
= &
\left(\boldsymbol{\beta} (s) -  \widehat{\boldsymbol{\beta}} (s)\right)^\top \widetilde{\mathbf{X}} (s)^\top
\widetilde{\mathbf{X}} (t) \left(\boldsymbol{\beta} (t) -  \widehat{\boldsymbol{\beta}} (t)\right) \\
& \ +
\boldsymbol{\epsilon} (s)^\top 
\widetilde{\mathbf{X}} (t) \left(\boldsymbol{\beta} (t) -  \widehat{\boldsymbol{\beta}} (t)\right)  \\
& \ + 
\left(\boldsymbol{\beta} (s) -  \widehat{\boldsymbol{\beta}} (s)\right)^\top \widetilde{\mathbf{X}} (s)^\top
\boldsymbol{\epsilon} (t) \\
& \ +
\boldsymbol{\epsilon} (s)^\top 
\boldsymbol{\epsilon} (t).
\end{align*}
Now using that $\boldsymbol{\beta} (s) -  \widehat{\boldsymbol{\beta}} (s) = -\left(\widetilde{\mathbf{X}} (t) ^\top \widetilde{\mathbf{X}} (t)\right)^{-1} \left(\widetilde{\mathbf{X}} (t)^\top \boldsymbol{\epsilon} (t)\right)$,
we have that the above expression is equal to
\begin{align*}
    & \left(\left(\widetilde{\mathbf{X}} (s) ^\top \widetilde{\mathbf{X}} (s)\right)^{-1} \widetilde{\mathbf{X}} (s)^\top \boldsymbol{\epsilon} (s)\right)^\top  \widetilde{\mathbf{X}} (s)^\top \widetilde{\mathbf{X}} (t) \left(\widetilde{\mathbf{X}} (t) ^\top \widetilde{\mathbf{X}} (t)\right)^{-1} \widetilde{\mathbf{X}} (t)^\top \boldsymbol{\epsilon} (t) \\ 
    &- \ \boldsymbol{\epsilon} (s)^\top 
\widetilde{\mathbf{X}} (t) \left(\widetilde{\mathbf{X}} (t) ^\top \widetilde{\mathbf{X}} (t)\right)^{-1} \widetilde{\mathbf{X}} (t)^\top \boldsymbol{\epsilon} (t) \\
&- \ \left(\left(\widetilde{\mathbf{X}} (s) ^\top \widetilde{\mathbf{X}} (s)\right)^{-1}\widetilde{\mathbf{X}} (s)^\top \boldsymbol{\epsilon} (s)\right)^\top \widetilde{\mathbf{X}} (s)^\top
\boldsymbol{\epsilon} (t) \\
&+ \
\boldsymbol{\epsilon} (s)^\top 
\boldsymbol{\epsilon} (t) \\
= & \ \boldsymbol{\epsilon} (s)^\top \widetilde{\mathbf{X}} (s) \left(\widetilde{\mathbf{X}} (s) ^\top \widetilde{\mathbf{X}} (s)\right)^{-1}  \widetilde{\mathbf{X}} (s)^\top \widetilde{\mathbf{X}} (t) \left(\widetilde{\mathbf{X}} (t) ^\top \widetilde{\mathbf{X}} (t)\right)^{-1} \widetilde{\mathbf{X}} (t)^\top \boldsymbol{\epsilon} (t) \\
&-\
\boldsymbol{\epsilon} (s)^\top 
\widetilde{\mathbf{X}} (t) \left(\widetilde{\mathbf{X}} (t) ^\top \widetilde{\mathbf{X}} (t)\right)^{-1} \widetilde{\mathbf{X}} (t)^\top \boldsymbol{\epsilon} (t) \\
&- \
\boldsymbol{\epsilon} (s)^\top \widetilde{\mathbf{X}} (s) \left(\widetilde{\mathbf{X}} (s) ^\top \widetilde{\mathbf{X}} (s)\right)^{-1} \widetilde{\mathbf{X}} (s)^\top
\boldsymbol{\epsilon} (t) \\
&+ \
\boldsymbol{\epsilon} (s)^\top 
\boldsymbol{\epsilon} (t).
\end{align*}
When $s=t$, we have the simplification
\begin{align*}
    & \widehat{\boldsymbol{\epsilon}} (s) ^\top \widehat{\boldsymbol{\epsilon}} (t)\\
 &=  \boldsymbol{\epsilon} (t)^\top \widetilde{\mathbf{X}} (t) \left(\widetilde{\mathbf{X}} (t) ^\top \widetilde{\mathbf{X}} (t)\right)^{-1} \widetilde{\mathbf{X}} (t)^\top \boldsymbol{\epsilon} (t) \\
  & \ -
  \boldsymbol{\epsilon} (t)^\top 
\widetilde{\mathbf{X}} (t) \left(\widetilde{\mathbf{X}} (t) ^\top \widetilde{\mathbf{X}} (t)\right)^{-1} \widetilde{\mathbf{X}} (t)^\top \boldsymbol{\epsilon} (t) \\
  & \ -
\boldsymbol{\epsilon} (t)^\top 
\widetilde{\mathbf{X}} (t) \left(\widetilde{\mathbf{X}} (t) ^\top \widetilde{\mathbf{X}} (t)\right)^{-1} \widetilde{\mathbf{X}} (t)^\top \boldsymbol{\epsilon} (t) \\
  & \ +
  \boldsymbol{\epsilon} (t)^\top 
\boldsymbol{\epsilon} (t) \\
&= 
\boldsymbol{\epsilon} (s)^\top 
\boldsymbol{\epsilon} (t) - \boldsymbol{\epsilon} (t)^\top 
\widetilde{\mathbf{X}} (t) \left(\widetilde{\mathbf{X}} (t) ^\top \widetilde{\mathbf{X}} (t)\right)^{-1} \widetilde{\mathbf{X}} (t)^\top \boldsymbol{\epsilon} (t).
\end{align*}
Therefore, our estimator 
$$
\widehat{C}(s, t) = \frac{\widehat{\boldsymbol{\epsilon}} (s) ^\top \widehat{\boldsymbol{\epsilon}} (t)}{N-m},
$$
has the bias
\begin{align*}
    & \ \Expec\left[\frac{\boldsymbol{\epsilon} (s)^\top \widetilde{\mathbf{X}} (s) \left(\widetilde{\mathbf{X}} (s) ^\top \widetilde{\mathbf{X}} (s)\right)^{-1}  \widetilde{\mathbf{X}} (s)^\top \widetilde{\mathbf{X}} (t) \left(\widetilde{\mathbf{X}} (t) ^\top \widetilde{\mathbf{X}} (t)\right)^{-1} \widetilde{\mathbf{X}} (t)^\top \boldsymbol{\epsilon} (t)}{N-m} \right]\\
-&\
\Expec\left[\frac{\boldsymbol{\epsilon} (s)^\top 
\widetilde{\mathbf{X}} (t) \left(\widetilde{\mathbf{X}} (t) ^\top \widetilde{\mathbf{X}} (t)\right)^{-1} \widetilde{\mathbf{X}} (t)^\top \boldsymbol{\epsilon} (t)}{N-m}\right] \\
-& \
\Expec\left[\frac{\boldsymbol{\epsilon} (s)^\top \widetilde{\mathbf{X}} (s) \left(\widetilde{\mathbf{X}} (s) ^\top \widetilde{\mathbf{X}} (s)\right)^{-1} \widetilde{\mathbf{X}} (s)^\top
\boldsymbol{\epsilon} (t)}{N-m}\right].
\end{align*}
Likewise, the pointwise error variance estimator
$$
\widehat{C}(t, t) = \frac{\widehat{\boldsymbol{\epsilon}} (t) ^\top \widehat{\boldsymbol{\epsilon}} (t)}{N-m},
$$
has the bias
$$
- \Expec\left[\frac{\boldsymbol{\epsilon} (t)^\top 
\widetilde{\mathbf{X}} (t) \left(\widetilde{\mathbf{X}} (t) ^\top \widetilde{\mathbf{X}} (t)\right)^{-1} \widetilde{\mathbf{X}} (t)^\top \boldsymbol{\epsilon} (t)}{N-m}\right].
$$

\subsubsection{Effect of Updating the Residuals}
It is worth understanding the effect of updating the residuals on the estimated residual covariance function $\widehat{C}(s, t)$, or equivalently $\widehat{\boldsymbol{\epsilon}}(s)^\top \widehat{\boldsymbol{\epsilon}}(t)$. The bias-corrected residual is
$$
\widehat{\boldeps} \bc (t) = D^m \boldx (t) - \widetilde{\boldX}(t) \widehat{\boldbeta} \bc (t),
$$
where $\widehat{\boldbeta} \bc (t) = \widehat{\boldbeta} (t) - \widehat{\Bias}\left({\widehat{\boldbeta}}(t)\right)$. Re-arranging gives
\begin{align*}
    \widehat{\boldeps} \bc (t) =& \  D^m \boldx (t)
- \widetilde{\boldX}(t) (\widehat{\boldbeta} (t) - \widehat{\Bias}\left({\widehat{\boldbeta}}(t)\right)) \\
=& \  \underbrace{D^m \boldx (t)
- \widetilde{\boldX}(t) \widehat{\boldbeta} (t)}_{=\widehat{\boldeps}(t)} + \widetilde{\boldX}(t) \widehat{\Bias}\left({\widehat{\boldbeta}}(t)\right) \\
=& \  \widehat{\boldeps}(t) + \widetilde{\boldX}(t) \widehat{\Bias}\left({\widehat{\boldbeta}}(t)\right).
\end{align*}
Then
\begin{align*}
    \widehat{\boldeps} \bc (s)^\top \widehat{\boldeps} \bc (t) =& \  
\left(\widehat{\boldeps}(s) + \widetilde{\boldX}(s) \widehat{\Bias}\left({\widehat{\boldbeta}}(s)\right)\right)^\top \left(\widehat{\boldeps}(t) + \widetilde{\boldX}(t) \widehat{\Bias}\left({\widehat{\boldbeta}}(t)\right)\right) \\
=& \ 
\left(\widehat{\boldeps}(s)^\top + \left(\widetilde{\boldX}(s) \widehat{\Bias}\left({\widehat{\boldbeta}}(s)\right)\right)^\top\right) \left(\widehat{\boldeps}(t) + \widetilde{\boldX}(t) \widehat{\Bias}\left({\widehat{\boldbeta}}(t)\right)\right) \\
 =& \  
\left(\widehat{\boldeps}(s)^\top + \widehat{\Bias}\left({\widehat{\boldbeta}}(s)\right)^\top \widetilde{\boldX}(s)^\top \right) \left(\widehat{\boldeps}(t) + \widetilde{\boldX}(t) \widehat{\Bias}\left({\widehat{\boldbeta}}(t)\right)\right) \\
=& \ 
\widehat{\boldeps}(s)^\top \widehat{\boldeps}(t) 
+
\widehat{\boldeps}(s)^\top \widetilde{\boldX}(t) \widehat{\Bias}\left({\widehat{\boldbeta}}(t)\right) 
+
\widehat{\Bias}\left({\widehat{\boldbeta}}(s)\right)^\top \widetilde{\boldX}(s)^\top \widehat{\boldeps}(t) \\
& 
+ \widehat{\Bias}\left({\widehat{\boldbeta}}(s)\right)^\top \widetilde{\boldX}(s)^\top \widetilde{\boldX}(t) \widehat{\Bias}\left({\widehat{\boldbeta}}(t)\right).
\end{align*}
Now, plugging in $\widehat{\Bias}\left({\widehat{\boldbeta}}(t)\right) =\left(\widetilde{\boldX}(t)^\top \widetilde{\boldX}(t)\right)^{-1} \widehat{\Expec}[\widetilde{\boldX}(t)^\top \boldeps (t)]$, we have
\begin{align*}
    \widehat{\boldeps} \bc (s)^\top \widehat{\boldeps} \bc (t) =& \  
    \widehat{\boldeps}(s)^\top \widehat{\boldeps}(t) \\
    &+ \ \widehat{\boldeps}(s)^\top \widetilde{\boldX}(t) \left(\widetilde{\boldX}(t)^\top \widetilde{\boldX}(t)\right)^{-1} \widehat{\Expec}[\widetilde{\boldX}(t)^\top \boldeps (t)] \\
    &+ \ 
\left(\left(\widetilde{\boldX}(s)^\top \widetilde{\boldX}(s)\right)^{-1} \widehat{\Expec}[\widetilde{\boldX}(s)^\top \boldeps (s)]\right)^\top \widetilde{\boldX}(s)^\top \widehat{\boldeps}(t) \\
&+ \
\left(\left(\widetilde{\boldX}(s)^\top \widetilde{\boldX}(s)\right)^{-1} \widehat{\Expec}[\widetilde{\boldX}(s)^\top \boldeps (s)]\right)^\top \widetilde{\boldX}(s)^\top \widetilde{\boldX}(t) \left(\widetilde{\boldX}(t)^\top \widetilde{\boldX}(t)\right)^{-1} \\
& \ \ \ \ 
\widehat{\Expec}[\widetilde{\boldX}(t)^\top \boldeps (t)].
\end{align*}
Some manipulation gives
\begin{align*}
    \widehat{\boldeps} \bc (s)^\top \widehat{\boldeps} \bc (t) 
    =& \  \widehat{\boldeps}(s)^\top \widehat{\boldeps}(t) \\
    &+ \ \widehat{\boldeps}(s)^\top \widetilde{\boldX}(t) \left(\widetilde{\boldX}(t)^\top \widetilde{\boldX}(t)\right)^{-1} \widehat{\Expec}[\widetilde{\boldX}(t)^\top \boldeps (t)] \\
    &+ \ 
\widehat{\Expec}[\boldeps (s) ^ \top \widetilde{\boldX}(s)]
\left(\widetilde{\boldX}(s)^\top \widetilde{\boldX}(s)\right)^{-1}\widetilde{\boldX}(s)^\top \widehat{\boldeps}(t) \\
&+\ 
\widehat{\Expec}[\boldeps (s) ^ \top \widetilde{\boldX}(s)]
\left(\widetilde{\boldX}(s)^\top \widetilde{\boldX}(s)\right)^{-1} \widetilde{\boldX}(s)^\top \widetilde{\boldX}(t) \left(\widetilde{\boldX}(t)^\top \widetilde{\boldX}(t)\right)^{-1} \\
& \ \ \ \  \widehat{\Expec}[\widetilde{\boldX}(t)^\top \boldeps (t)].
\end{align*}
\section{Numerical Implementation}\label{sec:numerical-implementation}\label{sec:pda-numerical}

Algorithm 1 details the computation of the bias-reduction algorithm and some of its key components are described in more detail below.

\begin{algorithm}
\footnotesize
\LinesNumbered
\label{alg-bias-reduction}
\caption{Bias Reduction Algorithm}
\KwData{
\begin{itemize}
    \item Grid of $D$ points $\mathbf{t} = (t_1, \dots, t_D)$.
    \item Function data and derivative evaluations on the grid $\mathbf{x}(\mathbf{t}), D\mathbf{x}(\mathbf{t}), \dots, D^{m} \mathbf{x}(\mathbf{t})$.
    \item Number of iterations $J$ of the bias correction.
\end{itemize}
}
\KwResult{Bias-reduced estimates $\widehat{\boldsymbol{\beta}}^{(J+1)} (\mathbf{t}) = \left[\beta_0^{(bc)}(\mathbf{t}) \lvert \dots \rvert \beta_{m-1}^{(bc)}(\mathbf{t}) \right]$}

\For{$t_{i} = t_1$ \KwTo $t_D$} 
{  
$\widetilde{\mathbf{X}} (t_i) \gets \bigl[ \mathbf{x}(t_i) | \dots | D^{m-1} \mathbf{x}(t_i)\bigr]$ \Comment*[r]{Pointwise Design Matrix}
$\widehat{\boldsymbol{\beta}} (t_i) \gets \bigl(\widetilde{\mathbf{X}} (t_i)^\top\widetilde{\mathbf{X}} (t_i)\bigr)^{-1} \widetilde{\mathbf{X}} (t_i)^\top D^{m} \mathbf{x}(t_i)$ \Comment*[r]{Initial Pointwise OLS}
}
\Init{}{
$\widehat{\boldsymbol{\beta}}^{(1)} (\mathbf{t}) \gets \widehat{\boldsymbol{\beta}} (\mathbf{t})$
}
\For{$j = 1$ \KwTo $J$\Comment*[r]{Loop through Iterations of Bias Correction}} {
\For{$m_{i} = 0$ \KwTo $m-1$}
{   
\Fn{$\widehat{\beta}^{(j)}_{m_{i}} (t_{arg})$}{
 $\texttt{Interpolate}(\mathbf{t}, \widehat{\beta}^{(j)}_{m_{i}} (\mathbf{t}), t_{arg})$\Comment*[r]{Interpolate Current Estimates}} 
}
\For{$t_{i} = t_1$ \KwTo $t_D$} 
{   
$\widehat{\boldsymbol{\epsilon}}^{(j)}(t_i) \gets D^{m} \mathbf{x}(t_i) - \widetilde{\mathbf{X}} (t_i) \widehat{\boldsymbol{\beta}}^{(j)}(t_i)$ \Comment*[r]{Update Residuals}
}
$\widehat{C}^{(j)}(\mathbf{t}, \mathbf{t}) \gets \widehat{\boldsymbol{\epsilon}}(\mathbf{t})^\top \widehat{\boldsymbol{\epsilon}}(\mathbf{t}) / (N-1)$\Comment*[r]{Update Residual Covariance}
\Fn{$\widehat{C}^{(j)} (s_{arg}, t_{arg})$}{
 $\texttt{Interpolate\_2d}(\mathbf{t}, \mathbf{t}, \widehat{C}^{(j)}(\mathbf{t}, \mathbf{t}), s_{arg},t_{arg})$\Comment*[r]{Interpolate Covariance}}
\Fn{$\widehat{\mathbf{C}}^{(j)} (s_{arg}, t_{arg})$}{
$\begin{pmatrix} 0 & 0 \\ 0 & \widehat{C}^{(j)} (s_{arg}, t_{arg}) \end{pmatrix}$\Comment*[r]{Matrix-Valued Covariance Function}
}
\Fn{$\boldsymbol{\Phi}^{(j)} (\text{start\_time}, \text{end\_time})$}{
$initial\_conditions \gets \mathbf{I}_m$\ \Comment*[r]{Define S.T.M. Function}
$\texttt{ODE\_solve}(start\_time, end\_time, initial\_conditions, \widehat{\beta}_{1}^{(j)} (t_{arg}), \dots, \widehat{\beta}_{m-1}^{(j)} (t_{arg}))$;
}

\Fn{$\boldsymbol{\rho}^{(j)} (s, t)$}{
$[\boldsymbol{\Phi}^{(j)} (\text{start\_time} = s, \text{end\_time} = t) \widehat{\mathbf{C}}^{(j)} (s_{arg} = s, t_{arg} = t)]_{\cdot, m}$\Comment*[r]{Integrand}}

\For{$t_{i} = t_1$ \KwTo $t_D$} 
{   
$\widehat{\Expec}^{(j)}[\widetilde{\mathbf{X}}(t_i)^\top \boldsymbol{\epsilon}(t_i)] \gets \texttt{NumInt}(\boldsymbol{\rho}^{(j)} (s, t = t_i), lower = 0, upper = t_i)$;

$\widehat{\Bias}^{(j)} (\widehat{\boldsymbol{\beta}}(t_i)) \gets 
\bigl(\widetilde{\mathbf{X}} (t_i)^\top\widetilde{\mathbf{X}} (t_i)\bigr)^{-1} \widetilde{\mathbf{X}} (t_i)^\top \widehat{\Expec}^{(j)}[\widetilde{\mathbf{X}}(t_i)^\top \boldsymbol{\epsilon}(t_i)]$;

$\widehat{\boldsymbol{\beta}}^{(j+1)} (t_i) \gets \widehat{\boldsymbol{\beta}} (t_i) - \widehat{\Bias}^{(j)} (\widehat{\boldsymbol{\beta}}(t_i))$\Comment*[r]{Subtract current bias estimate from initial OLS Estimate}
}
\Return 
$\widehat{\boldsymbol{\beta}}^{(J+1)} (\mathbf{t})$
}
\end{algorithm}

\begin{enumerate}
    \item \textbf{One-dimensional function interpolation} is used (line {\footnotesize \textbf{11}}) to convert the pointwise regression coefficient estimates to functions so that they can be passed as arguments to the numerical ODE solver. The functions we are working with are smooth, so given a reasonably dense grid we opt for linear interpolation using the base \proglang{R} function \texttt{approxfun()}, which has been recommended for use with the \pkg{deSolve} \proglang{R} package \parencite[p. 57]{soetaert_solving_2012}.
    \item \textbf{Two-dimensional function interpolation} is performed to convert the discretised covariance estimate to a bivariate function (line {\footnotesize \textbf{18}}) that can be evaluated by the numerical integration function (line {\footnotesize \textbf{27}}). There are two options doing this. The first is to interpolate each residual function first using a set of cubic B-spline basis functions $\{\psi_k(t)\}_{k=1}^K$, so that we have the representation
    $$
    \widehat{\epsilon}_i (t) = \sum_k^K \widehat{e}^*_{ik} \psi_k (t).
    $$
    This means that the $N$-vector containing the residual functions can be written as 
    $$
    \widehat{\boldsymbol{\epsilon}}(t) = \widehat{\mathbf{E}}^* \boldsymbol{\psi} (t)
    $$
    where $\boldsymbol{\psi} (t) = (\psi_1(t), \dots, \psi_K(t))^\top$ and 
    $$
    \widehat{\mathbf{E}}^* = 
    \begin{pmatrix}
        \widehat{e}^*_{11} &  \hdots & \widehat{e}^*_{1K} \\
        \vdots &\vdots \vdots \vdots & \vdots \\
        \widehat{e}^*_{N1} &  \hdots & \widehat{e}^*_{NK}
    \end{pmatrix}.
    $$
    This representation allows evaluations of $\widehat{C}(s, t)$ to be obtained by evaluating the basis functions at $s$ and $t$ as follows\footnote{We use a denominator of $N-1$, rather than $N-m$ as is standard when computing the error variance in regression, because of conventions in the software. As we are dealing with PDA models of low order (e.g., $m=2$) and reasonably large sample sizes, the difference should be negligible}
    $$
    \widehat{C} (s, t) = \frac{\widehat{\boldsymbol{\epsilon}}(s)^\top  \widehat{\boldsymbol{\epsilon}}(t)}{N-1}= \frac{\boldsymbol{\psi} (s)^\top  \widehat{\mathbf{E}}^{*\top} \widehat{\mathbf{E}}^* \boldsymbol{\psi} (t)}{N-1} = \boldsymbol{\psi} (s)^\top \widehat{\mathbf{C}}^* \boldsymbol{\psi} (t)
    $$
    where
    $$
    \widehat{\mathbf{C}}^* = \frac{\widehat{\mathbf{E}}^{*\top} \widehat{\mathbf{E}}^*}{N-1}.
    $$
    The \texttt{smooth.basis()}, \texttt{var.fd()} and  \texttt{eval.bifd()} functions in the \pkg{fda} \proglang{R} package \parencite{ramsay_fda_2020} are used to obtain the basis representation and to compute and evaluate the covariance function.

    The second option is to use bi-linear interpolation of the matrix $\widehat{C}(\mathbf{t},\mathbf{t}) = \widehat{\boldsymbol{\epsilon}}(\mathbf{t})^\top \widehat{\boldsymbol{\epsilon}}(\mathbf{t})$. For this, we use the \texttt{interp2()} function in the \proglang{R} package \pkg{pracma} \parencite{borchers_pracma_2022}.
    \item \textbf{ODE Solving}: The state transition matrix $\fundmats$ is defined on line {\footnotesize \textbf{21}}, and is part of the integrand which is integrated numerically to obtain $\widehat{\Expec}[\widetilde{\mathbf{X}}(t) \boldsymbol{\epsilon} (t)]$ (line {\footnotesize \textbf{27}}). 

    For linear time-varying systems, an analytic expression for $\fundmats$ is not available and therefore numerical methods are used to estimate it as the solution of the matrix ODE
    $$
    D \fundmats = \mathbf{B} (t) \fundmats
    $$
    with initial conditions $\mathbf{\Phi}(s, s) = \mathbf{I}_m$. In other words, the homogeneous part of the ODE is solved $m$ times with starting time $s$ and initial values $(1, \dots, 0)^T, \dots, (0, \dots, 1)^T$ to produce the columns of the state transition matrix.    

    We use the numerical ODE solver \texttt{lsoda()} from the \pkg{deSolve} \proglang{R} package \parencite{soetaert_desolve_2023}. As detailed in the documentation, \texttt{lsoda()} switches automatically between stiff and non-stiff methods to choose the most appropriate approach.

    Because the state transition matrix will be called many times as part of the integrand during numerical integration, we can reduce the computational burden by an intermediate step which avoids calling the numerical ODE solver repeatedly:
    \begin{itemize}
        \item \textbf{Step 1}: We evaluate the state transition matrix on a grid of points, i.e., $\boldsymbol{\Phi} (\mathbf{t}, \mathbf{t})$.
        \item \textbf{Step 2}: Create the function $\boldsymbol{\Phi} (start\_time, end\_time)$ by bi-linear interpolation of $\boldsymbol{\Phi} (\mathbf{t}, \mathbf{t})$.
    \end{itemize}
    We only ever need to know $\boldsymbol{\Phi} (start\_time, end\_time)$ for $start\_time \leq end\_time$ to compute $\widehat{\Expec}[\widetilde{\mathbf{X}}(t) \boldsymbol{\epsilon} (t)]$, but we can easily obtain $\boldsymbol{\Phi} (start\_time, \allowbreak end\_time)$ for $start\_time > end\_time$ by using the identity $\boldsymbol{\Phi} (start\_time, end\_time) = \boldsymbol{\Phi}^{-1} (end\_time, start\_time)$
    
    Note that this speed-up may not be as numerically stable as using \texttt{lsoda()} to define the state transition matrix function, so we only employ it in cases where the computational overhead is prohibitive and use the standard approach if it fails.
    
    \item \textbf{Numerical Integration}: 
    For numerical integration, we use the base \proglang{R} function \texttt{integrate()}. From the description
    \begin{quote}
        For a finite interval, globally adaptive interval subdivision is used in connection with extrapolation by Wynn's Epsilon algorithm, with the basic step being Gauss--Kronrod quadrature \parencite{r_core_team_r_2022}.
    \end{quote}
    We have not experimented with any other numerical integration techniques. The numerical integration is performed at each $t_i \in \mathbf{t}$, making it an \emph{embarrassingly parallel} task. Therefore, we make use of the \texttt{mclapply()} function in the base \proglang{R} package \pkg{parallel} \parencite{r_core_team_r_2022}.
\end{enumerate}

\section[Additional Harmonic Motion Simulations]{Additional Harmonic Motion Simulations}\label{sec:additional-hm-simulations}

\subsection{Simple Harmonic Motion}\label{sec:shm-full-smulation}

\begin{figure}
    \centering
    \includegraphics[page=3, width = 0.6\textwidth]{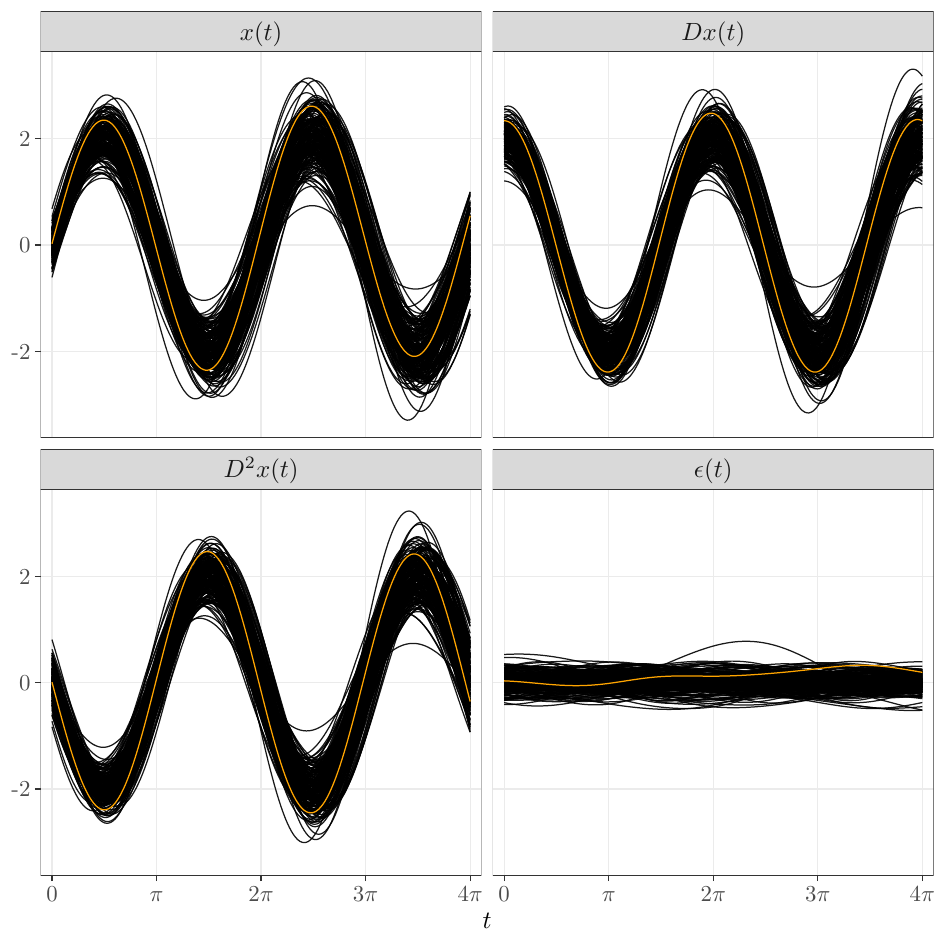}
    \caption{The first simulated dataset from the simulation to demonstrate the general bias correction algorithm using the SHM model. The collection of $N = 200$ functions are shown in black with a single observation overlaid in gold.}
    \label{fig:shm-general-dataset}
\end{figure}

To demonstrate the general bias reduction, we re-use the SHM model from the main paper, but this time fit the full model
$$
D^2 x_i(t) = \alpha (t) + \beta_0 (t) x_i(t) + \beta_1 (t) D x_i(t) +\epsilon_i (t),
$$
i.e., we do not know a priori that $\alpha (t) = \beta_1 (t) = 0$. We set $\sigma = 0.4$ and $l = 1$ for this demonstration, because these parameter values produced the most severe bias in the first SHM simulation. We consider $t \in [0, 4\pi]$ so that the functions run for approximately two periods. For initial conditions, we use $\boldsymbol{\mu}_0 = (0, 2)^\top$ and $\boldsymbol{\Sigma}_0 = \diag\{0.05, 0.05\}$. We generate 50 simulated datasets and apply 10 iterations of the bias-reduction algorithm each time. The first simulated dataset is shown in Figure \ref{fig:shm-general-dataset}.

\begin{figure}
    \centering
    \includegraphics[width = 1\textwidth]{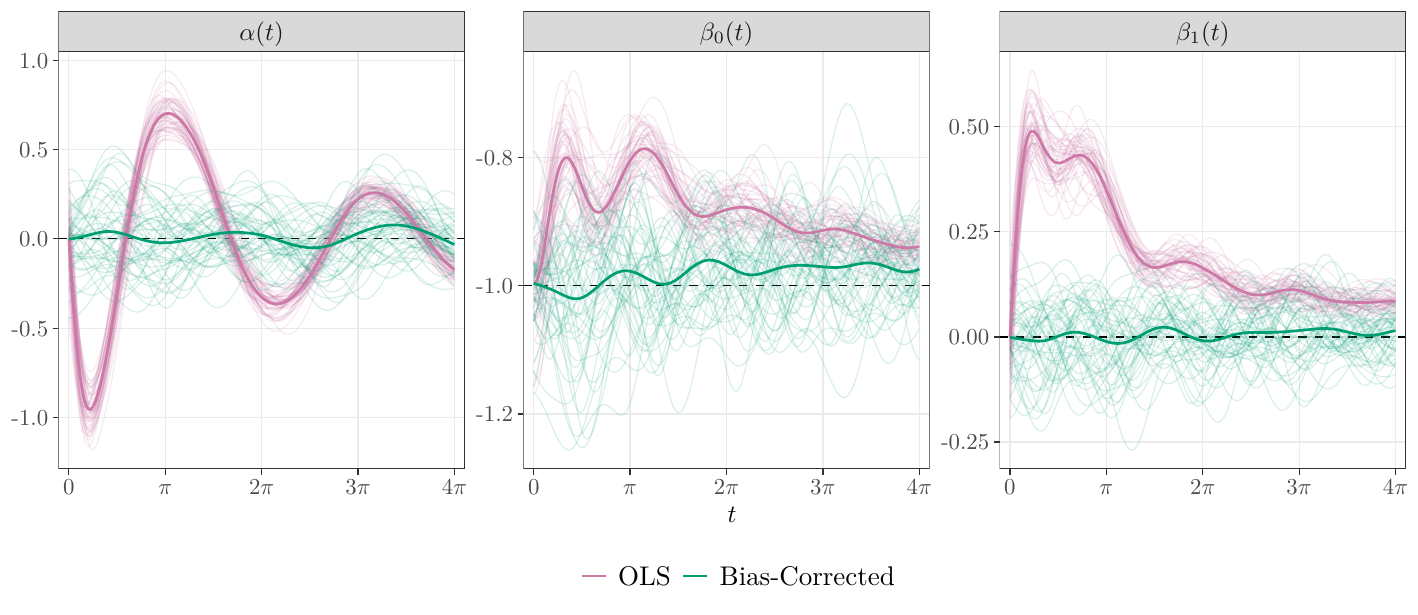}
    \caption{Results of the simulation to demonstrate the general bias correction algorithm using the SHM model.}
    \label{fig:shm-general-results}
\end{figure}

The results of the simulation are displayed in Figure \ref{fig:shm-general-results}. The individual OLS (pink) and bias-reduced (green) estimates from the simulation replicates are shown by light thin lines, and their respective averages are overlaid as darker solid lines. The true parameters, the constant functions $\alpha (t) = 0$, $\beta_0 (t) = - 1$ and $\beta_1 (t) = 0$ are indicated as dashed black lines. The bias-reduction algorithm successfully eliminates the bias in the estimated parameters. It does, however, come at the expense of increased variance in the bias-reduced parameter estimates, evidenced by the greater spread in the green lines.

\subsection[Damped Harmonic Motion with Constant Damping]{Damped Harmonic Motion with Constant Damping}\label{sec:dhm-full-smulation}

\begin{figure}
    \centering
    \includegraphics[page=5, width = 0.6\textwidth]{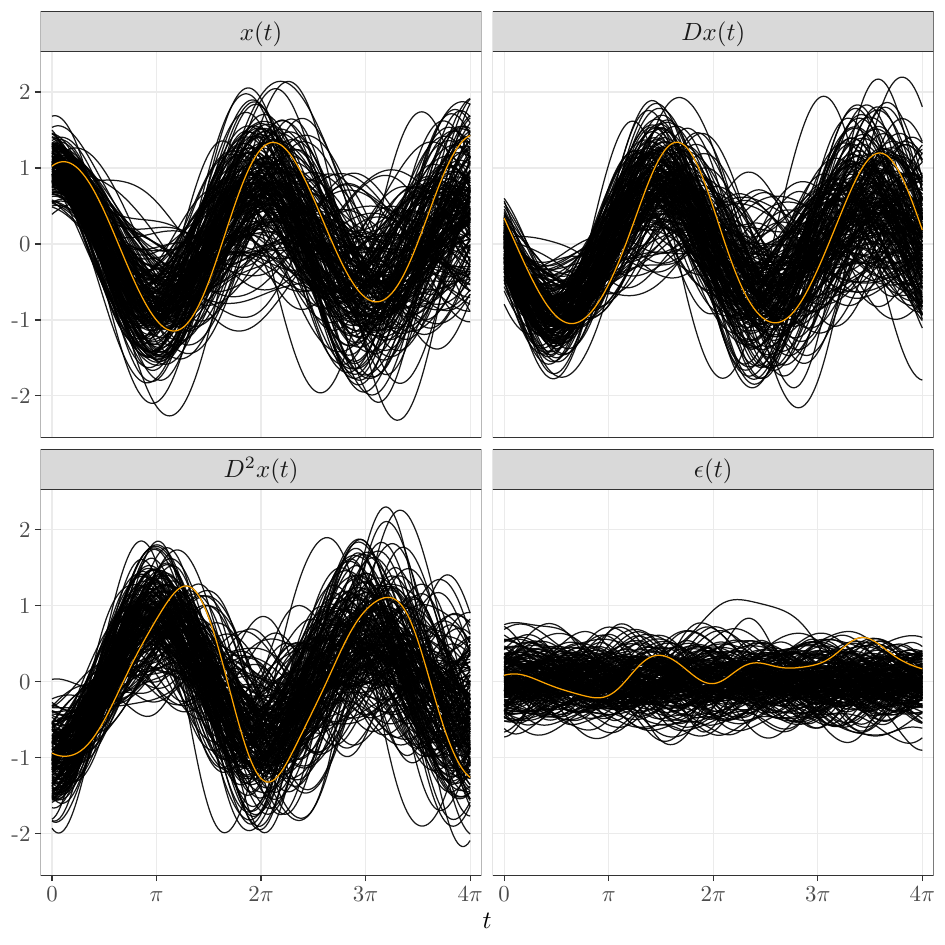}
    \caption{The first simulated dataset from the simulation to demonstrate the general bias correction algorithm using the damped harmonic motion model. The collection of $N = 200$ functions are shown in black with a single observation overlaid in gold.}
    \label{fig:dhm-general-dataset}
\end{figure}

\begin{figure}
    \centering
    \includegraphics[width = 1\textwidth]{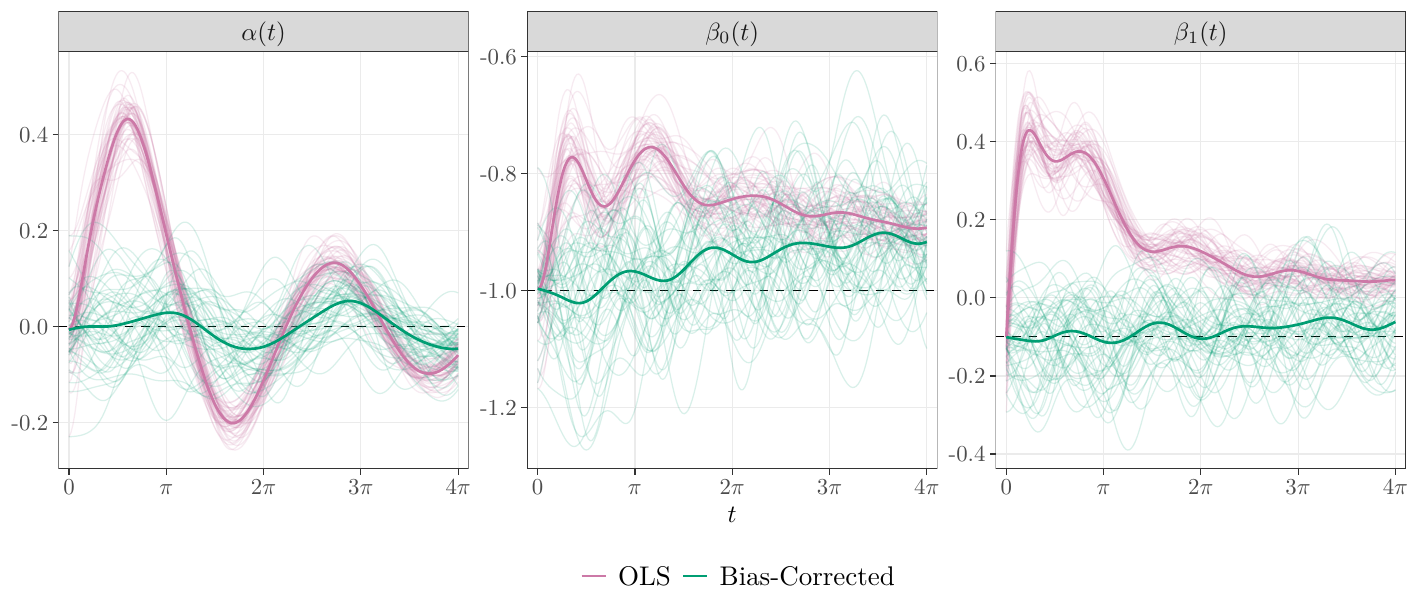}
    \caption{Results of the simulation to demonstrate the general bias correction algorithm using the damped harmonic motion model.}
    \label{fig:dhm-general-results}
\end{figure}

Next, we consider adding a constant damping term to the generative SHM model
$$
D^2 x_i(t) = - x_i(t) - 0.1 D x_i(t) +\epsilon_i (t).
$$
Once again, we fit the full model so the targets of the estimation are $\alpha(t)=0$, $\beta_0 (t) = -1$ and $\beta_1 (t) = -0.1$. The other data-generating conditions from Section \ref{sec:shm-full-smulation} remain unchanged, except that we set $\boldsymbol{\mu}_0 = (1, 0)^\top$. The first simulated dataset is shown in Figure \ref{fig:dhm-general-dataset}. The results are shown in Figure \ref{fig:dhm-general-results}. The bias-reduction algorithm works well again, but fails to fully eliminate the bias in $\beta_0 (t)$ for $t > 2 \pi$. It is possible that the system is being damped heavily at this stage and that the forcing is the main source of variation, hence the bias is larger and more difficult to correct. It may also simply be that more than 10 iterations are needed to fully reduce the bias. Overall, however, the results are satisfactory.

\subsection[Damped Harmonic Motion with Time-Varying Damping]{Damped Harmonic Motion with Time-Varying Damping}

\begin{figure}
    \centering
    \includegraphics[page=7, width = 0.6\textwidth]{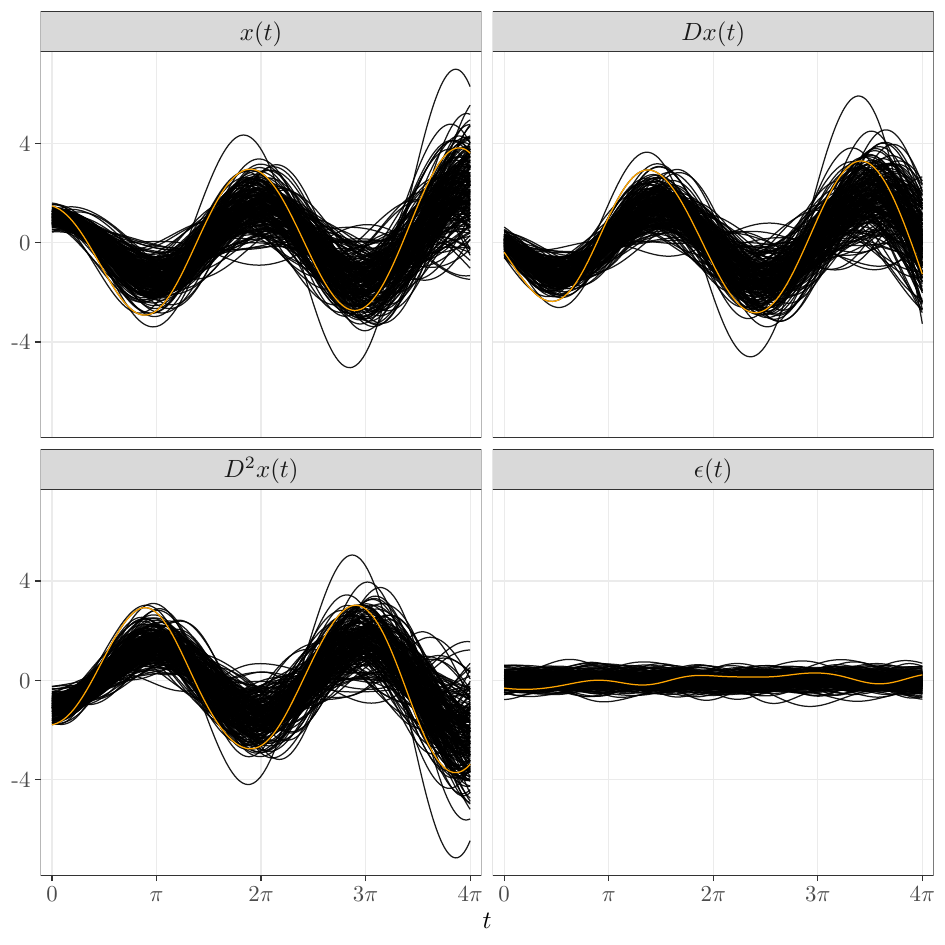}
    \caption{The first simulated dataset from the simulation to demonstrate the general bias correction algorithm using the damped harmonic motion model with time-varying $\beta_1 (t)$. The collection of $N = 200$ functions are shown in black with a single observation overlaid in gold.}
    \label{fig:dhm-tv-general-dataset}
\end{figure}

\begin{figure}
    \centering
    \includegraphics[width = 1\textwidth]{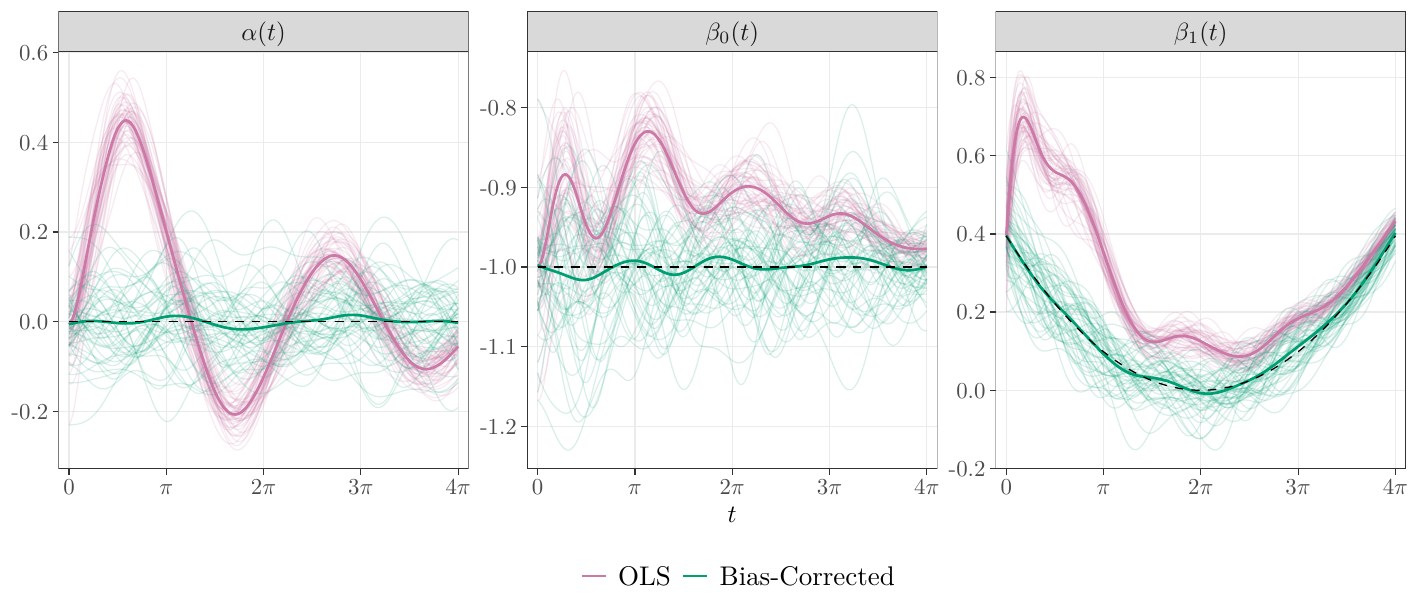}
    \caption{Results of the simulation to demonstrate the general bias correction algorithm using the damped harmonic motion model with time-varying $\beta_1 (t)$.}
    \label{fig:dhm-tv-general-results}
\end{figure}

Finally, we consider adding a \emph{time-varying} parameter to the SHM model
$$
D^2 x_i(t) = - x_i(t) + \beta_1(t) D x_i(t) +\epsilon_i (t), \quad \beta_1 (t) = 0.01 (t - 2 \pi)^2.
$$
Otherwise, the settings are the same as in Section \ref{sec:dhm-full-smulation}. The first simulated dataset from this model is shown in Figure \ref{fig:dhm-tv-general-dataset} and the results are shown in Figure \ref{fig:dhm-tv-general-results}. The bias-reduction algorithm works well to reduce the bias in the estimated parameters; in particular the shape of the time-varying coefficient $\beta_1 (t)$ is recovered well.

\section[Additional Details of the Van der Pol Analysis]{Additional Details of the Van der Pol Analysis}\label{sec:additional-vdp}

This section contains additional analysis of the Van der Pol model.
It contains an expanded description of the data-generating model (Section \ref{sec:additional-vdp-data-gen}) and estimation procedure (Section \ref{sec:additional-vdp-estim}) as well as simulation results of varying the forcing amplitude, varying the forcing lengthscale and comparing with non-parametric modelling of the full dynamical system (Section \ref{sec:additional-vdp-simu}). 

The Van der Pol (VdP) equation was initially formulated by \textcite{van_der_pol_frequency_1927} to describe electrical circuits employing vacuum tubes \parencite[see][p. 7]{RJ-2010-013}. It is a second-order, non-linear, time-invariant ODE 
$$
D^2 x(t) - \mu (1 - x(t)^2) D x(t) + x(t) = 0, \quad \mu \geq 0,
$$
and can be understood as a non-linearly damped oscillator with the parameter $\mu$ controlling the amount of non-linear damping.

\subsection{Data-Generating Model}\label{sec:additional-vdp-data-gen}

 To construct a data-generating process from the VdP equation, we first convert the second-order ODE to a system of two coupled first order ODEs
 \begin{align*}
     Dx(t) &= \mu \left(x(t) - \frac{x(t)^3}{3} - y(t) \right) \\
     D y(t) &= \frac{x(t)}{\mu},
 \end{align*}
which uses the transformation $y(t) = x(t) - \frac{x(t)^3}{3} + \frac{Dx(t)}{\mu}$ for $\mu > 0$. 
We consider values of $t \in [0, 13]$ which corresponds to approximately two periods of oscillations when $\mu$ is small ($\mu \leq 1$), and use an equally-spaced grid of length $200$ on this interval for discretisation. 
We then add a smooth stochastic disturbance to generate replicate observation pairs $(x_i(t), y_i (t)), \ i = 1, \dots, N$, as solutions of the stochastically-forced system
\begin{align*}
    Dx_i (t) &= \mu \left(x_i(t) - \frac{x_i (t) ^3}{3} - y_i(t)\right) + \epsilon_i^{(x)} (t)  \\ \label{true-Dyi}
    Dy_i(t) &= \frac{x_i(t)}{\mu} + \epsilon_i^{(y)} (t),
\end{align*}
where $\epsilon_i^{(y)} (t)$ and $\epsilon_i^{(x)} (t)$ are independent copies of a mean-zero Gaussian process with covariance function $C(s, t) =  \sigma^2\phi(l|t-s|)$, where $\phi$ is a standard Gaussian density. In other words, we add a smooth Gaussian stochastic disturbance to each dimension independently.

As a baseline scenario, we set  $\mu = 1$, \ $\sigma = 0.1$ and $l = 2$.
In all simulations, we draw initial conditions as $(x_i(0), y_i(0))^\top \sim \mathcal{N}_2 (\boldsymbol{\mu}_0, \boldsymbol{\Sigma}_0)$ where $\boldsymbol{\mu}_0 \sim (1.99, -0.91)^\top$, which is a point lying approximately on the stable limit cycle of the deterministic ODE. We use 
$$
\boldsymbol{\Sigma}_0 
=
\begin{pmatrix}
    0.025 & 0 \\
    0 & 0.025
\end{pmatrix}.
$$

\subsection{Linear PDA Regression Model}\label{sec:additional-vdp-estim}

A first-order Taylor expansion of $Dx(t)$ about an operating trajectory $(x_0(t), y_0 (t))$, is given by
\begin{align*}
    Dx(t) \approx \ &\mu \left(x_0(t) - \frac{1}{3} x_0(t)^3 - y_0(t)\right) + \mu \left(1 - x_0(t)^2\right) \left(x(t) - x_0(t)\right) \\
    &+ \mu \left(y(t) - y_0 (t)\right) + R_2(x(t), y(t)),
\end{align*}
where $R_2$ denotes the remainder containing terms of order 2 and higher. These terms are
$$
R_2(x(t), y(t)) = - x_0 (t) \left(x(t)- x_0 (t)\right)^2 -  \frac{1}{3} \left(x(t)- x_0 (t)\right)^3.
$$
Re-arranging the Taylor expansion gives
\begin{align*} 
        Dx(t) &\approx \underbrace{\mu \left(x_0(t) - \frac{1}{3} x_0(t)^3 - y_0(t)\right) - \mu \left(1 - x_0(t)^2 \right) x_0(t) + \mu y_0 (t)}_{\approx \ \alpha^{(x)} (t)}  \\ 
    &\ + \underbrace{\mu \left(1 - x_0(t)^2 \right)}_{\beta_{xx} (t)} x(t) \underbrace{- \mu}_{\beta_{xy} (t)} y(t)  \\
    &\ + R_2(x(t), y(t)),
\end{align*}
which suggests an approximate linear PDA model for $Dx(t)$ with coefficients $\beta_{xx} (t) =  \mu (1 - x_0(t)^2)$ and $\beta_{xy} (t) = - \mu$. Therefore, the following linear PDA regression models are fitted:
\begin{align*}
    Dx_i (t) &= \alpha^{(x)} (t) + \beta_{xx} (t) x_i(t) + \beta_{xy} (t) y_i(t)  + \widetilde{\epsilon}_i^{(x)} (t)  \\
    Dy_i(t) &= \alpha^{(y)}(t) + \beta_{yx} (t) x_i(t) + \beta_{yy} (t) y_i(t) + \epsilon_i^{(y)} (t).
\end{align*}
We note that the intercept term $\alpha^{(x)} (t)$ absorbs the constant (w.r.t. $x$ and $y$) terms in  the remainder $R_2(x(t), y(t))$, and that $\widetilde{\epsilon}_i^{(x)} (t) \neq  \epsilon_i^{(x)} (t)$ as $\widetilde{\epsilon}_i^{(x)} (t)$ contains some non-constant terms from the remainder $R_2(x(t), y(t))$. Since we are working under the assumption that perturbations around the operating point are small, these terms should be negligible. It is also worth noting that the model for $Dy_i(t)$ is correctly specified, i.e., the true model is linear  with $\alpha^{(y)}(t) = 0$, $\beta_{yx} (t) = 1/\mu$ and $\beta_{yy} (t) = 0$. However, there still exists dependence among the covariate and error terms.

PDA for the coupled system involves fitting the time-varying \emph{multivariate} linear regression model
$$
\underbrace{
    \begin{pmatrix}
    Dx_1 (t) & Dy_1 (t) \\
    \vdots & \vdots \\
    Dx_N (t) & Dy_N (t)
\end{pmatrix}
}_{D \mathbf{X}(t)}
= 
\underbrace{
\begin{pmatrix}
    1 & x_1 (t) & y_1 (t) \\
    \vdots & \vdots & \vdots \\
    1 & x_N (t) & y_N (t)
\end{pmatrix}}_{\mathbf{Z}(t)}
\underbrace{
\begin{pmatrix}
    \alpha^{(x)} & \alpha^{(y)} \\
    \beta_{xx} (t) & \beta_{yx} (t) \\ 
    \beta_{xy} (t) & \beta_{yy} (t)
\end{pmatrix}}_{\mathbf{B}(t)}
+
\underbrace{
\begin{pmatrix}
    \widetilde{\epsilon}_1^{(x)} (t) & \epsilon_1^{(y)} (t) \\
    \vdots & \vdots \\
    \widetilde{\epsilon}_N^{(x)} (t) & \epsilon_N^{(y)} (t)
\end{pmatrix}}_{\mathbf{E}(t)}.
$$
The OLS parameter estimates are given by
$$
\widehat{\mathbf{B}} (t) = \left(\mathbf{Z}(t)^\top \mathbf{Z}(t)\right)^{-1} \mathbf{Z}(t) D \mathbf{X}(t),
$$
and the bias is
$$
\Expec\left[\widehat{\mathbf{B}} (t)\right] - \mathbf{B} (t) = \Expec\left[(\mathbf{Z}(t)^\top \mathbf{Z}(t))^{-1} \mathbf{Z}(t)^\top \mathbf{E}(t)\right].
$$
Expanding the ``numerator" in the bias gives
\begin{align*}
\Expec[\mathbf{Z}(t)^\top \mathbf{E}(t)]
&=
\Expec\left[
\begin{pmatrix}
    \sum_{i=1}^N \widetilde{\epsilon}^{(x)}_i(t) & \sum_{i=1}^N \epsilon^{(y)}_i(t) \\
    \sum_{i=1}^N x_i (t) \widetilde{\epsilon}^{(x)}_i(t) & \sum_{i=1}^N x_i(t)\epsilon^{(y)}_i(t) \\
    \sum_{i=1}^N y_i (t) \widetilde{\epsilon}^{(x)}_i(t) & \sum_{i=1}^N y_i(t)\epsilon^{(y)}_i(t)
    \end{pmatrix}\right]\\
    & = 
N \cdot
\begin{pmatrix}
    \Expec[\widetilde{\epsilon}^{(x)}(t)] & \Expec[\epsilon^{(y)}(t)] \\
    \Expec[x(t) \widetilde{\epsilon}^{(x)}(t)] & \Expec[x(t) \epsilon^{(y)}(t)] \\
    \Expec[y(t) \widetilde{\epsilon}^{(x)}(t)] & \Expec[y(t) \epsilon^{(y)}(t)]
\end{pmatrix},
\end{align*}
which, assuming the linear approximation is reasonable, can be estimated using an extension of the correction in Section \ref{sec:general-bias-correction} to coupled ODEs.

\subsection{Simulations}\label{sec:additional-vdp-simu}

For all simulation scenarios we generate $50$ datasets of size $N = 200$. We apply 10 iterations of the bias-reduction algorithm and compare the initial OLS estimates with the final bias-reduced estimates. We compare the known parameters with their true values
$$
\beta_{xy} (t) =  - \mu, \quad \alpha^{(y)}(t) = 0 , \quad \beta_{yx} (t) = \frac{1}{\mu} \quad \text{and} \quad \beta_{yy} (t) = 0.
$$
However, the parameters
$$
\beta_{xx} (t) =  \mu \left(1 - x_0(t)^2\right),
$$
and 
$$
\alpha^{(x)} (t) = \mu \left(x_0(t) - \frac{1}{3} x_0(t)^3 - y_0(t)\right) - \mu \left(1 - x_0(t)^2 \right) x_0(t) + \mu y_0 (t),
$$
require the choice of trajectory $(x_0 (t), y_0(t))^\top$ to use as a reference when computing the ``true values". We choose $(x_0 (t), y_0(t))^\top$ to be the stable limit cycle trajectory for the deterministic VdP model with initial values $(x_0 (0), y_0(0))^\top = \boldsymbol{\mu}_0$ at $t = 0$, because we are mainly concerned with PDA as describing behaviour around this trajectory.

\subsubsection{Forcing Amplitude} \label{sec:vary-sigma}

\begin{figure}
    \centering
    \includegraphics[page=11, width = 0.9\textwidth]{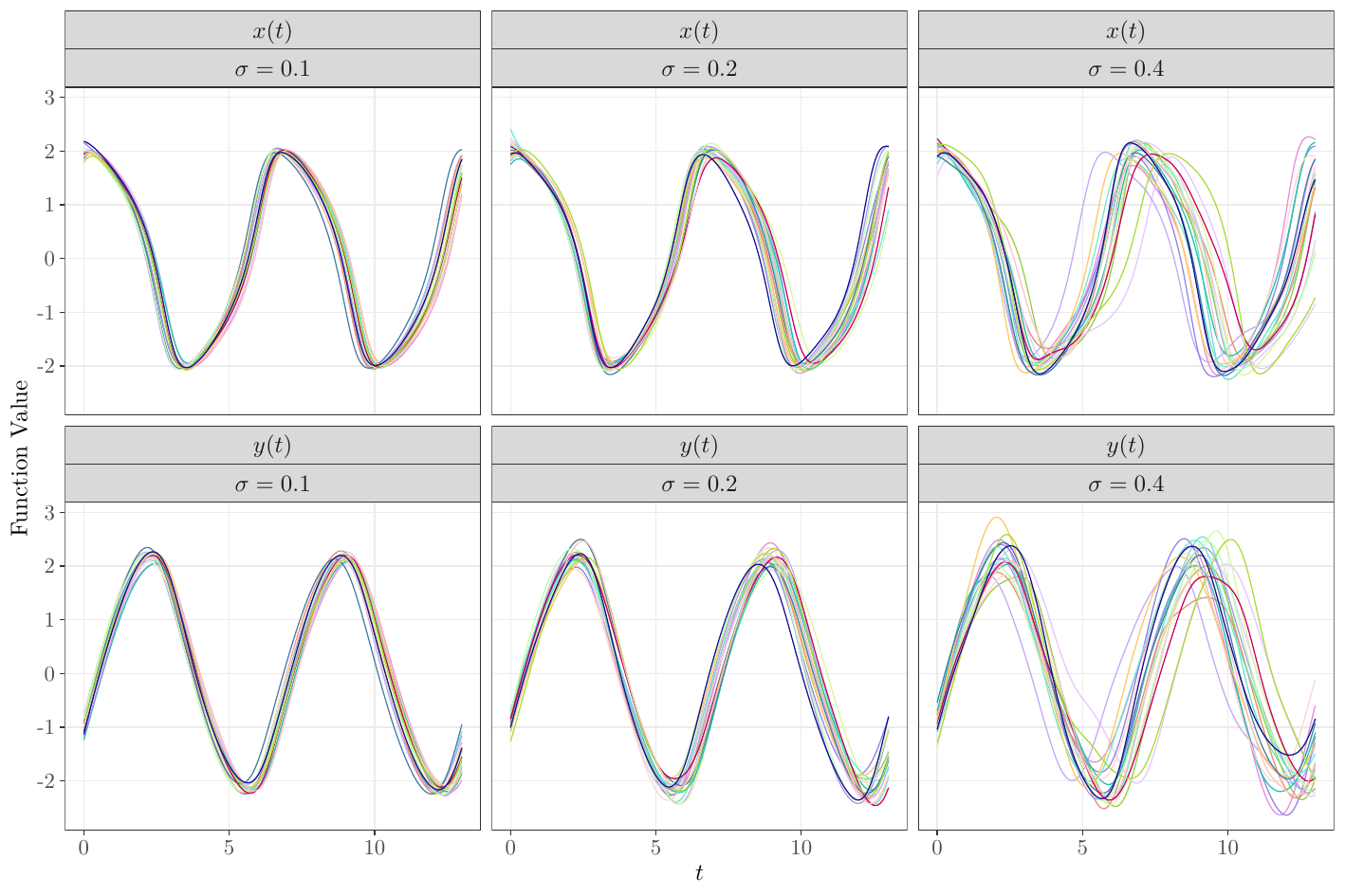}
    \caption{20 generated function pairs $(x_i(t), y_i(t))^\top$ at the amplitude values $\sigma = 0.1$ (low amplitude, baseline), $\sigma = 0.2$ (medium amplitude) and $\sigma = 0.4$ (high amplitude).
    The parameters $\mu$ and $l$ are held at their baseline values of $1$ and $2$, respectively.}
    \label{fig:vary-sigma-example-dataset}
\end{figure}

\begin{figure}
    \centering
    \includegraphics[page=12, width = 0.9\textwidth]{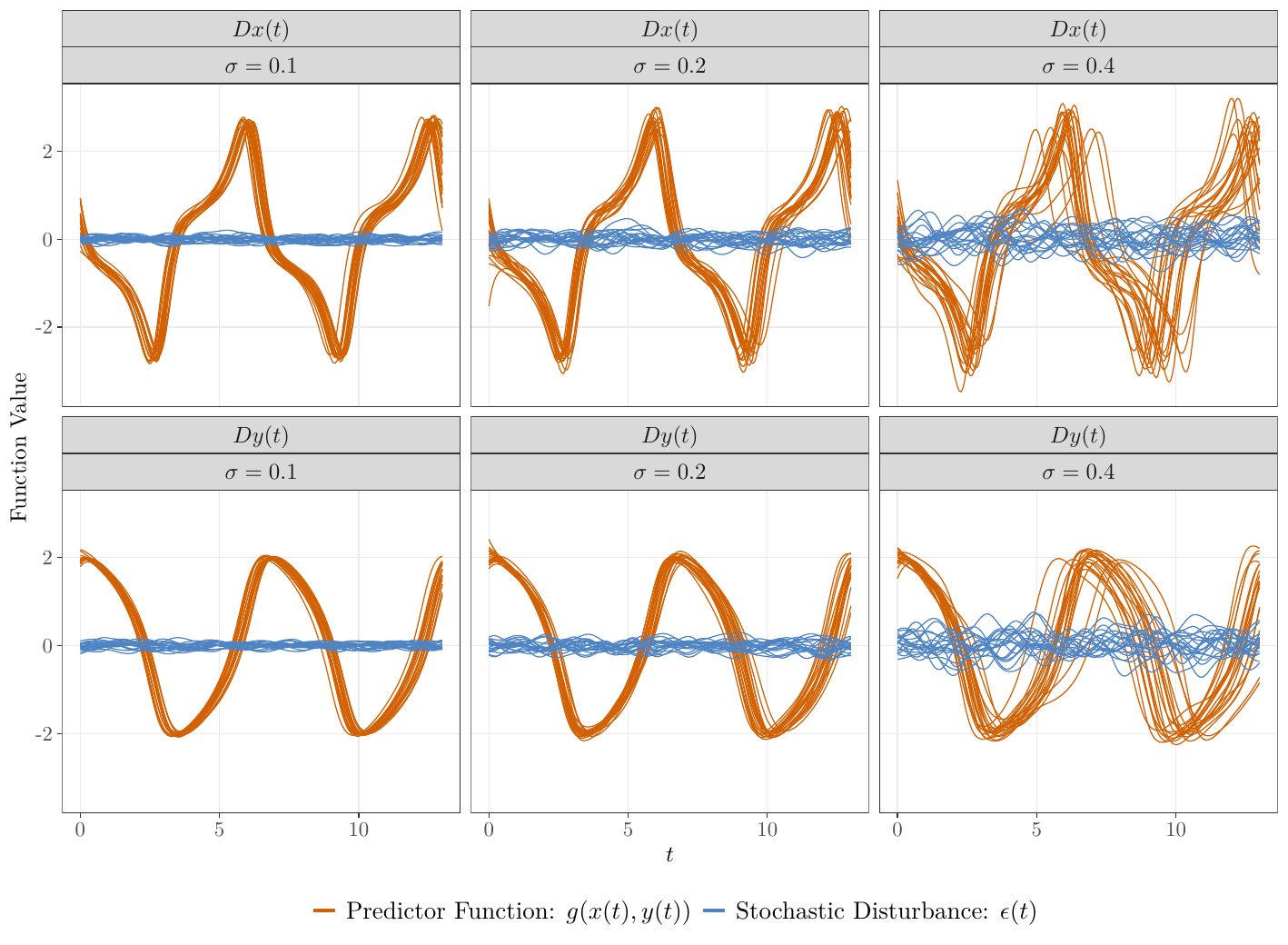}
    \caption{Comparison of the predictor function $g(x(t), y(t))$ and the stochastic disturbance $\epsilon(t)$ for the amplitude values $\sigma = 0.1$ (low amplitude, baseline), $\sigma = 0.2$ (medium amplitude) and $\sigma = 0.4$ (high amplitude).
    The parameters $\mu$ and $l$ are held at their baseline values of $1$ and $2$, respectively.}
    \label{fig:vary-sigma-SNR}
\end{figure}

The first parameter we vary is the forcing amplitude $\sigma$, while fixing the other parameters at their baseline values $l = 2$ and $\mu = 1$. Increasing $\sigma$ should negatively affect parameter estimation for three reasons:
\begin{enumerate}
    \item Increasing $\sigma$ increases covariate and error dependence and hence increases bias in the estimated parameters. Because the bias-reduction algorithm relies on initial (biased) parameter estimates to estimate the bias, it is reasonable to assume that the performance of the algorithm will also be negatively affected.
    \item Increasing $\sigma$ increases the scale of perturbations to $(x(t), y(t))^\top$, which means that the linear approximation of the non-linear model will become less accurate.
    \item Increasing the scale of the perturbations will induce phase variation to the functions and cause some to ``run away", meaning that the mean function might be dampened due to misalignments and may not represent the limit cycle trajectory. This phenomenon is illustrated in more detail in Figure \ref{fig:beta_xx_limit_cycle}.
\end{enumerate}
We consider three values for the forcing amplitude: $\sigma = 0.1$ (low amplitude, baseline), $\sigma = 0.2$ (medium amplitude) and $\sigma = 0.4$ (high amplitude). Figure \ref{fig:vary-sigma-example-dataset} shows 20 generated function pairs $(x_i(t), y_i(t))^\top$ at each of the amplitude values. To give a clearer illustration of the effect of $\sigma$ on estimation, $Dx_i(t)$ and $Dy_i(t)$ are split into their predictor and error components in Figure \ref{fig:vary-sigma-SNR}. This can be broadly understood as visualising the signal-to-noise ratio (SNR) for the estimation problem. Note, however, that the SNR cannot be interpreted in the traditional sense as simply the ratio of the variance of the (non-)linear predictor to the error variance \parencite[e.g.,][p. 401, Eq. (11.18)]{hastie_elements_2009} because of the covariate-error dependence in the data-generating model.

\begin{figure}
    \centering
    \includegraphics[width = 0.75\textwidth]{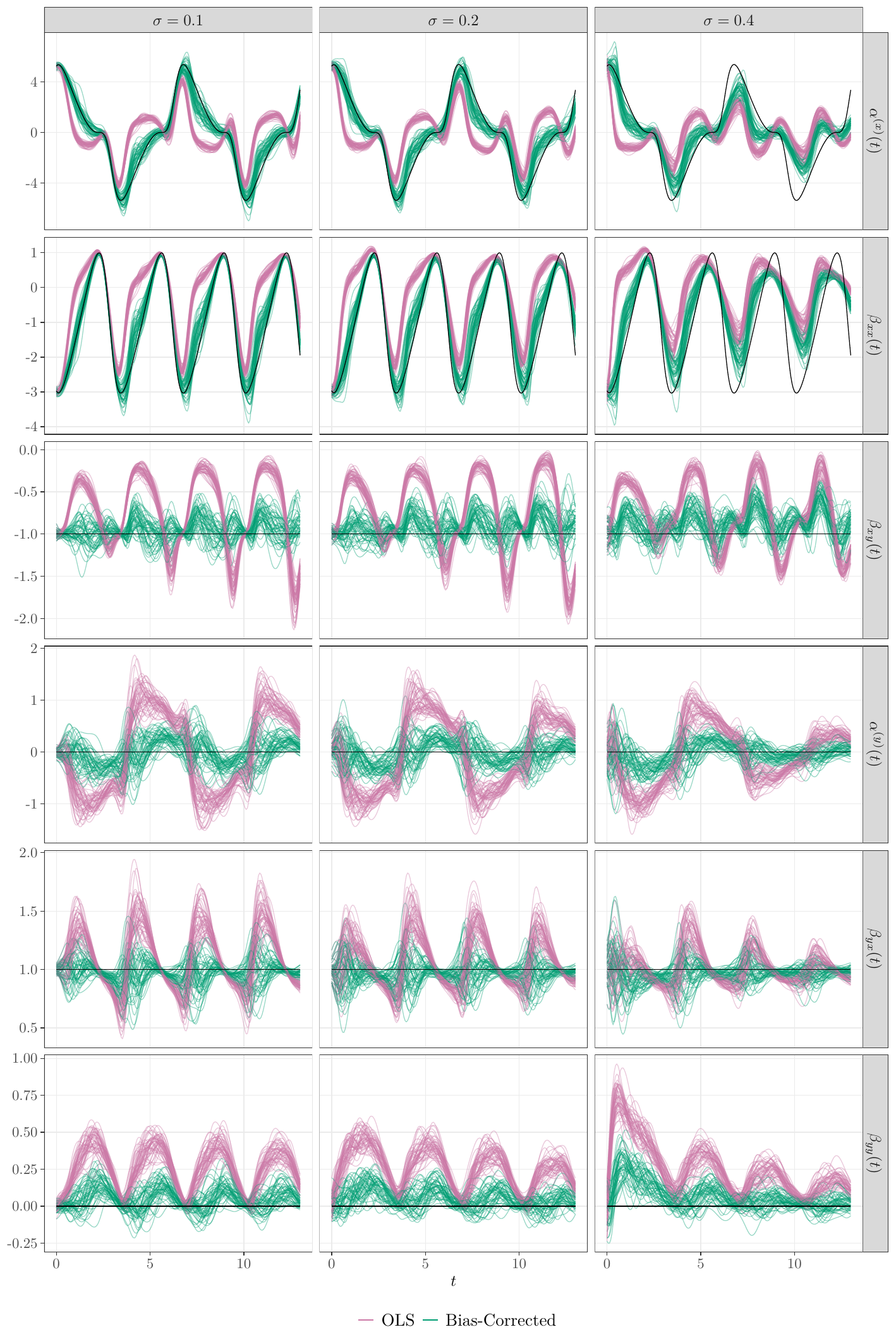}
    \caption{Results of the simulation varying the forcing amplitude $\sigma$. Each row corresponds to a different parameter and each column corresponds to a different value of $\sigma$.
    The parameters $\mu$ and $l$ are held at their baseline values of $1$ and $2$, respectively.
    The results from all 50 simulation replicates are shown, with the initial OLS estimates coloured pink and the bias-corrected estimates coloured dark green. The solid black lines indicate the true parameter values.}
    \label{fig:vary-sigma-results}
\end{figure}

\begin{figure}
    \centering
    \includegraphics[page=2, width = 0.45\textwidth]{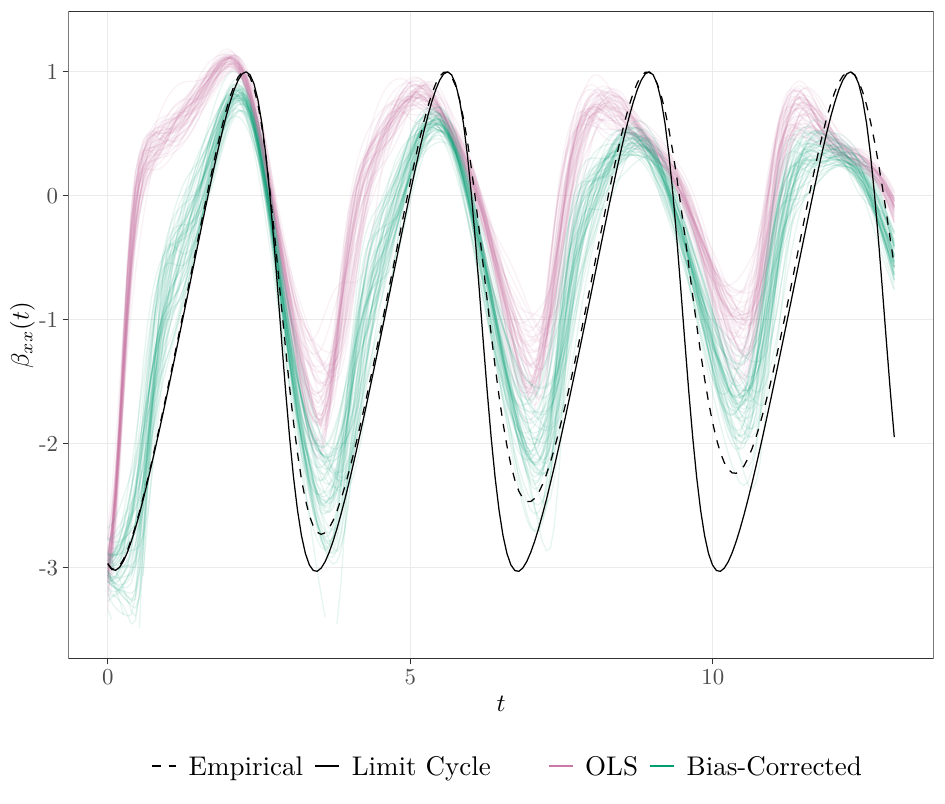}
    \caption{Estimation of the parameter $\beta_{xx}(t)$ in the simulation at $\sigma = 0.4$. The pink lines show the initial OLS estimates and the green lines show the bias-reduced estimates after applying 10 iterations of the bias-reduction algorithm. The solid black line shows the truth $\beta_{xx} (t) =  \mu (1 - x_0(t)^2)$ calculated using the $x$-coordinate of the limit-cycle trajectory as $x_0(t)$. The dashed black line uses the $x$-coordinate of the sample mean trajectory, calculated using \num{20000} samples as $x_0(t)$.}
    \label{fig:beta_xx_limit_cycle}
\end{figure}

Figure \ref{fig:vary-sigma-results} displays the results of the simulation. Each row corresponds to a different parameter and each column corresponds to a different value of $\sigma$. The results from all 50 simulation replicates are shown\footnote{Two of the 50 replicates failed to complete the full number of iterations of the bias-reduction algorithm at the setting $\sigma=0.4$ due to numerical errors. This was investigated further and it turned out to be a result of the grid approximations used to increase computational efficiency. We re-ran these simulation replicates with the standard non-optimised code successfully without any errors, so they are included in the results.}, with the initial OLS estimates coloured pink and the bias-corrected estimates coloured dark green. The solid black lines indicate the true parameter values. At $\sigma = 0.1$ and $\sigma = 0.2$, the bias-reduction algorithm appears to work well, i.e., bias-reduction step moves the OLS estimates towards being centered approximately at the true parameter values. However, estimation   appears to work less well for $\sigma = 0.4$. In addition, the increased forcing amplitude causes phase variation in the generated functions so that the mean function cannot be interpreted as the limit cycle trajectory, and hence comparisons with the true value of $\beta_{xx}(t)$ calculated based on the limit cycle trajectory are unreasonable (Figure \ref{fig:beta_xx_limit_cycle}). However, the bias-reduction algorithm still fails to reconstruct the true value of $\beta_{xx}(t)$ calculated using the empirical mean trajectory $(\Bar{x}(t), \Bar{y}(t))^\top$ as a reference operating trajectory (Figure \ref{fig:beta_xx_limit_cycle}, dashed line).
It can be concluded that the interpretation of PDA as estimating the Jacobian of this model and the performance of the bias-reduction algorithm are reasonable when the amplitude of the forcing is low to moderate, but are less so when the forcing amplitude is large.

\subsubsection{Forcing Lengthscale}\label{sec:vary-l}

\begin{figure}
    \centering
    \includegraphics[page=8, width = 0.9\textwidth]{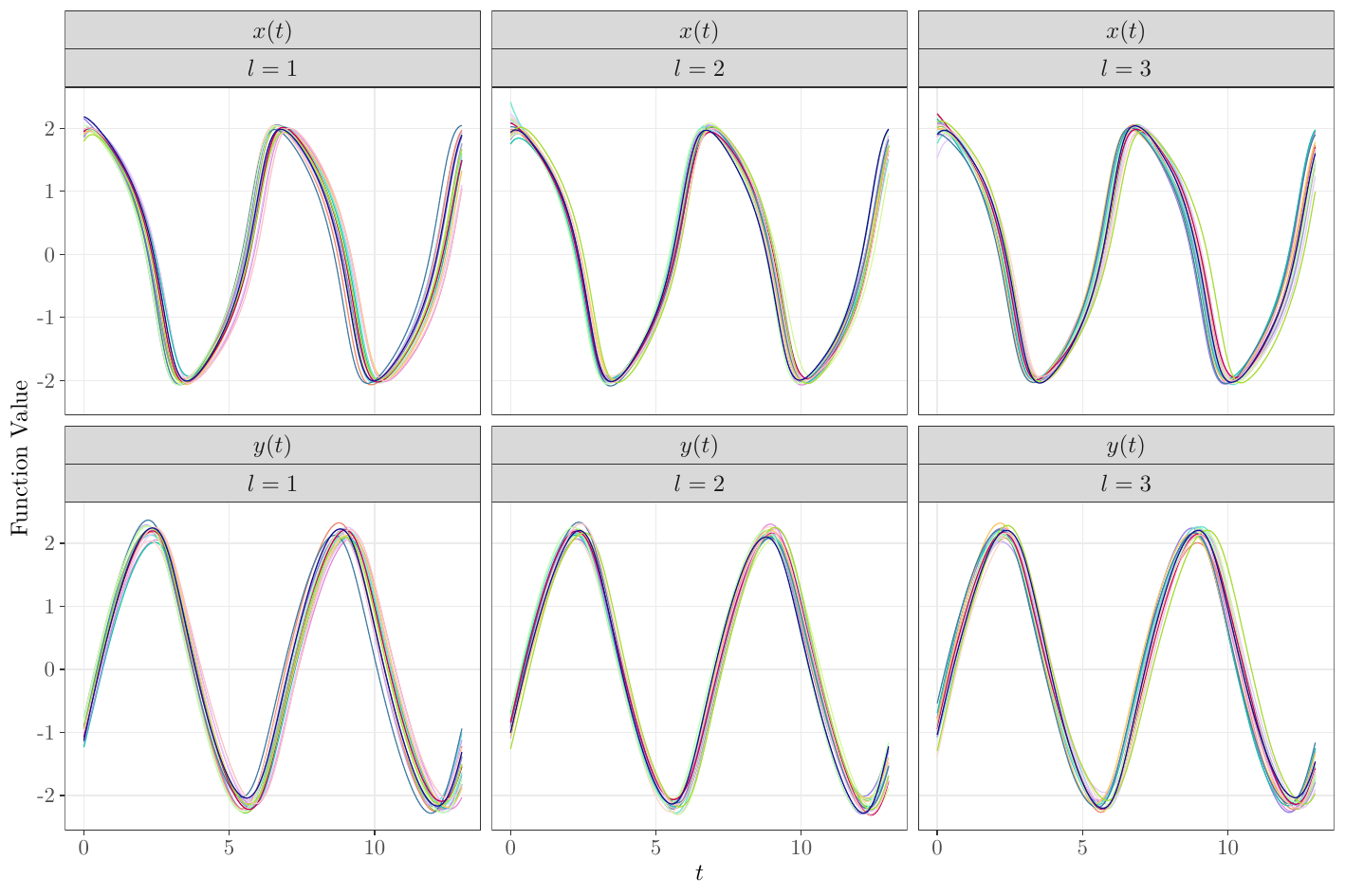}
    \caption{20 generated function pairs $(x_i(t), y_i(t))^\top$ at the lengthscale values $l = 1$ (small), $l = 2$ (medium, baseline) and $l = 3$ (large). The parameters $\mu$ and $\sigma$ are held at their baseline values of $1$ and $0.1$, respectively.}
    \label{fig:vary-l-example-dataset}
\end{figure}

\begin{figure}
    \centering
    \includegraphics[width = 0.8\textwidth]{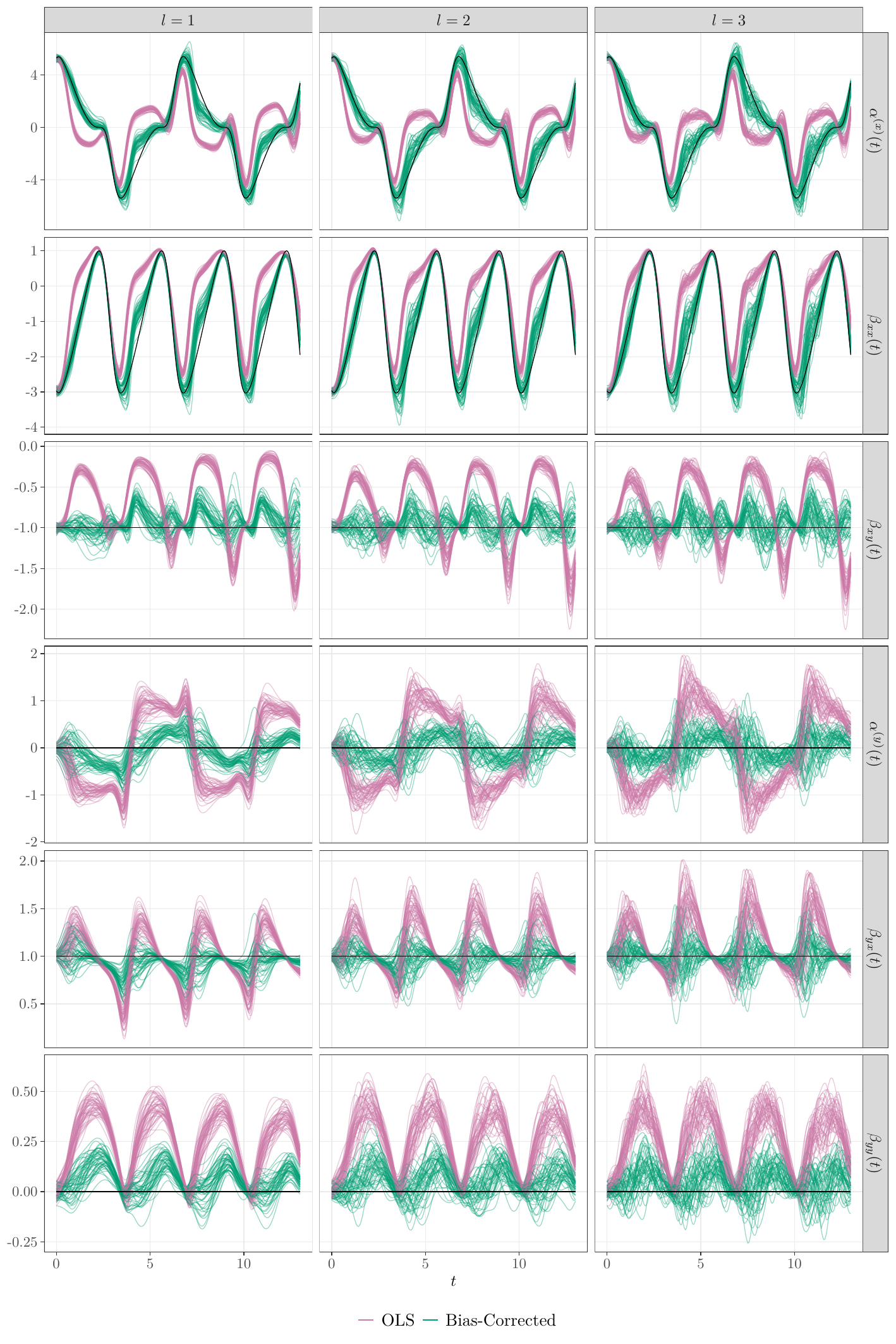}
    \caption{Results of the simulation varying the forcing lengthscale $l$. Each row corresponds to a different parameter and each column corresponds to a different value of $l$.
    The parameters $\mu$ and $\sigma$ are held at their baseline values of $1$ and $0.1$, respectively.
    The results from all 50 simulation replicates are shown, with the initial OLS estimates coloured pink and the bias-corrected estimates coloured dark green. The solid black lines indicate the true parameter values.}
    \label{fig:vary-l-results}
\end{figure}

Fixing the forcing amplitude at the baseline (low) level $\sigma = 0.1$ and fixing $\mu=1$, we investigate the effect of varying the lengthscale $l$ of the stochastic disturbance. Figure \ref{fig:vary-l-example-dataset} displays 20 generated function pairs at the lengthscale values $l = 1$ (small), $l = 2$ (medium, baseline) and $l = 3$ (large). Differences in the datasets among the levels of $l$ are less obvious than for the different levels of $\sigma$ in Figure \ref{fig:vary-sigma-example-dataset}. However, there does appear to be slightly more variation in the generated functions for $l = 1$. Our intuition for this effect is that smaller values of $l$ lead to smoother stochastic disturbances and hence allow longer periods of (or more ``sustained") forcing. The results in Figure \ref{fig:vary-l-results} suggest that varying the lengthscale parameter has a small effect on the parameter estimation (when $\mu$ and $\sigma$ are fixed at $1$ and $0.1$, respectively). In particular, the bias-reduction algorithm appears to work less well in reducing the bias in $\alpha^{(y)}(t)$ and $\beta_{yy}(t)$ for $l=1$ than it does for $l =2$ or $l = 3$.

\subsubsection{Comparison with a non-parametric approach}\label{sec:compare-nonpar}

\begin{figure}
    \centering
    \includegraphics[width = 0.915\textwidth]{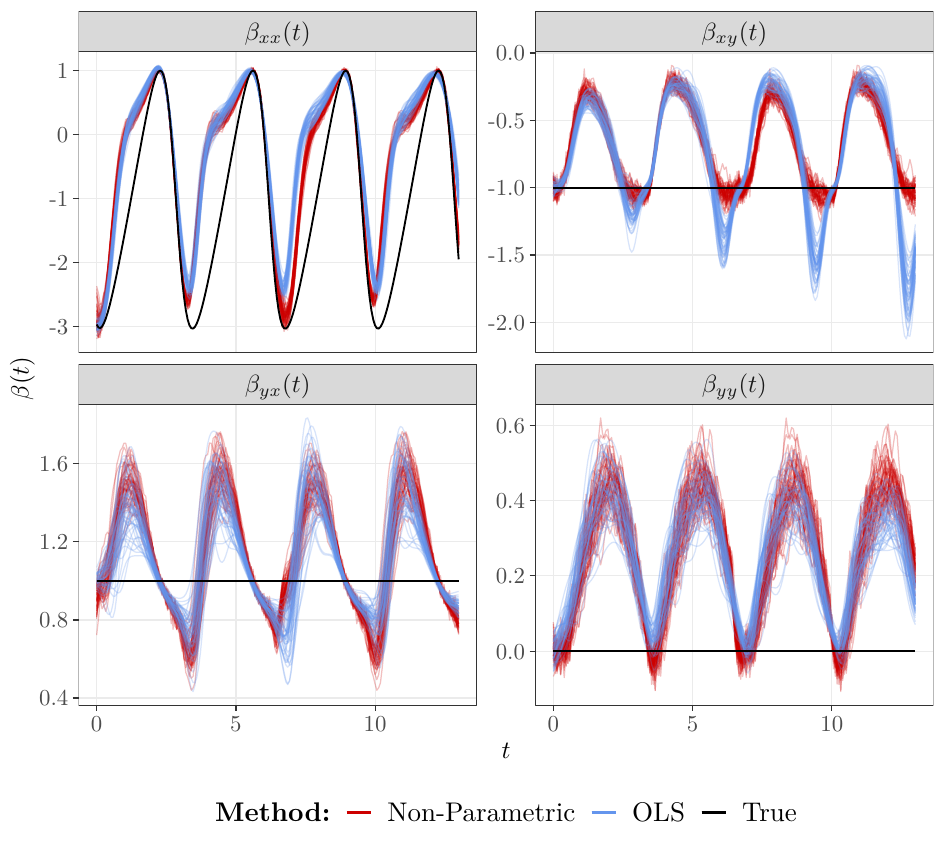}
    \caption{Results of a simulation to compare the initial biased PDA parameter estimates with the partial derivatives of time-invariant non-parametric estimates of $Dx(t)$ and $Dy(t)$. The baseline simulation values of $\mu=1$, $\sigma=0.1$ and $l=2$ were used to generate 50 simulated datasets from the Van der Pol model for the demonstration. 
    The blue lines represent the OLS estimates (i.e., the initial biased estimates from PDA). The red lines represent the partial derivatives of non-parametric estimates calculated around the mean $(x(t), y(t))^\top$ trajectory.
    The solid black lines indicate the true parameter values.}
    \label{fig:non-par-partials}
\end{figure}

As a short experiment, we compare the initial PDA parameter estimates with the results of modelling the full time-invariant system non-parametrically. We generate 50 simulated datasets from the two-dimensional Van der Pol model with the baseline simulation parameter values $\mu=1$, $\sigma=0.1$ and $l=2$. We then use the \texttt{pffr()} function in the \pkg{refund} \proglang{R} package \parencite{goldsmith_refund_2020} to model $Dx(t)$ and $Dy(t)$ as smooth non-parametric functions of $x(t)$ and $y(t)$ using a tensor product of $K=120$ B-spline basis functions. Given the fitted functions, we calculate the partial derivatives around the mean trajectory numerically to compare to our initial PDA estimates. The results are shown in Figure \ref{fig:non-par-partials}. The parameter estimates from this alternative approach to estimating the same model exhibit similar bias.

\section{Full Analysis of the Running Data}\label{sec:additional-running}

\subsection{Data Processing and Preparation}
The data were captured by a three-dimensional motion analysis system (Vantage, Vicon, Oxford, UK) and the kinematic parameters were extracted at a sampling rate of \si{200\hertz}.
To prepare the dataset for analysis, the long time series from the treadmill run was segmented into individual strides based on the event of touch down (i.e., the instant that the foot touches the ground).
The discrete kinematic measurements of the centre of mass (CoM) trajectory from each stride were converted to functions by expanding them on a quintic (sixth-order) B-spline basis.
The quintic basis was chosen because smooth estimates of acceleration were required. 
To avoid artefacts of time warping on derivative estimates, the strides were not time normalised.
Rather, the minimum collection of sampling points (or frames) common to all strides was used.
The strides from the same individual were similar in lengths, so this step simply amounted to discarding a small number of final frames for some strides.
The unit of measurement used for the time variable $t$ was $\text{seconds} \times 10$ (or equivalently $\text{millseconds} \times 10^{-2}$) so that domain lengths would be roughly comparable to those in the SHM and VdP examples.
We also centered the CoM trajectory around its overall average value, so that it can be understood as relative vertical displacement rather than absolute position.
The raw measurements, which were in millimetres (\si{\mm}), were converted to metres (\si{m}) for analysis.
The maximum possible number of basis functions was chosen to represent the data and the basis coefficients were estimated by penalised least squares, with a penalty on the integrated squared fourth derivative. 
The fourth-order penalty was chosen to ensure smoothness of the acceleration (i.e., second derivative) estimates. 
The roughness-penalty parameter, $\lambda$, was chosen by generalised cross-validation (GCV) \parencite[pp. 97-98]{ramsay_functional_2005}. 
Velocity and acceleration estimates were calculated directly from the basis representation of the CoM trajectory using the \texttt{deriv.fd()} function in the \pkg{fda} \proglang{R} package.

\subsection{Methodology}
We fitted the second-order PDA model
$$
D^2 x_i(t) = \alpha(t) + \beta_0 (t) x_i(t) + \beta_1 (t) Dx_i(t), \quad i = 1, \dots, N.
$$
We first obtained initial PDA estimates and then applied $10$ iterations of the bias-reduction algorithm. We included some regularisation to stabilise the calculations. On each iteration, we post-smoothed the estimated regression coefficient functions with the \texttt{gam()} function in the \pkg{mgcv} \proglang{R} package \parencite{wood_fast_2011}, using $50$ basis functions with the smoothness level chosen via restricted maximum likelihood (REML). We also regularised the residual covariance estimate used in the bias-reduction algorithm. We used the fast additive covariance estimator for functional data \parencite[FACE;][]{xiao_fast_2016} in the \pkg{refund} \proglang{R} package \parencite{goldsmith_refund_2020} with $50$ knots to estimate a smoothed covariance and then projected the residuals on to the first $R$ functional principal components that explained $99\%$ of the variance. We used the covariance function of these smoothed residuals in the computation of the bias. We also computed non-parametric estimates of the function $D^2 x_i(t) = g(x_i(t), Dx_i(t))$ using the \texttt{pffr()} function in the \pkg{refund} \proglang{R} package \parencite{goldsmith_refund_2020} and calculated the partial derivatives of $g$ numerically by finite differences. 
We compared these to the initial, rather than bias-reduced, PDA estimates because both should exhibit similar bias (as shown in Figure \ref{fig:non-par-partials}).

\subsection{Results}

\begin{figure}
    \centering
    \includegraphics[width = 0.75\textwidth]{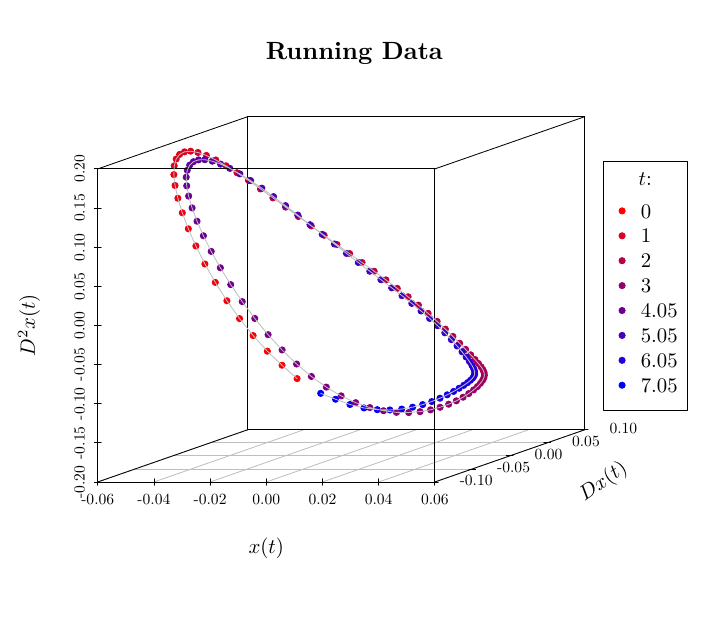}
    \caption{The mean of the functions $x(t)$, $Dx(t)$ and $D^2x(t)$ displayed on a three-dimensional phase-plane plot, where the points are coloured according to the time $t$.}
    \label{fig:subject-01-coloured-3d}
\end{figure}

\begin{figure}
    \centering
    \includegraphics[width = 0.85\textwidth]{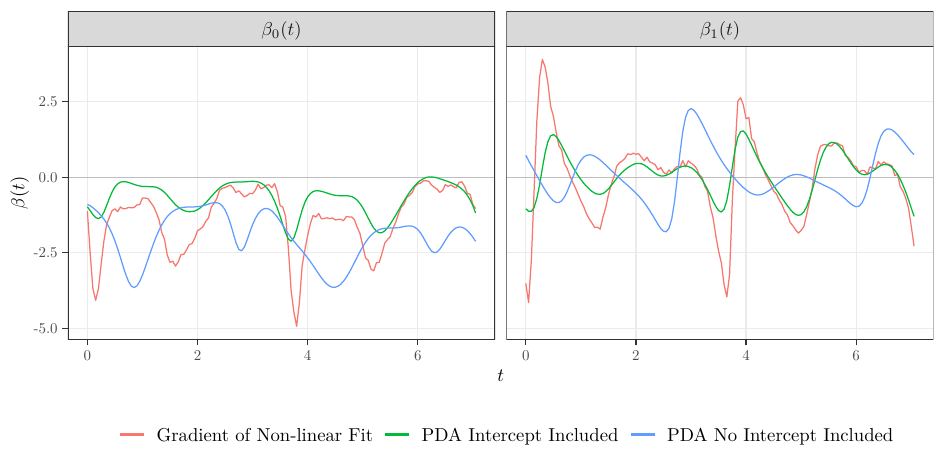}
    \caption{Comparison of the initial PDA estimates with the partial derivatives of a non-parametric fit. The initial parameter estimates from a model without an intercept are also shown for comparison.}
    \label{fig:pda-intercept-estimates}
\end{figure}

There were $82$ unique strides available for the single subject chosen for the analysis. 
The number of common sampling points for each curve was $142$ frames, so the time domain ranged from $0$ to $710$ milliseconds ($t \in [0, 7.1]$).
Figure \ref{fig:subject-01-coloured-3d} displays the mean trajectory for this subject on a three-dimensional phase-plane plot, where the points are coloured according to time $t$.
Figure \ref{fig:pda-intercept-estimates} displays a comparison of the initial PDA estimates of $\beta_0(t)$ and $\beta_1(t)$ with the partial derivatives of a non-parametric estimate. 
For comparison, initial parameter estimates from a model without an intercept are also displayed.
We see that the general shape of the non-parametric and initial estimates from our PDA model with an intercept are similar throughout, though the non-parametric estimates are rougher and spikier.
Both sets of estimates differ substantially from those of the PDA model without an intercept, where the timings of peaks and dips of the functions occur at different times.
The difference between the estimates from the models with and without an intercept re-enforces the importance of correctly formulating the data-generating model in PDA.

\begin{figure}
    \centering
    \includegraphics[width = 0.75\textwidth]{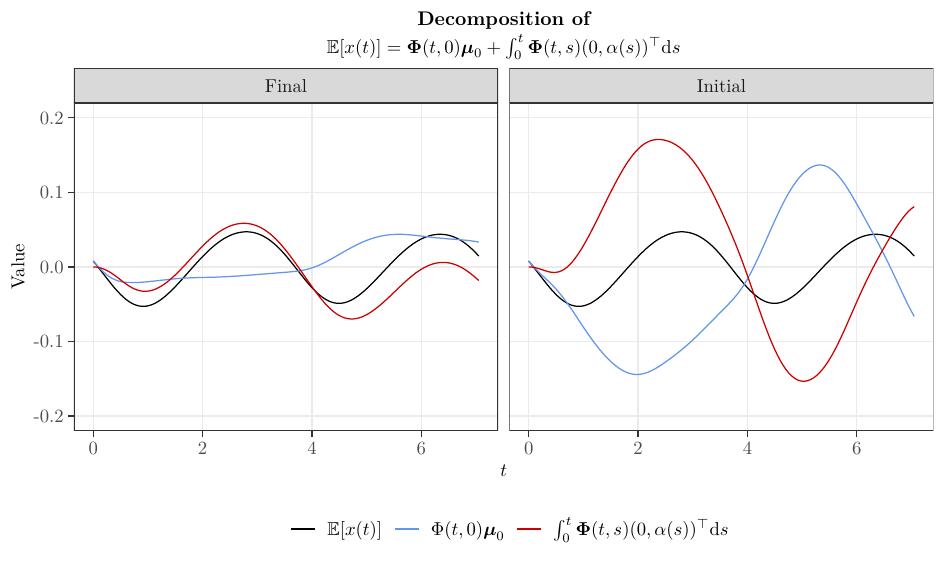}
    \caption{Decomposition of the mean function $\Expec[x(t)]$ (black line) into its two component parts (blue and red lines) based on the generative statistical model. The decomposition based the final bias-reduced parameters is contained in the left plot. For comparison, the decomposition based on the initial OLS parameter estimates is contained in the right plot.}
    \label{fig:subject-01-mean-decomposition}
\end{figure}

In addition to the PDA basis functions and variance decomposition presented in Section \ref{examples}, we also present a decomposition of the mean function.
From the generative model, we have
$$
\Expec \left[
\begin{pmatrix}
    x(t) \\
    Dx(t)
\end{pmatrix}
\right]
=
\boldsymbol{\Phi} (t, 0) \boldsymbol{\mu}_0 +
\int_0^t \boldsymbol{\Phi} (t, s) 
\begin{pmatrix}
    0 \\
    \alpha(s)
\end{pmatrix}
\mathrm{d}s,
$$
where $\boldsymbol{\mu}_0 = \Expec \left[(x(0), Dx(0))^\top\right]$.
Figure \ref{fig:subject-01-mean-decomposition} illustrates this decomposition graphically, based on the final bias-reduced parameters (left) and the initial parameters (right). The interpretation of this decomposition differs depending on the parameters used -- using the bias-reduced parameters emphasises the term $\int_0^t \boldsymbol{\Phi} (t, s) (0, \alpha(s))^\top \mathrm{d}s$ (red line) in reconstructing the mean function with only a small contribution due to initial conditions (blue line). In contrast, using the initial parameter estimates, the two terms oscillate in opposite directions to produce the mean function.

\begin{figure}
    \centering
    \includegraphics[width = 1\textwidth]{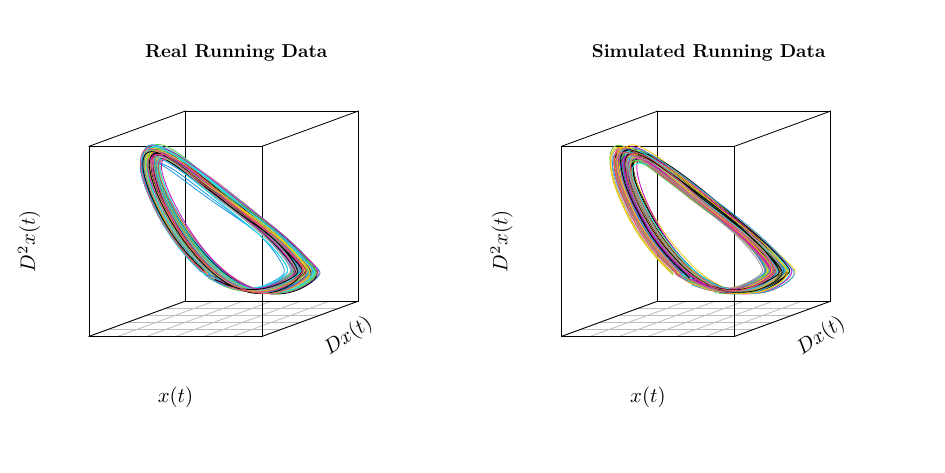}
    \caption{\textbf{(a)} The running data used in the PDA analysis. \textbf{(b)} A dataset obtained by simulating data from the PDA model fitted to the running data in \textbf{(a)}.}
    \label{fig:subject-01-real-vs-simulated}
\end{figure}

\begin{figure}
    \centering
    \includegraphics[width = 0.75\textwidth]{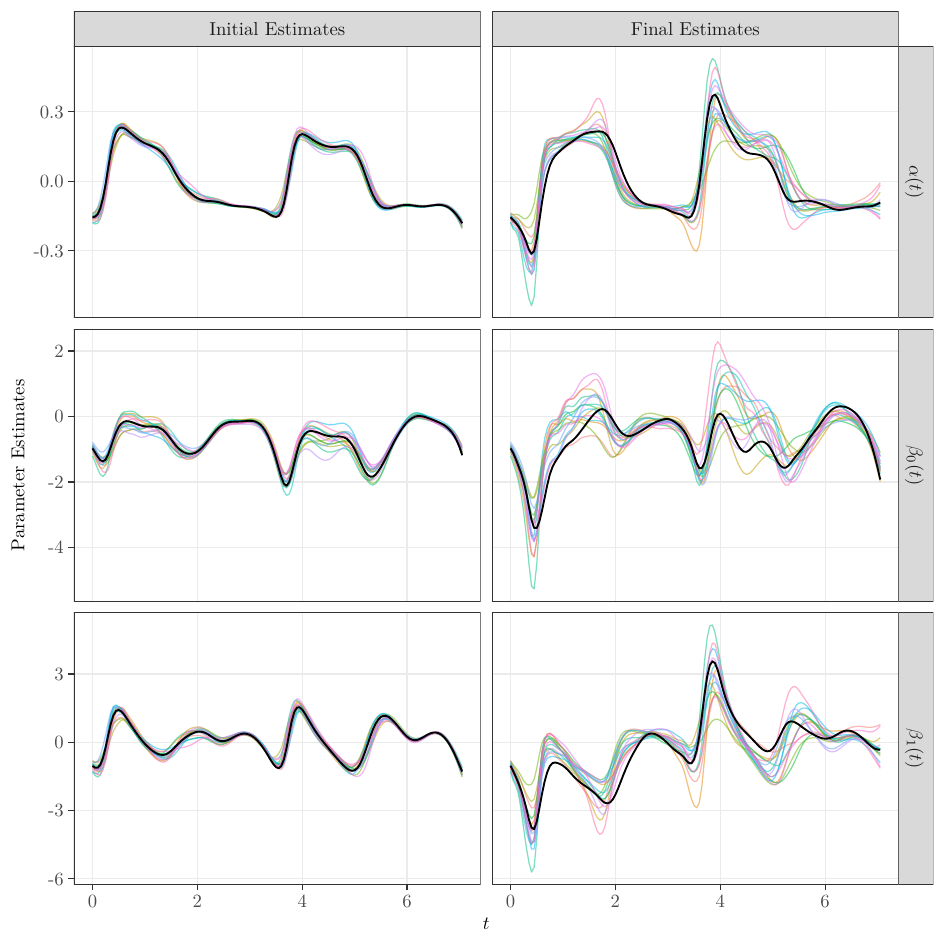}
    \caption{The results of the simulation using the bias-reduced parameters from the data application to generate replicate running datasets. 
    The left panel displays the initial OLS estimates from each replicate of the simulation as coloured lines, with the initial estimate from the real data analysis overlaid in black.
    The right panel displays the final bias-reduced estimates from each replicate of the simulation as coloured lines with the final bias-reduced estimate from the real data analysis overlaid in black.
    }
    \label{fig:subject-01-simulation-results}
\end{figure}

As final exploratory exercise, we use the final bias-reduced parameters and residual covariance function estimate to simulate data from our proposed data-generating model.
This allows us to address two questions:
\begin{enumerate}
    \item Do the estimated parameters produce datasets that resemble our observed data?
    \item How do parameter estimates from the simulated datasets (both initial and bias-reduced) compare to our empirical estimates? 
\end{enumerate}
The second question is similar to the idea of the parametric bootstrap, though it is less formal as we only inspect the parameter estimates qualitatively.
Figure \ref{fig:subject-01-real-vs-simulated} displays the real dataset and one simulated dataset of the same size on a three-dimensional phase-plane plot. The characteristics of the simulated data (e.g., shape and variability) closely resemble those of the real data. 
Figure \ref{fig:subject-01-simulation-results} displays the results of generating 15 such datasets and then estimating the parameters using $10$ iterations of the bias-reduction algorithm.
The left panel of Figure \ref{fig:subject-01-simulation-results} displays the initial estimates, which exhibit very little variation and track the initial estimate from the empirical dataset (black line) closely.
The bias-reduced estimates, displayed in the right column of Figure \ref{fig:subject-01-simulation-results}, exhibit greater variation, emphasising that applying the bias-reduction algorithm increases the variance of the parameter estimates.
The final bias-reduced estimates from the simulation track the final estimates from the empirical dataset (black line) closely, though there are some points at which the simulated estimates are not perfectly centered on the estimate from the real dataset.
Overall, however, we can conclude that the parameter estimates from the simulated datasets (both initial and bias-reduced) qualitatively resemble the empirical estimates.

\clearpage
\printbibliography

\end{document}